\documentclass{aa}  
\bibpunct{(}{)}{;}{a}{}{,} 
\usepackage{graphicx,amssymb,amsmath}

\usepackage{natbib}
\usepackage{verbatim} 
\usepackage{rotating}
\usepackage[table]{xcolor}
\usepackage{multicol}
\usepackage{multirow}
\usepackage{txfonts}
\usepackage{wasysym}
\usepackage{pifont}
\usepackage{color}
\usepackage{pxfonts}
\usepackage{ifsym}
\usepackage{latexsym}
 
\usepackage{xcolor}

\def\lsun{\hbox{$\rm ~L_{\odot}$}}
\def\msun{\hbox{$\rm ~M_{\odot}$}}

\def\dg{^{\circ}}

\def\H0{{\rm ~km~s^{-1}~Mpc^{-1}}}

\def\850{850$\mu$m}
\def\550{550$\mu$m}
\def\350{350$\mu$m}
\def\l850{L$_{\nu 850}$}

\def\.25{0.25 keV\thinspace}

\def\mh2{\hbox{M$_{H2}$}}

\def\mum{\hbox{$\mu$m}}

\def\h2{\hbox{H${2}$}}

\begin{document}

   \title{Merging galaxies in isolated environments}
   \subtitle{I. Multiband photometry, classification, stellar masses, and star formation rates}

   \author{Calder\'on-Castillo P.\inst{1}
          \and
          Nagar N. M.\inst{2}
          \and
          Yi S.K.\inst{3}
          \and
          Chang Y.-Y.\inst{4}
          \and
          Leiton R.\inst{5} 
          \and
          Hughes T. M. \inst{6,5,7,8}
          }
   \offprints{P. Calder\'on-Castillo}

   \institute{Departamento de Física, Universidad Técnica Federico Santa María, Avenida Vicuña Mackenna 3939, San Joaquín, Santiago de Chile \\ 
              \email{pau.astro.cc@gmail.com}
              \and Astronomy Department, Universidad de Concepci\'on, Casilla 160-C, Concepci\'on, Chile. 
             \and Department of Astronomy and Yonsei University Observatory, Yonsei University, Seoul 03722, Republic of Korea.
             \and Academia Sinica Institute of Astronomy and Astrophysics, No.1 Section 4 Roosevelt Rd., 11F of Astro-Math Building, Taipei 10617, Taiwan.  
             \and Instituto de Física y Astronomía, Universidad de Valparaíso, Avda. Gran Bretaña 1111, Valparaiso, Chile. 
             \and Chinese Academy of Sciences South America Center for Astronomy, China-Chile Joint Center for Astronomy, Camino El Observatorio \#1515, Las Condes, Santiago, Chile.
             \and CAS Key Laboratory for Research in Galaxies and Cosmology, Department of Astronomy, University of Science and Technology of China, Hefei 230026, China.
             \and School of Astronomy and Space Science, University of Science and Technology of China, Hefei 230026, China.
     }    
   \date{Received ; Accepted }

  \abstract
   {Extragalactic surveys provide significant statistical data for the study of crucial galaxy parameters 
   (e.g. stellar mass, M$_*$, and star formation rate ,SFR) used to constrain galaxy evolution under 
   different environmental conditions. These quantities are derived using manual or automatic methods for 
   galaxy detection and flux measurement in imaging data at different wavelengths. The reliability of these 
   automatic measurements, however, is subject to mis-identification and poor fitting due to the 
   morphological irregularities present in resolved nearby galaxies (e.g. clumps, tidal disturbances, star-
   forming regions) and its environment (galaxies in overlap).    
}
   {Our aim is to provide accurate multi-wavelength photometry (from the UV to the IR, including 
   GALEX, SDSS, and WISE) in a sample of $\sim 600$ nearby (z<0.1) isolated mergers, as well as 
   estimations of M$_*$ and SFR.
   }
   {We performed photometry following a semi-automated approach using SExtractor, confirming by visual 
   inspection that we successfully extracted the light from the entire galaxy, including tidal tails and 
   star-forming regions. 
   We  used the available SED fitting code MAGPHYS in order to estimate M$_*$ and SFR.
   }
   {We provide the first catalogue of isolated merging galaxies of galaxy mergers including 
   aperture-corrected photometry in 11 bands  (FUV, NUV, u, g, r, i, z, W1, W2, W3, and W4), 
   morphological classification, merging stage, M$_*$, and SFR. We found that SFR and M$_*$ 
   derived from automated catalogues can be wrong by up to three orders of magnitude as a result of 
   incorrect photometry.
} 
   {Contrary to previous methods, our semi-automated method can reliably  extract the flux of a 
   merging system completely. Even when the SED fitting often smooths out some of the differences in the 
   photometry, caution using automatic photometry is suggested as these measurements can lead to large 
   differences in M$_*$ and SFR estimations.
  }

   \keywords{galaxies: photometry - galaxies: star formation - method: SED fitting }
   \maketitle

\footnote{Full catalogue (PCCMC) is only available in electronic form at the CDS via anonymous ftp to cdsarc.u-strasbg.fr (130.79.128.5) or via http://cdsweb.u-strasbg.fr/cgi-bin/qcat?J/A+A/.}
%

\section{Introduction}

 Galaxies evolve through time, increasing their stellar mass (M$_*$) by forming stars at distinct rates 
 (SFR) according to their internal properties (e.g. mass, metallicity, gas and dust content) and the 
 environments they inhabit. The majority of star-forming galaxies appear to follow a relation 
 in the SFR-M$_*$ plane, called the main sequence (MS) of star formation \citep{Elbazetal07, Noeskeetal07, Daddietal07, Pannellaetal09, Pannellaetal15, Karimetal11, Schreiberetal15, Schreiberetal17, Wuytsetal11, Rodighieroetal14, Whitakeretal12, Whitakeretal14, RenzininPeng15}. The location of the MS evolves with redshift \citep{Noeskeetal07, Wuytsetal11, Whitakeretal12, Schreiberetal15} 
 in a manner consistent with secular star formation. At all 
 redshifts and values of  M$_*$, a small fraction of galaxies (known as  the 
starburst galaxies) show an increased SFR for a given M$_*$ with 
 respect to the MS \citep{Rodighieroetal11, Schreiberetal15} where the 
 majority of these objects show features typically associated with mergers (e.g. \citealt{Kartaltepeetal07}).

Observations and simulations have shown that merging galaxies can enhance their SFR while 
undergoing mergers \citep{LarsonnTinsley78, Parketal17}. 
As galaxies approach and pass through each other, tidal tails 
and bridges are produced by tidal forces and torques. As the gas loses momentum, it falls into the galaxy 
centre, where it forms new stars and also feeds, and activates, the super-massive black hole (SMBH). 
These processes occur throughout the entire merging process, and how they evolve depends on many 
parameters such as the ratio of the M$_*$ of the system components, initial gas and dust content, 
 morphologies, and also the collisional parameters of their orbits. According to simulations, galaxies in mergers 
may form stars $\sim$10-20 times faster than isolated galaxies on the MS, 
depending on the bulge-to-total luminosity ratio \citep{MihosnHernquist94a, MihosnHernquist94b}, 
the M$_*$ ratio \citep{DiMatteoetal08, Hopkinsetal08}, and the interacting galaxies orbits 
\citep{DiMatteoetal05}.  
At high redshifts, simulations show a mild enhancement in SFR depending on the gas fraction 
of the merging galaxies \citep{Fenschetal17}.

As  observed and derived quantities are critical in order to describe the evolution of a galaxy, 
it is important to determine these parameters in the most accurate, uniform, and automated way possible 
in order to obtain statistically significant samples.  
The accuracy 
and reliability of the resulting galaxy properties strongly depend on the ability of both automated and manual 
methods to correctly find, identify, and measure the whole light from a specific target, and to properly 
exclude the contribution from close neighbours. 
Large extragalactic surveys at different wavelengths have provided us with the photometric and morphological 
data to build various catalogues  of galaxy properties. For example, the Max Planck for Astrophysics and Johns Hopkins University 
(MPA-JHU\footnote{\url{https://www.sdss3.org/dr10/spectro/galaxy_mpajhu.php}}) catalogue, the 
NASA Sloan Atlas (NSA\footnote{\url{http://www.nsatlas.org/}}), and 
\citet[hereafter \textsc{Chang15}\footnote{\url{ http://www.asiaa.sinica.edu.tw/~yychang/sw.html}}]{Changetal15}, each use 
different methods to determine a galaxy's M$_*$ and SFR, basing their results on 
available survey data such as the Galaxy Evolution Explorer (GALEX), Sloan Digital Sky Survey (SDSS), 
and/or  Wide-field Infrared Survey Explorer (WISE), using imaging and/or spectroscopy 
 obtained for each galaxy in an automated and uniform manner. 

Automatic photometry methods applied on resolved galaxies must overcome difficulties related to the 
identification and measurement of targets due to the varying morphology of galaxies and the presence of 
nearby neighbours. Merging galaxies are a very challenging case for the automatic approach
since their complex morphologies are difficult to fit in a single aperture. Merger morphologies are perturbed 
(e.g. faint tidal features with non-smooth profiles) and show a distribution of star-forming regions that can 
be incorrectly identified as separate individual objects, removing their contribution from the total photometric 
budget of the merging system. The catalogues mentioned  have been 
not optimised for mergers, but for statistical studies. Since mergers are rare, it is not surprising that issues 
like the ones we find have not been  corrected.

To reduce the uncertainties derived from automatic photometry and classification, in this work we  apply a 
semi-automatic approach to visually classify a sample of nearby isolated mergers. 
We  performed the photometry by summing, after visual inspection, the light of each luminous 
component of a galaxy into its final magnitude. 

To study how different properties of merging galaxies are affected by the merging process, 
we  classified (by visual inspection) the mergers following criteria designed to distinguish 
galaxies according to the stage of merging. This is somewhat different to the 
classifications made by the Galaxy Zoo project \citep{Dargetal10} and by  \citet{Ellisonetal08, Ellisonetal11, Ellisonetal13}, where the  mergers were classified by separation. 
The classification by separation results in a greater mixing of galaxies at different stages through the 
merging process since the galaxies approach and separate many times before coalescing. 

To focus only on the effects that galaxies undergo through the merging process, we have excluded 
systems within groups (three or more components) and clusters. In this way we hope to significantly reduce environmental effects from the larger scale environment and  to focus on those effects induced by the binary interaction alone. 

Our sample is a compilation of $\sim$600 nearby ($z < 0.1$) isolated mergers, 
classified by morphology and merging stage, with 
available imaging from the GALEX, SDSS, and WISE surveys. We  chose these surveys as they span a wide range of wavelengths, 
tracing young stars (GALEX), old stars (SDSS), and the obscuration of young stars by dust (WISE). This allowed us 
to determine M$_*$ and SFR more accurately. 
In order to use this data to estimate the M$_*$ and the SFR of each merger, 
we used the spectral energy distribution (SED) fitting code MAGPHYS \citep{daCunhaetal2008}. We  also 
visually classified the sample by morphology and merging stage 
in order  to study the time evolving impact of the merging process on different galaxies SFR, and 
the activation of their SMBHs. We eventually hope to determine whether merging processes significantly increase 
the M$_*$  of galaxies and, if so, at what stage of the merger sequence this is most significant. 
This part of the study will be presented in an upcoming publication (Calder\'on-Castillo et al., in prep, 
hereafter Paper II).

In Sect. \ref{sec:samplendata} we present our sample selection and imaging/data analysis. 
We show the merger classification and the procedures to obtain M$_*$ and SFR  in Sect. \ref{sec:results}. 
In this section we also show the comparison of our M$_*$ and SFR results with results and estimators 
found in the literature. We show how mergers separate in the WISE colour-morphology relation. 
We also show the specific star formation rate--WISE colour correlation for mergers.  
In section \ref{sec:catalogue} we present the public catalogue of isolated merging galaxies. 
In section \ref{sec:discussion} we  discuss the accuracy of  our results compared with existing public 
catalogues, and potential uses for the data considering the associated biases and errors.
Finally, we present examples of our morphology classification, SED fit accuracy, and additional correlations between different properties (such as morphology, merging stage, and separation)  in the Appendix.

As a matter of notation, we use the word `merger' to denote a merging system of galaxies that includes one or more 
individual merging galaxies. Throughout this paper we adopt a flat cosmology with $\rm \Omega_m=0.3$ and 
$\rm H_0=72 ~km~s^{-1}Mpc^{-1}$.

\section{Sample and data}    \label{sec:samplendata}

\subsection{Sample selection} \label{sec:sampleselection}

Our sample was drawn primarily from five large samples: 
the Arp Catalog of Peculiar Galaxies\footnote{\url{http://arpgalaxy.com}} \citep{Arp96} (Arp 1966; ARP Galaxies) 
containing 338 peculiar galaxies;
The VV\footnote{\url{www.sai.msu.su/sn/vv}} Catalogue of Interacting Galaxies 
\citep{VVetal01} with 852 interacting systems; 
the mergers classified by Neil Nagar (co-author, private communication) containing 81 mergers with 
submillimetre and gas information; 
the mergers classified by citizen scientists in the Galaxy Zoo (GZ) 
Project\footnote{\url{http://data.galaxyzoo.org}} \citep[GZ mergers]{Holinchecketal16}  listing 
3373 mergers; 
and the mergers selected from the Great Observatories All-sky LIRG 
Survey\footnote{\url{Http://goals.ipac.caltech.edu}} \citep[GOALS]{Sandersetal03} with  629 (U)LIRGs. 

From these $\sim$4000 galaxies, 
we  selected $\sim$600 mergers, counting a merger only once even when it appears 
in multiple catalogues, obeying the following criteria:

- if the system contains two galaxies, both components must show a difference in redshift $\Delta z$ < 0.002 
($\Delta \rm v_{rel} < 500~ km/s$), which  excludes most fly-bys or unrelated galaxies, and selects  systems that are more likely to eventually coalesce;  

- the merger is not part of either a group or a cluster, and thus can be considered  an isolated binary system;

- in order to have well-sampled SEDs,  the selected mergers must have full photometric
coverage from imaging: 
far-UV and near-UV from GALEX; u, g, r, i, and z from SDSS; and W1, W2, W3, and W4 from WISE.

\begin{figure}[!htb]
\begin{center}
    \includegraphics[width=0.45\textwidth,clip]{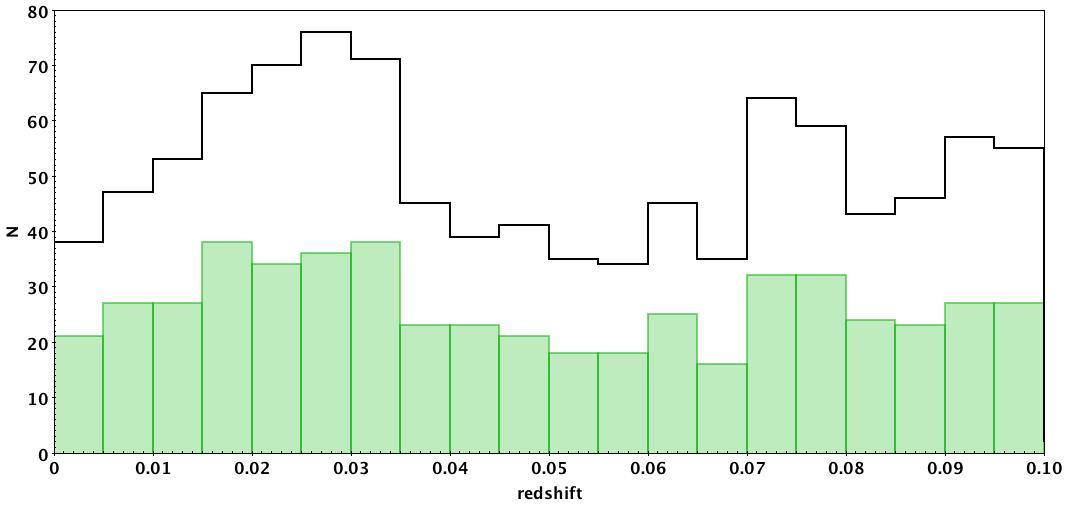}

  \caption{ Redshift distribution of the 540 merging systems (green filled histogram)  and 
    for all 919 individual galaxies in the merging systems (black histogram).  
}  
\label{fig:RedsfhitDistribution}
\end{center}
\end{figure}

\begin{figure}[!htb]
\begin{center}
  \includegraphics[width=0.45\textwidth,clip]{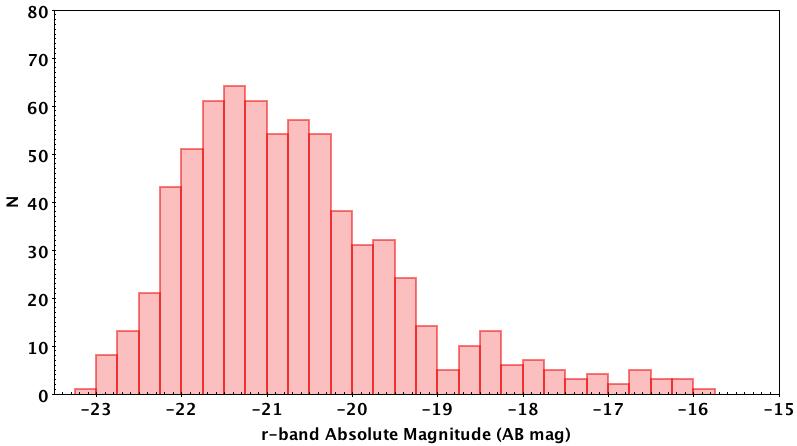}
  \caption{  SDSS r-band absolute magnitude distribution of all individual galaxies in the final sample.
}  
\label{fig:rbandDistribution}
\end{center}
\end{figure}

The final sample contains 919 galaxies in 540 mergers. Since our galaxies are found in the SDSS, 
the majority of the sample is at $z<0.1$, with a median value of $z=0.044\pm0.029$. 
In Fig. \ref{fig:RedsfhitDistribution} the redshift distribution of mergers is shown in green and for individual galaxies in black. 
The absolute-magnitude distribution of individual galaxies, shown in Fig. \ref{fig:rbandDistribution}, 
has a median of $-21.25\pm1.35 ~ \rm mag$ in the r band (SDSS). 
   
Since the mergers were classified based primarily on the SDSS imaging, where the sensitivity is 24 
$mag/arcsec^2$ in the r band, we may miss merging features fainter than this value. Therefore inevitably, 
we may miss mergers at early and late merging stages, overaccounting for merging 
systems at intermediate stages. 

A considerable bias of the GZ project is that the images they provide to the citizens for  visual inspection 
are relatively small, hence merging pairs with large separations are missed. 
Another bias that can be introduced in the GZ classification and  in our merging stage criteria is that some of the 
late merging stages containing only one galaxy could be  old interactions that did not result in 
a merger, where the secondary galaxy which passed by is too far away to be considered  a companion.

\subsection{Imaging data} \label{sec:imagingdata}

We compiled data from several surveys spanning the UV, optical, and IR wavelengths. 
We gathered fully reduced imaging for the FUV and NUV from 
GALEX\footnote{\url{http://galex.stsci.edu/data/}} (GR6/GR7), obtaining images of $\rm 1. 2 \dg$ 
(1450 pixels) in radius. 
For SDSS\footnote{\url{https://dr13.sdss.org/sas/dr13/}} (DR13), we obtained images for the following optical bands: 
u, g, r, i, and z  (10x13 $\rm arcmin^2$, which corresponds to 2048x1489 $\rm pixel^2$). 
Finally, we used $\rm 18.3x18.3 arcmin^2$ (800x800 $\rm pixel^2$) images for W1, W2, W3, and W4 
from WISE\footnote{\url{http://irsa.ipac.caltech.edu/applications/wise/}}. 

\section{Results} \label{sec:results}
In this section we  present our semi-automated photometry approach, the complications that arise from  automated photometry, and possible 
solutions. We  show the results of our photometry and compare them with available catalogues. 
Finally, we  compare our measurements of M$_*$ and SFR with those in the catalogues, and discuss the possible biases. 

\subsection{Galaxy and merger sequence classification} \label{sec:class}

\begin{figure}[!htb]
\begin{center}
   \includegraphics[width=0.45\textwidth,clip]{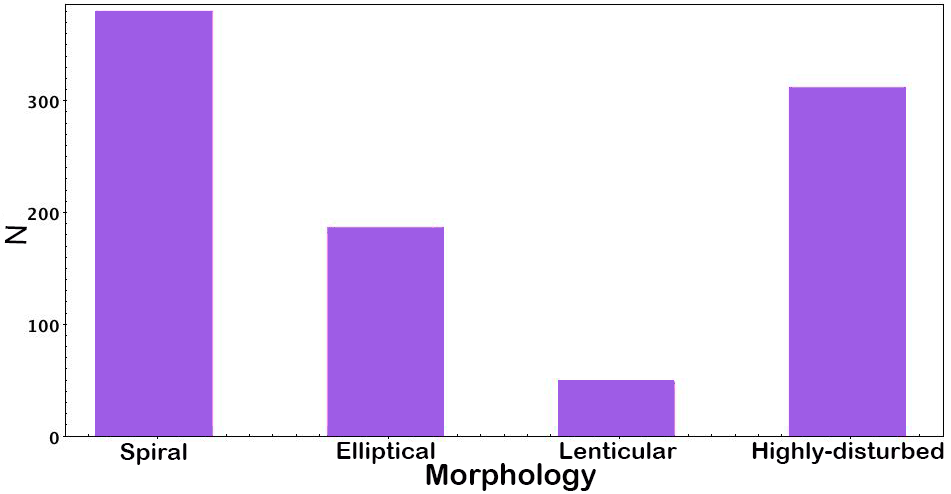}

  \caption{ Distribution of galaxy morphology for all individual galaxies in the sample: 
  classifications are spiral, elliptical, lenticular (S0), and highly disturbed (i.e. galaxies that are too disturbed to be included in the other classifications).
}  
\label{fig:morph_hist}
\end{center}
\end{figure}

\begin{figure*}[!htb]
\begin{center}
   \includegraphics[width=1.02\textwidth]{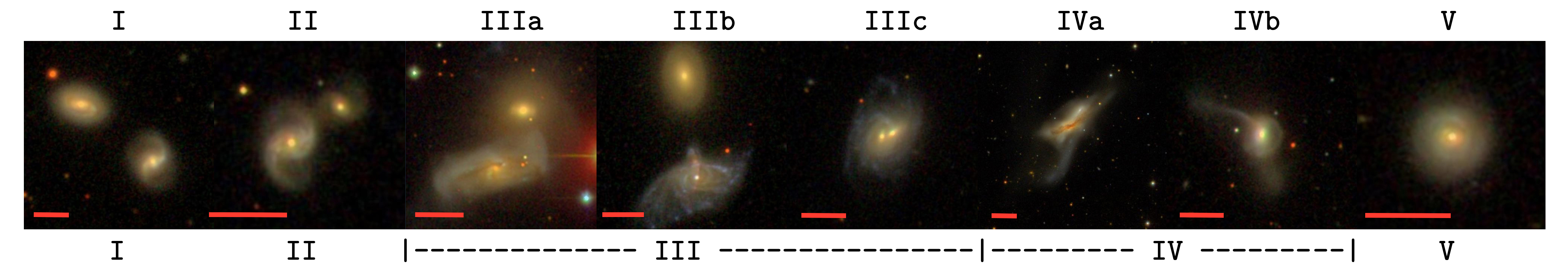}

  \caption{ SDSS images of an example merger from each merging stage defined in this study (top), 
  and defined in V02 (bottom). Red lines represent 20" in each image. 
}  
\label{fig:mrgseq}
\end{center}
\end{figure*}

The 540 mergers of the final sample were classified by morphology and merging stage 
based on a visual inspection of the SDSS images. The morphological classes of the 
individual galaxies in the systems are 
spiral, elliptical, lenticular (S0), and highly disturbed. The last classification includes all galaxies that are 
too disturbed to be clearly included in any of the other three classifications. An example of each classification 
is shown in Fig. \ref{fig:Morphs}. 
Figure \ref{fig:morph_hist} shows the distribution of the morphology of the merging galaxies. 
A large fraction of the sample (34\%) is highly disturbed. Most of these galaxies 
are likely to have been identified as spirals in the past as they presently show very disturbed tidal tails and 
nuclear regions. Other galaxies show shells, features often associated with mergers involving elliptical galaxies.

Some studies classify mergers based on their separation 
\citep{Ellisonetal08, Ellisonetal11, Ellisonetal13, Dargetal10}. 
However, applying this criterion based on component 
separation does not necessarily imply that mergers are ordered in the correct merging time considering 
that the distance between the merger components depends on the mass ratio of the components and 
the particular orbital parameters of each merging system (e.g. relative speeds).

\begin{figure}[h]
\begin{center}
 \includegraphics[bb=30 10 1330 650,width=0.5\textwidth,clip]{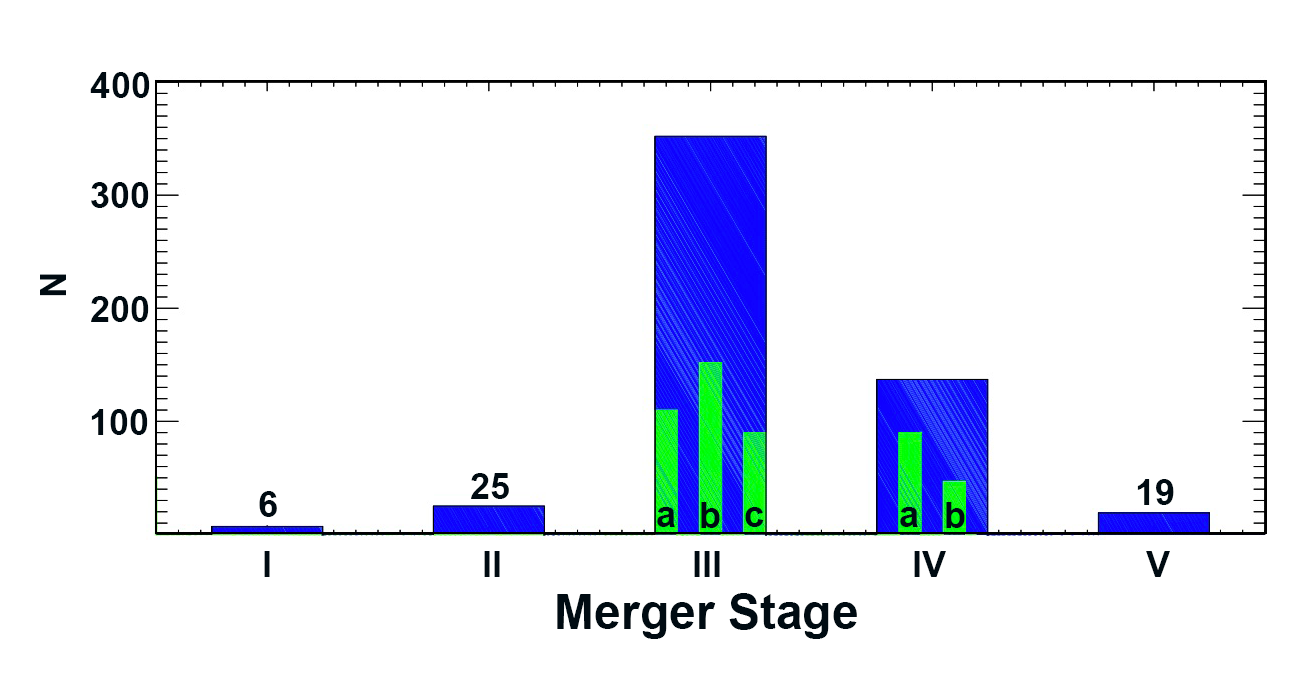}

  \caption{ Distribution of merger stages following the classification in V02 (blue), 
  and our own classification (blue plus green). 
}  
\label{fig:mrgstg_histo}
\end{center}
\end{figure}

We thus based our classification on a more timeline-like merging sequence. 
Firstly, we applied the classification prescription in \citet[hereafter V02]{Veilleuxetal02}, 
and then  developed a subclassification using a new criterion  defined in this work. 
\citet{Veilleuxetal02} 
separate the merging sequence into five merging stages: I (\textit{First Approach}), 
where the two galaxies are clearly separated, but are on course to collide. Here we added the additional constraint that the velocity separation is 
$\Delta z<0.002$ ($\Delta v<500 km/s$); II (\textit{First Contact}), where the galaxies are overlapping, 
but show no clear signs of disturbance in their morphology; III (\textit{Pre-Merger}), where galaxies show 
strong tidal tails, bridges, and/or shells, but there are still two nuclei clearly observed; 
IV (\textit{Merger}), where there is only one nucleus visible (diffuse or compact) and the resulting galaxy 
shows a very disturbed morphology; and V (\textit{Old Merger}), where there is only one galaxy with no visible 
tidal tails, but it shows a disturbed central morphology. Figure \ref{fig:mrgseq} shows the V02 
classification scheme for a complete merging sequence.

Our new merging sequence is based on the  V02 sequence with additional separations tracing 
a more detailed timeline of the merging process allowing us to explore possible 
dependences in more detail. We separate the merging stage III into three  substages: 
IIIa (\textit{overlap}), where the two galaxies overlap and show disturbances; 
IIIb (\textit{disturbed}), where the two galaxies 
show strong tidal tails, bridges, and/or shells, but they are clearly separated (not overlapping); 
IIIc (\textit{double-nucleus}), and an intermediate stage between  IIIb and IV, where only one galaxy is observed as highly perturbed and shows two clear 
nuclei. 
We also separated the merging stage IV into two different stages following the V02 description, 
but in a more visual way since our galaxies are all in the nearby Universe. 
Merging stage IVa (\textit{diffuse-nucleus}) shows a diffuse centre, and IVb (\textit{compact-nucleus}) 
shows a compact nucleus. 
Figure \ref{fig:mrgseq} shows an example of our classification.

Figure \ref{fig:mrgstg_histo} shows the distribution of merging stages defined by V02 (blue) 
with our additional definitions (green inserts). Most of the mergers are classified as merging stage III;  
this is the easiest stage in which to detect a merger because the merger features can be seen most clearly 
at this stage.

\subsection{Galaxy photometry: problems and solutions} \label{sec:photo}

\begin{figure}[h]
\begin{center}
  \includegraphics[bb=0 0 430 350,width=0.45\textwidth,clip]{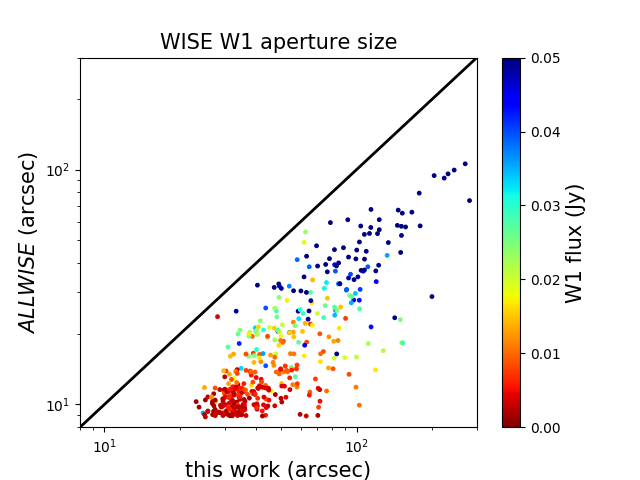}
  \caption{ Comparison of our aperture radius value,  required to measure all of a galaxy's light, in the WISE W1 images ({\it x}-axis; see text for details) with the $rsemi$ of the AllWISE catalogue. 
  The line of equality is shown with a solid black line and the galaxy symbols are colour-coded according to the total W1 flux (in Jy).
}  
\label{fig:rad_comp2}
\end{center}
\end{figure}

When visualising the images using the WISE interactive website
\footnote{https://irsa.ipac.caltech.edu/applications/wise/}, it was immediately clear that the galaxies 
were often much larger than the radii listed in the WISE website tables. 
As a first check, we measured the sizes of the mergers using the measuring tool available in the 
interactive WISE website. 
Figure \ref{fig:rad_comp2} compares the radii we measured using the interactive WISE website for W1 
with the semi-major axis ($rsemi$) tabulated in the AllWISE extended sources catalogue. Their $rsemi$ are 
always heavily underestimated.
These smaller radius measurements are likely linked to the low sensitivity of 2MASS, which was used 
to estimate the apertures for the AllWISE catalogue. This immediately shows that the  photometry of mergers listed in  the AllWISE catalogue   is not accurate, and that the total luminosity is underestimated. For this reason, we decided to perform our own photometry  across all the filters we use, not just on the WISE imaging.

Another  difficulty arising in mergers is that they can be so disturbed 
that the usual parameters used when making automated photometric measurements do not extract all the light from the galaxy. 
Disturbed morphologies, faint tidal tails, and luminous star-forming regions all have to be taken into account 
when measuring a merging galaxy's luminosity. Also, since the mergers do not show the same features along the 
merging sequence, or often with each other, there is no unique set of parameters that can be used to perform 
the photometry automatically from system to system. Therefore, we are forced to perform the photometry for all 
the systems, and in all bands, almost completely manually. However, to aid efficiency, we  developed a 
semi-automatic procedure that we   describe here.

\begin{figure}[h]
\begin{center}
  \includegraphics[width=0.4\textwidth,clip]{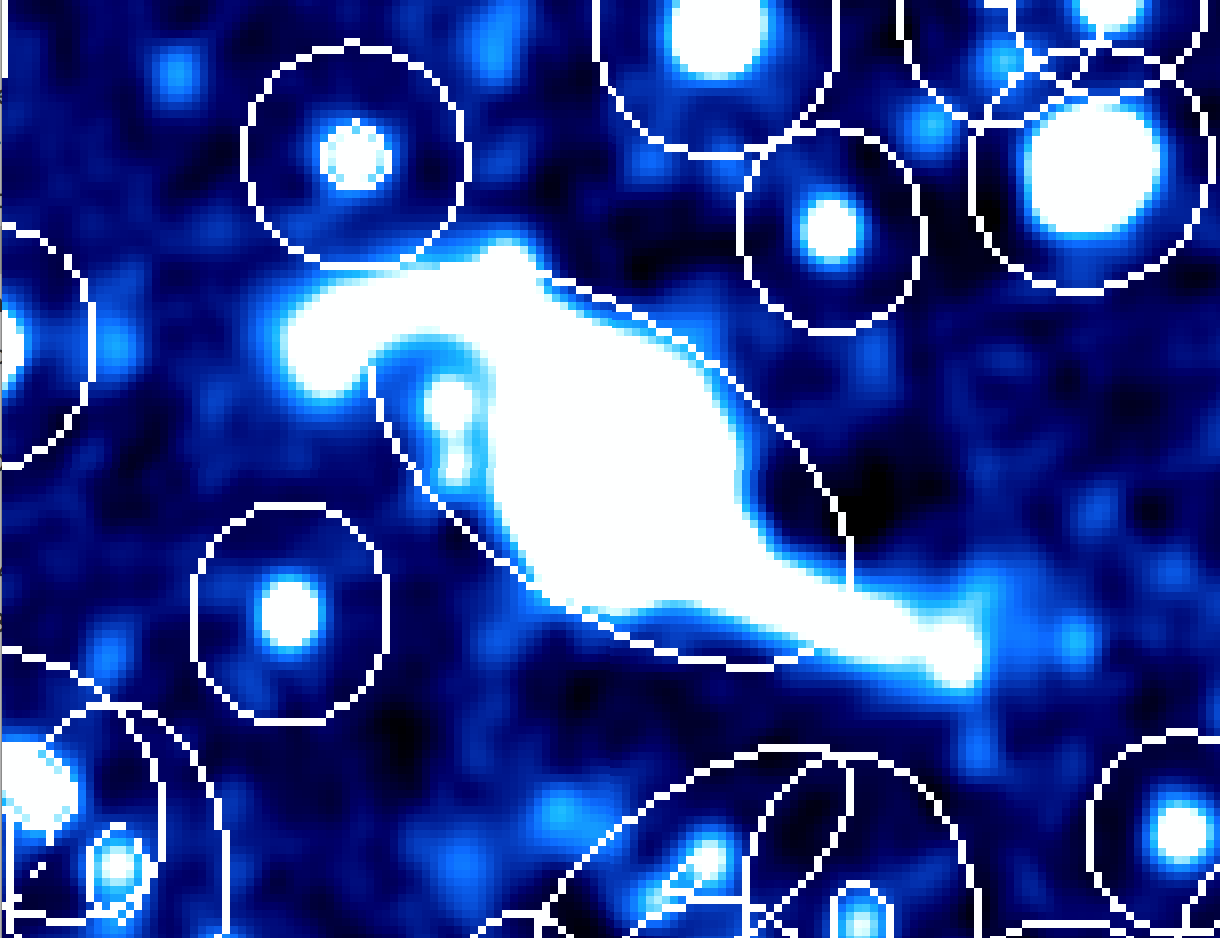}
  \includegraphics[width=0.4\textwidth,clip]{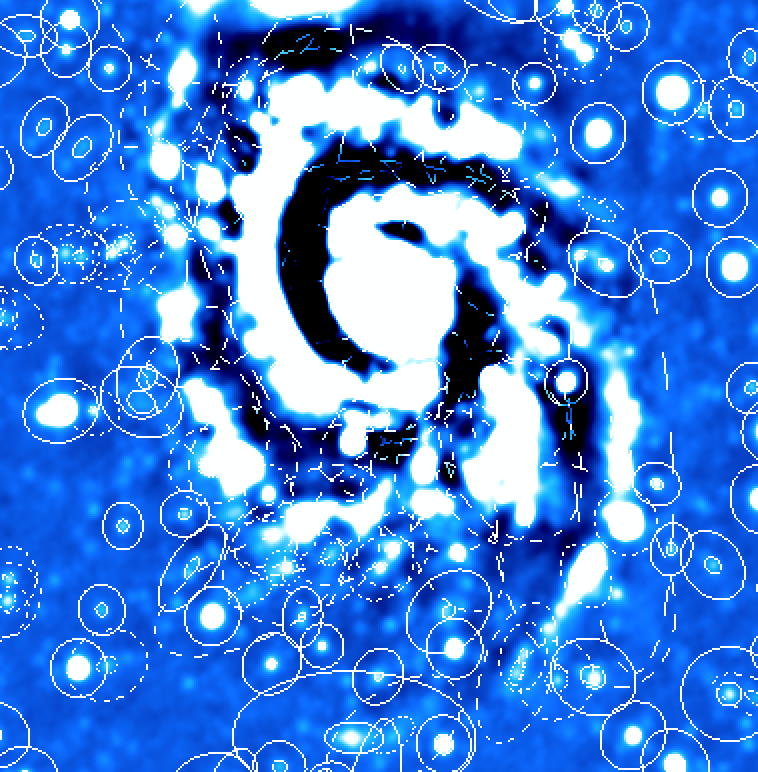}
  \caption{ Example of the danger of using the same automated SExtractor photometry on WISE images 
  of the full sample. SExtractor was run with the same parameters on both examples shown. Top:  SExtractor finds an aperture that is smaller than the galaxy. 
  Bottom: SExtractor detects the different star-forming regions as separate galaxies and not as  a single galaxy. 
}  
\label{fig:imgsSmallAp_SFRegions}
\end{center}
\end{figure}

We started by performing the photometry in the WISE images, using SExtractor \citep{BertinnArnouts96}.  
For example, as we apply commonly used values for the sky-threshold ($\rm 3 \sigma $) and 
deblending-threshold (n-deblending = 4), we saw that 
not all the mergers were included completely within the aperture, hence not all the light was 
extracted from the source by SExtractor (see Fig. \ref{fig:imgsSmallAp_SFRegions}, top panel). 
Also, for some galaxies the same SExtractor set-up extracts only 
the light of a very bright star-forming region within the merging galaxy 
(see Fig. \ref{fig:imgsSmallAp_SFRegions}, bottom panel). 
Thus, we experimented with various values for the main SExtractor 
parameters, until we obtained a matrix of possible reasonable parameters (sky-threshold ($\rm \sigma$): 
1.5, 3, and 5  and n-deblending: 2, 4, 8, 16, and 32).
We then repeatedly ran SExtractor on each of our sources, applying each pair of values from the 
matrix of SExtractor parameters and, in the process, obtaining 15 images per galaxy per filter. 

\begin{figure}[h]
\begin{center}
  \includegraphics[width=0.4\textwidth,clip]{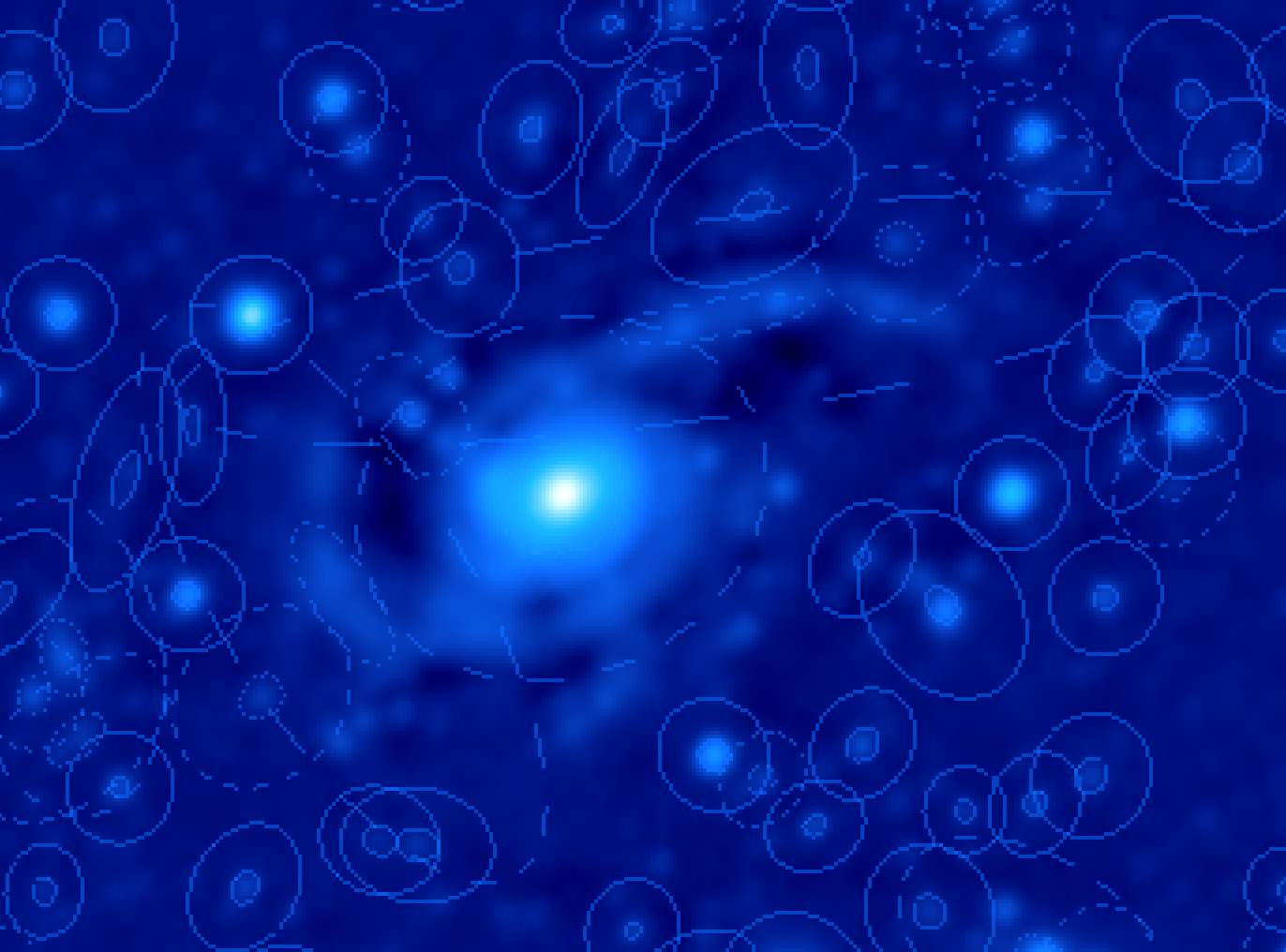}
  \includegraphics[width=0.4\textwidth,clip]{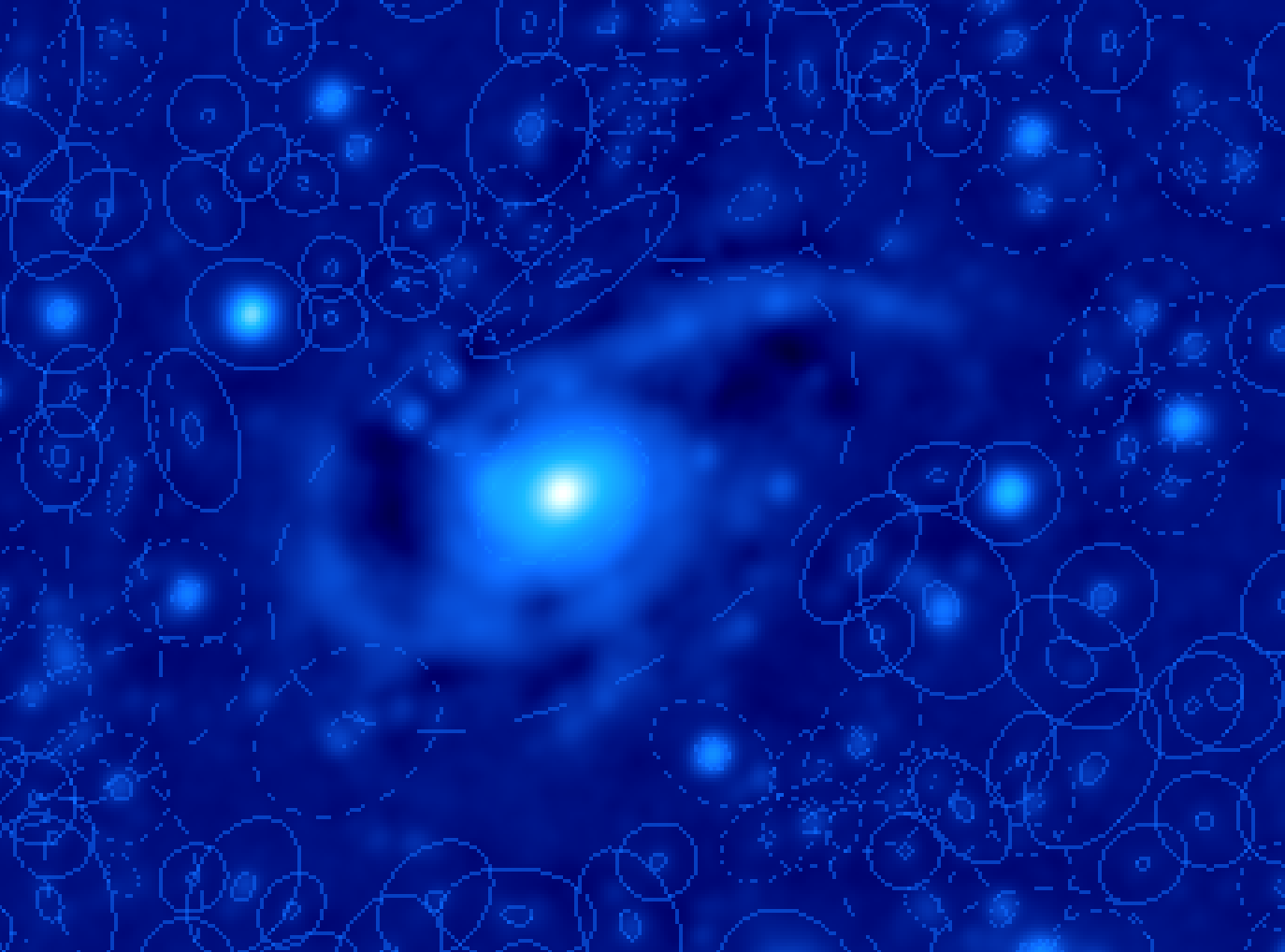}     
  \includegraphics[width=0.4\textwidth,clip]{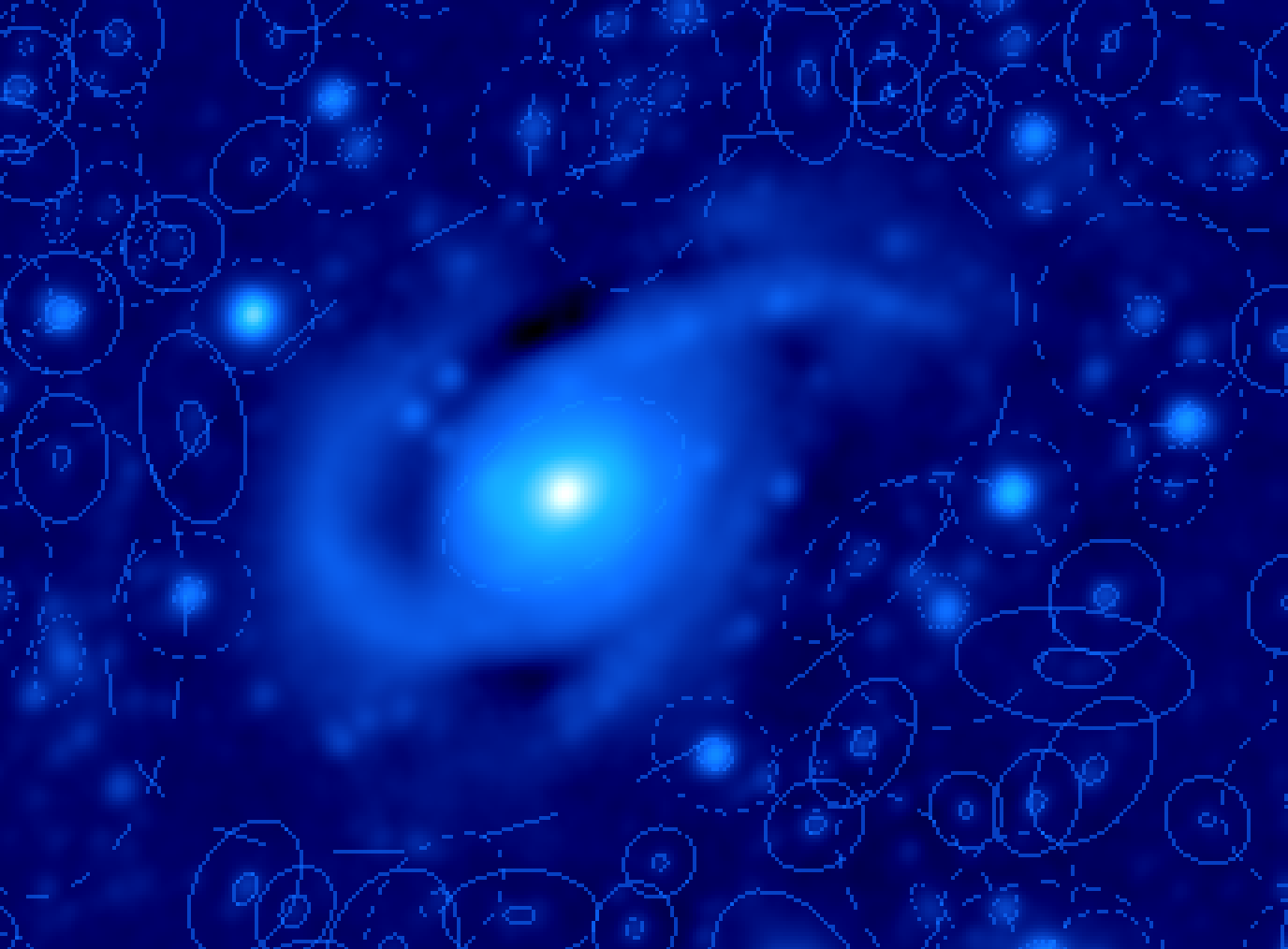}
    \caption{ Examples of running SExtractor, with different parameters (changing sky- and 
    deblending-thresholds) on the same galaxy. See text for details.
}  
\label{fig:Photparams}
\end{center}
\end{figure}

\begin{figure*}[h]
\begin{center}
  \includegraphics[width=1\textwidth,clip]{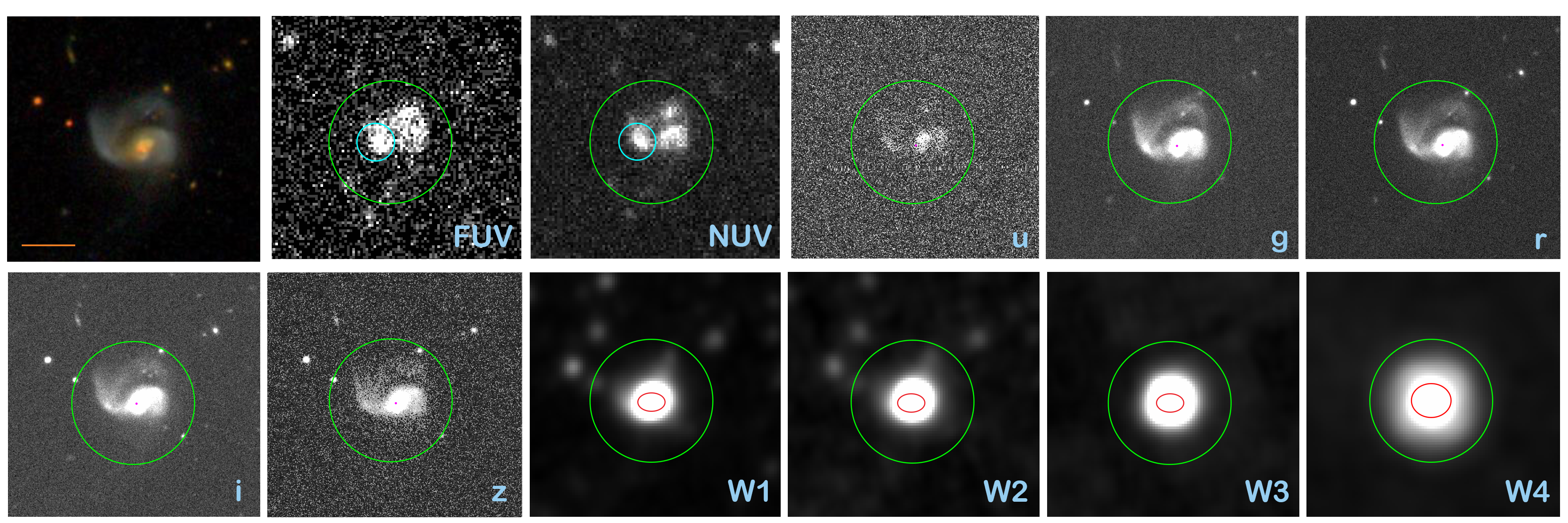}
  \caption{ Example of the different apertures used by different surveys and our measurement in 
  all the filters. The first image shows the galaxy in the SDSS $ugriz$ bands. The orange line represents 20".  
  This SDSS image is followed by FUV and NUV from GALEX; u, g, r, i, and 
  z from SDSS; and W1, W2, W3, and W4 from WISE. Cyan circles show the apertures shown on 
  the GALEX catalogue. Magenta apertures show the small aperture listed in SDSS tables, 
  which is barely seen in the figure. 
  Red apertures are the listed ellipses in AllWISE tables. 
  The aperture we use for all filters, in this case measured from the W4 image, is shown in green. 
}  
\label{fig:Merger_ALLaps}
\end{center}
\end{figure*}

\begin{figure}[h]
\begin{center}
  \includegraphics[bb=50 0 1325 800,width=0.48\textwidth,clip]{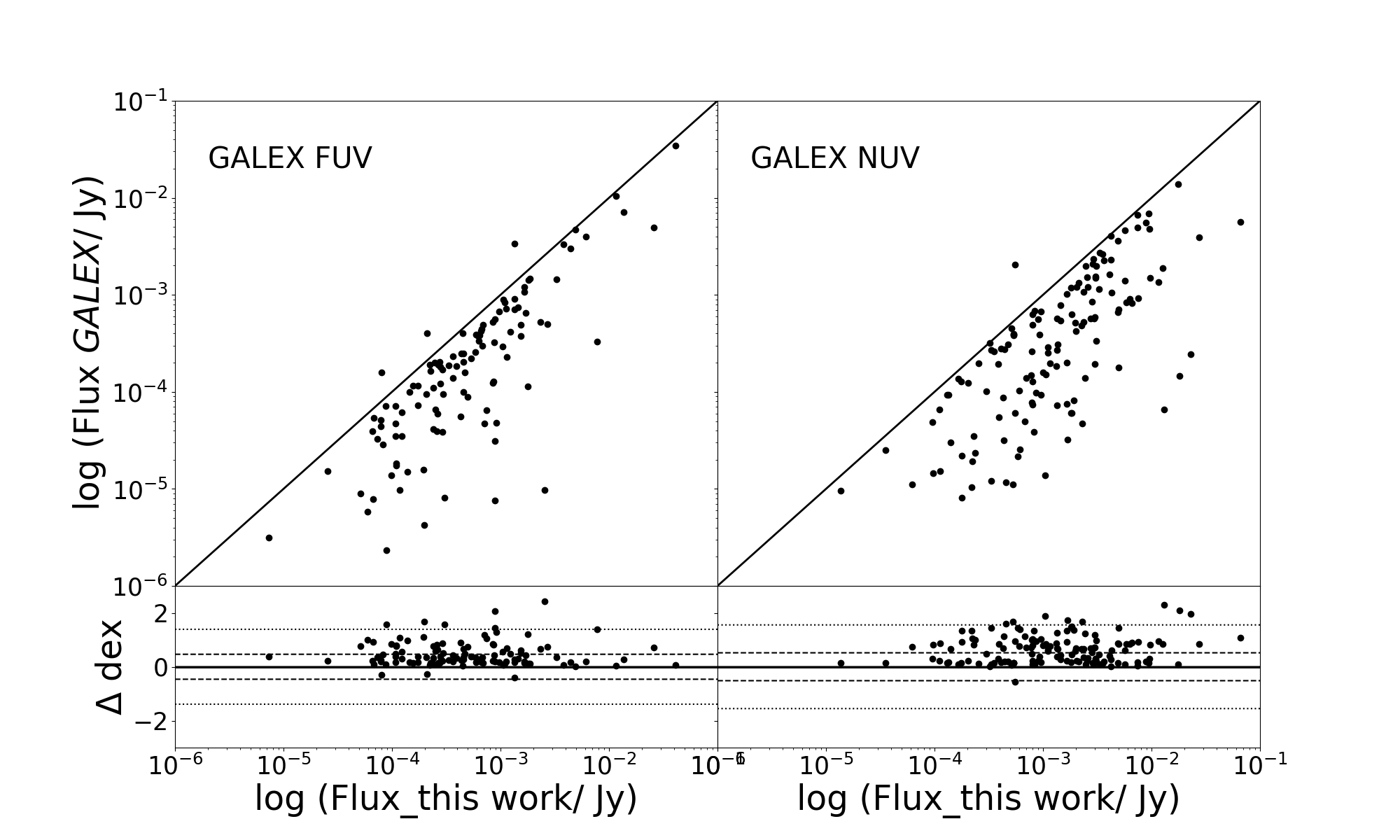}
  \caption{ Flux comparison between our measured photometry and the values in  the GALEX table. 
  The black line shows equality. The {\it y}-axes show the GALEX FUV-NUV flux. The dashed and 
  dotted lines show the 1- and 3-$\sigma$ deviation from the one-to-one relation, respectively. 
}  
\label{fig:FiltersFluxGALEX_comps}
\end{center}
\end{figure}

\begin{figure*}[h]
\begin{center}
  \includegraphics[width=0.3\textwidth,clip]{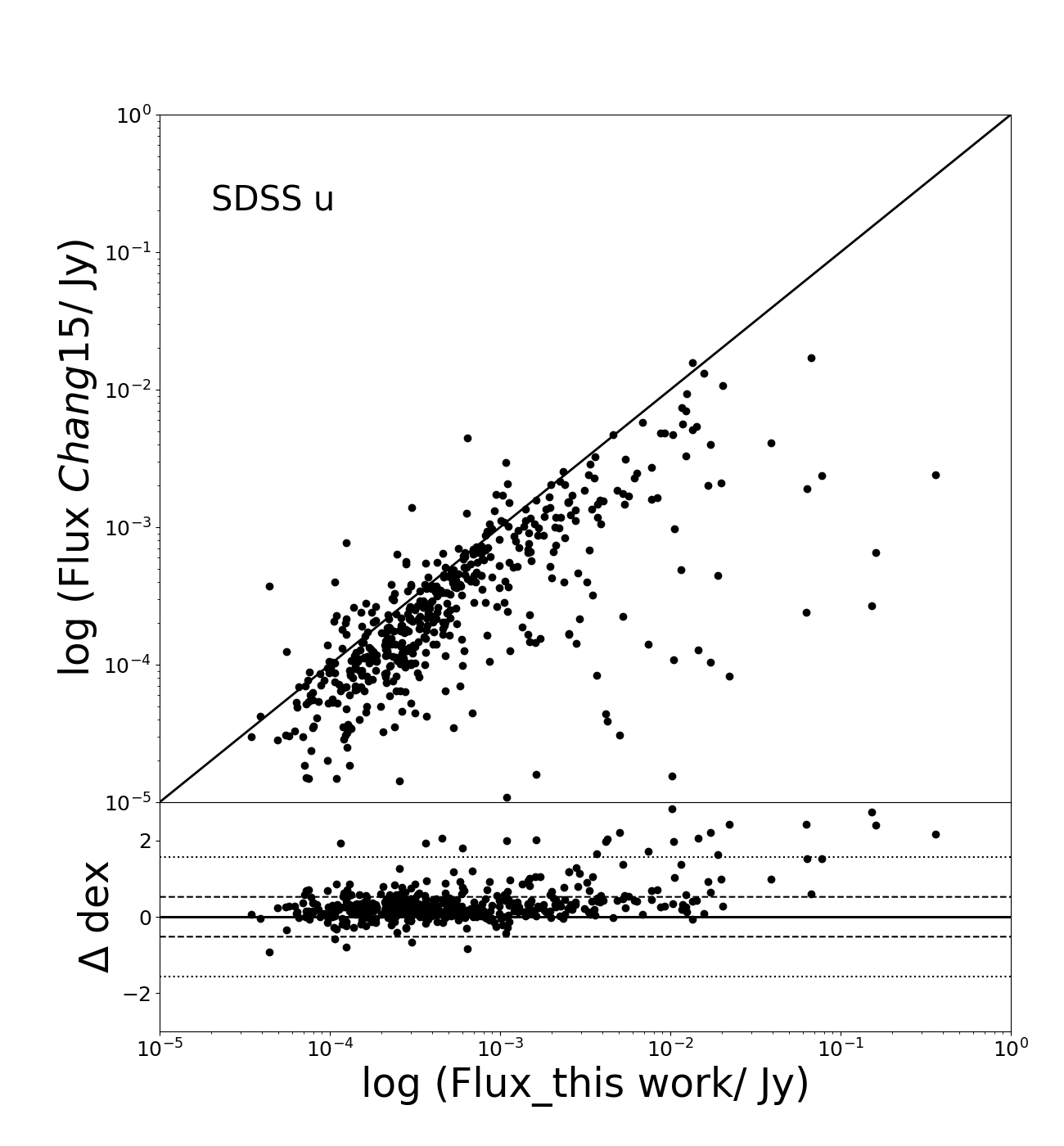}
  \includegraphics[width=0.3\textwidth,clip]{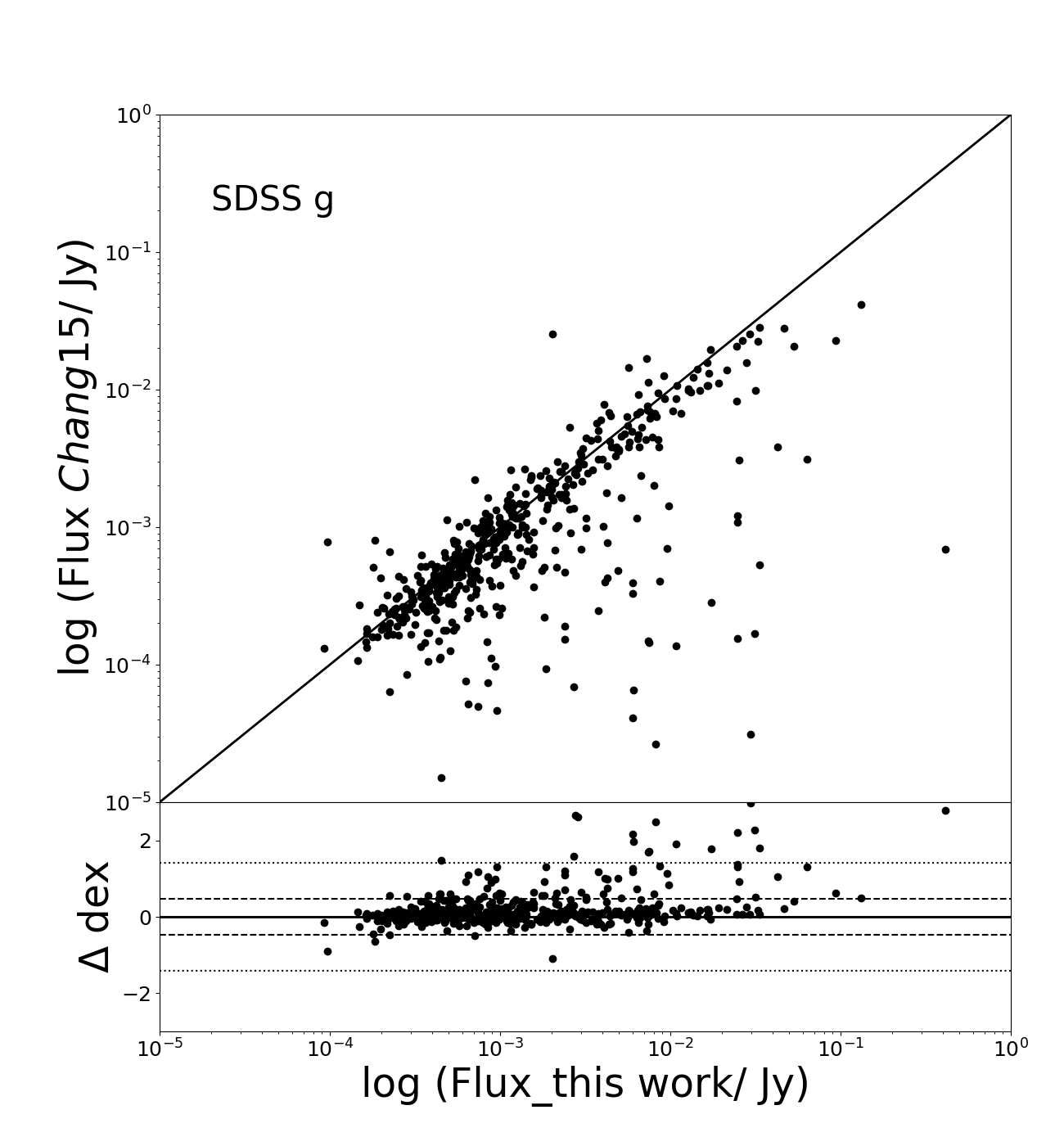}
  \includegraphics[width=0.3\textwidth,clip]{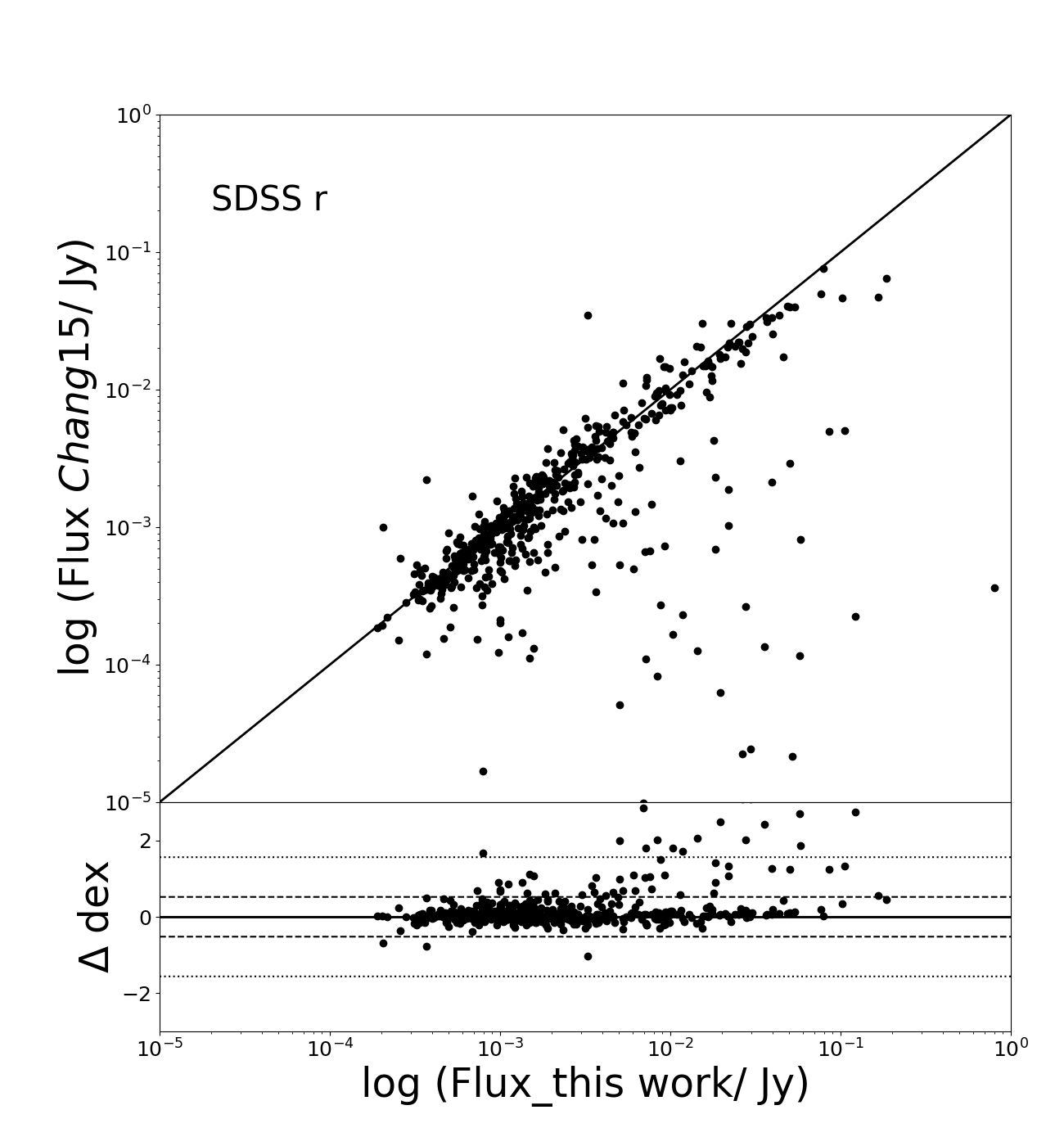}
  \includegraphics[width=0.3\textwidth,clip]{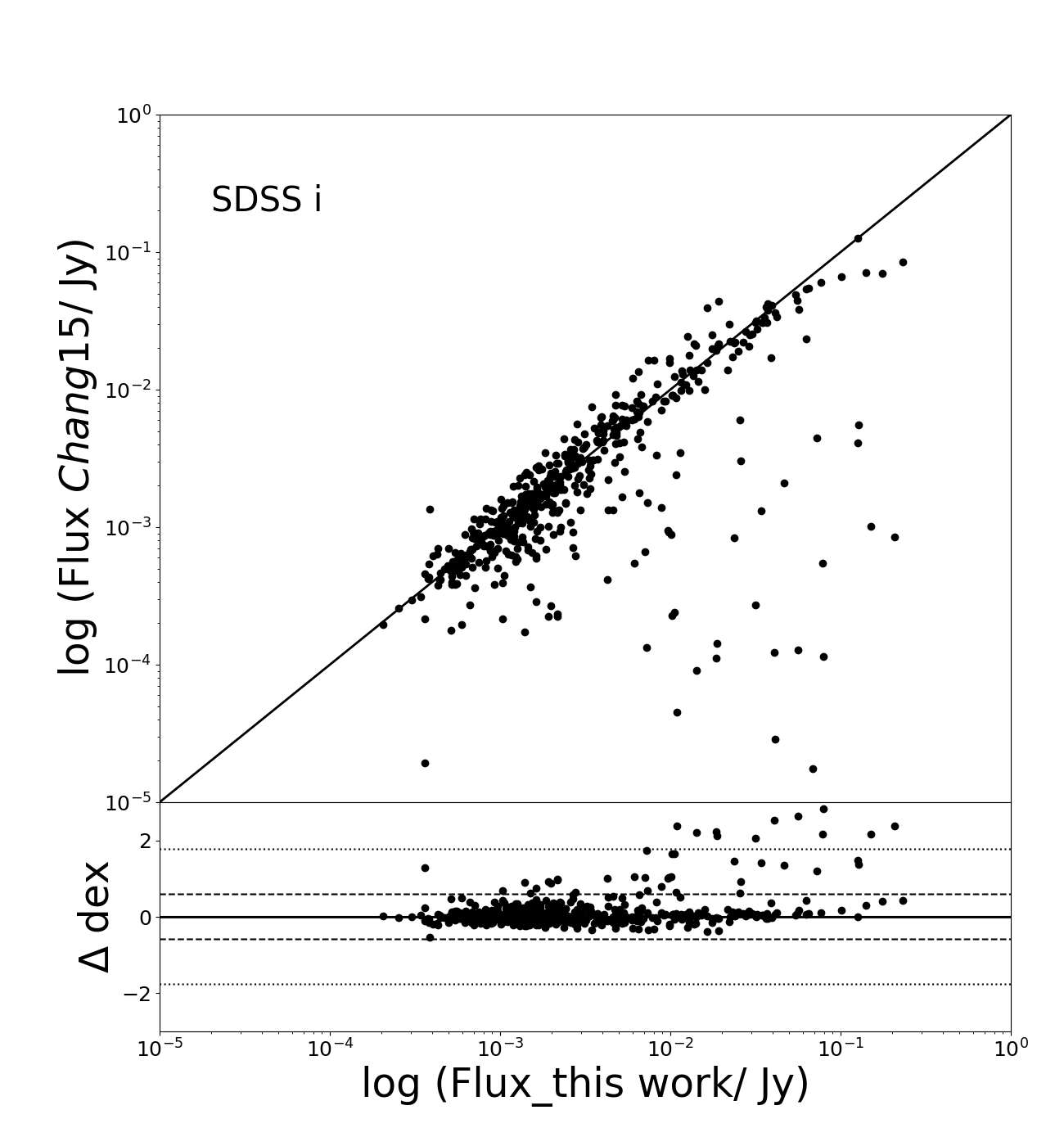}
  \includegraphics[width=0.3\textwidth,clip]{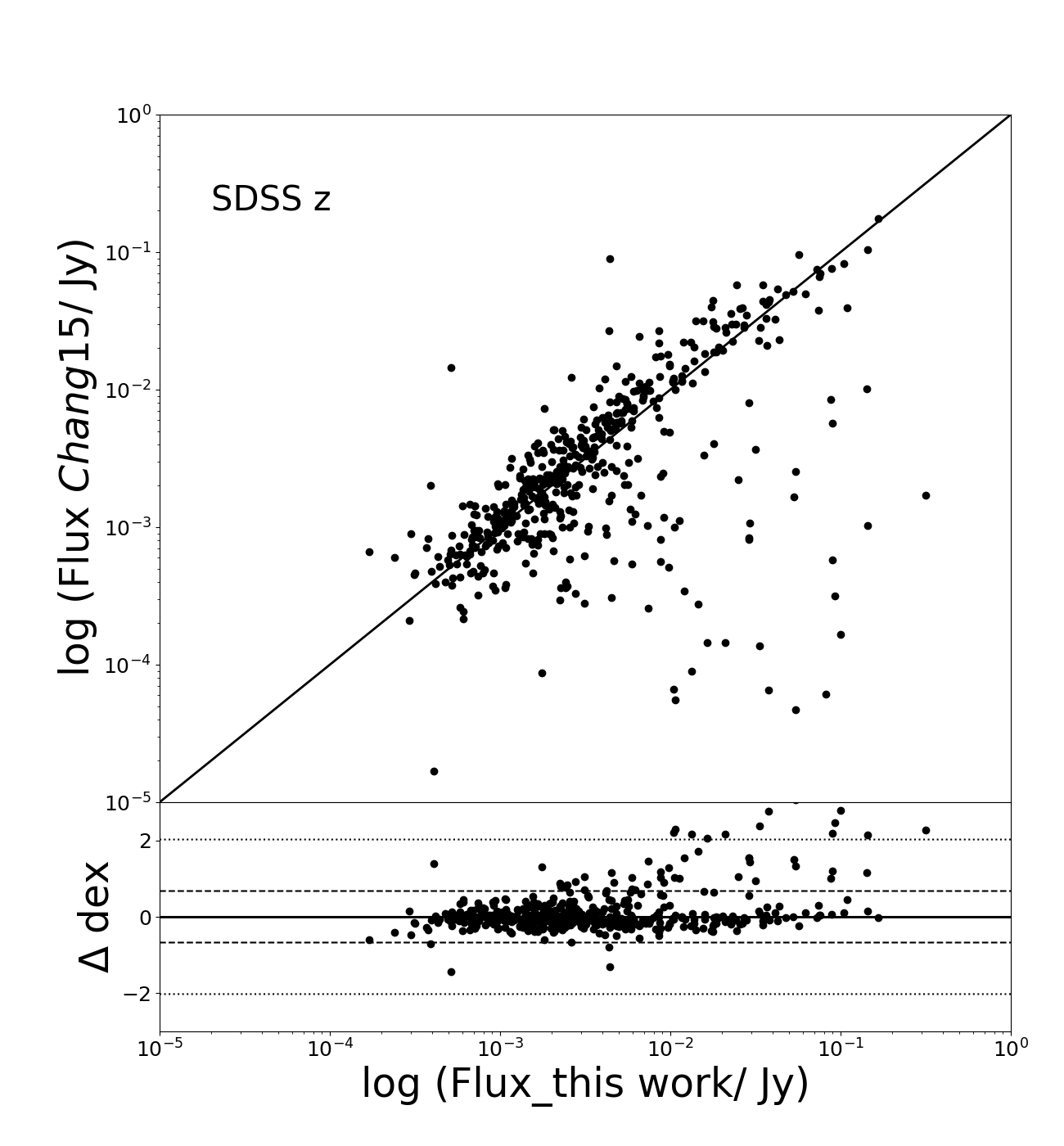}
  \includegraphics[width=0.3\textwidth,clip]{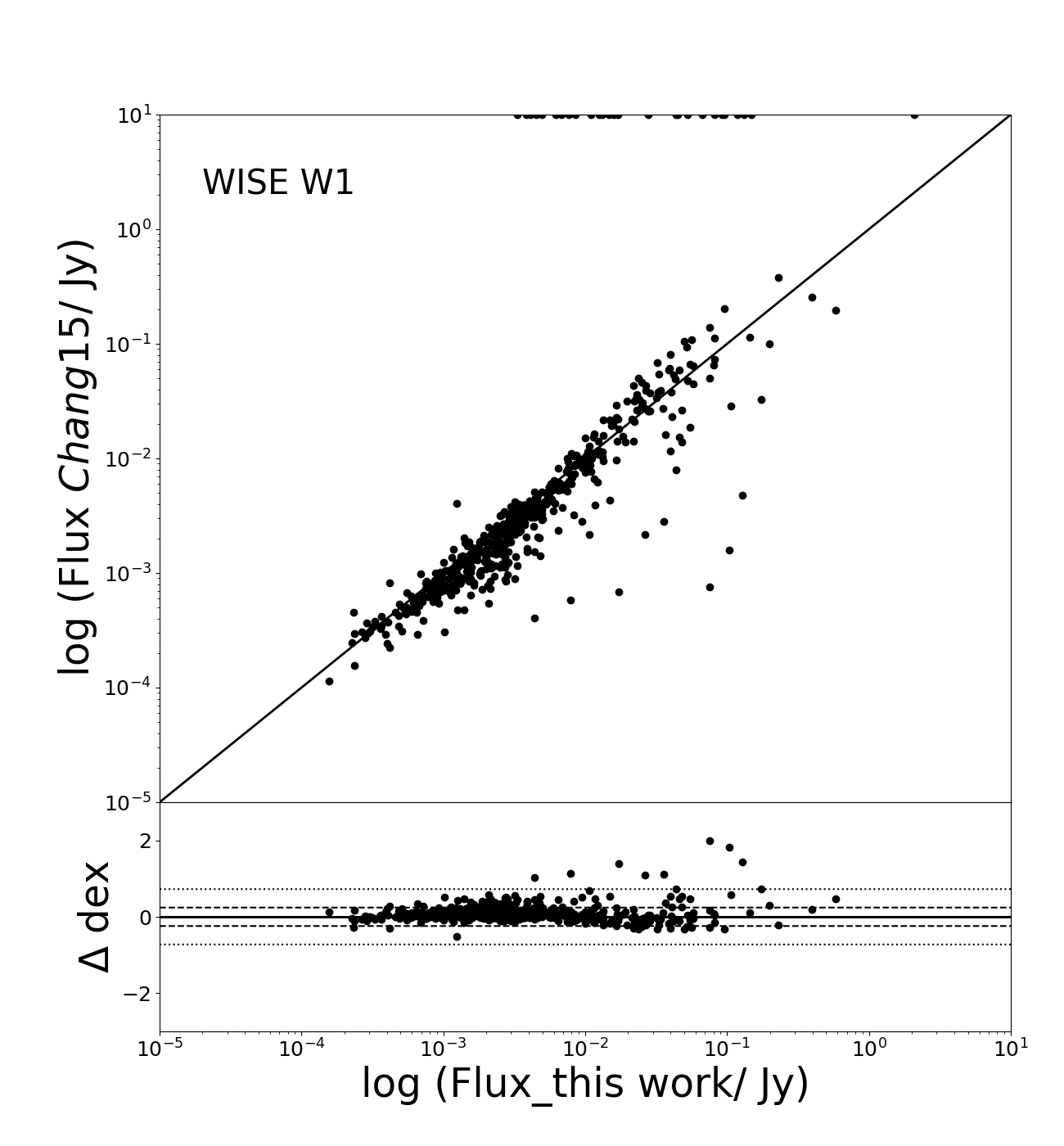}
  \includegraphics[width=0.3\textwidth,clip]{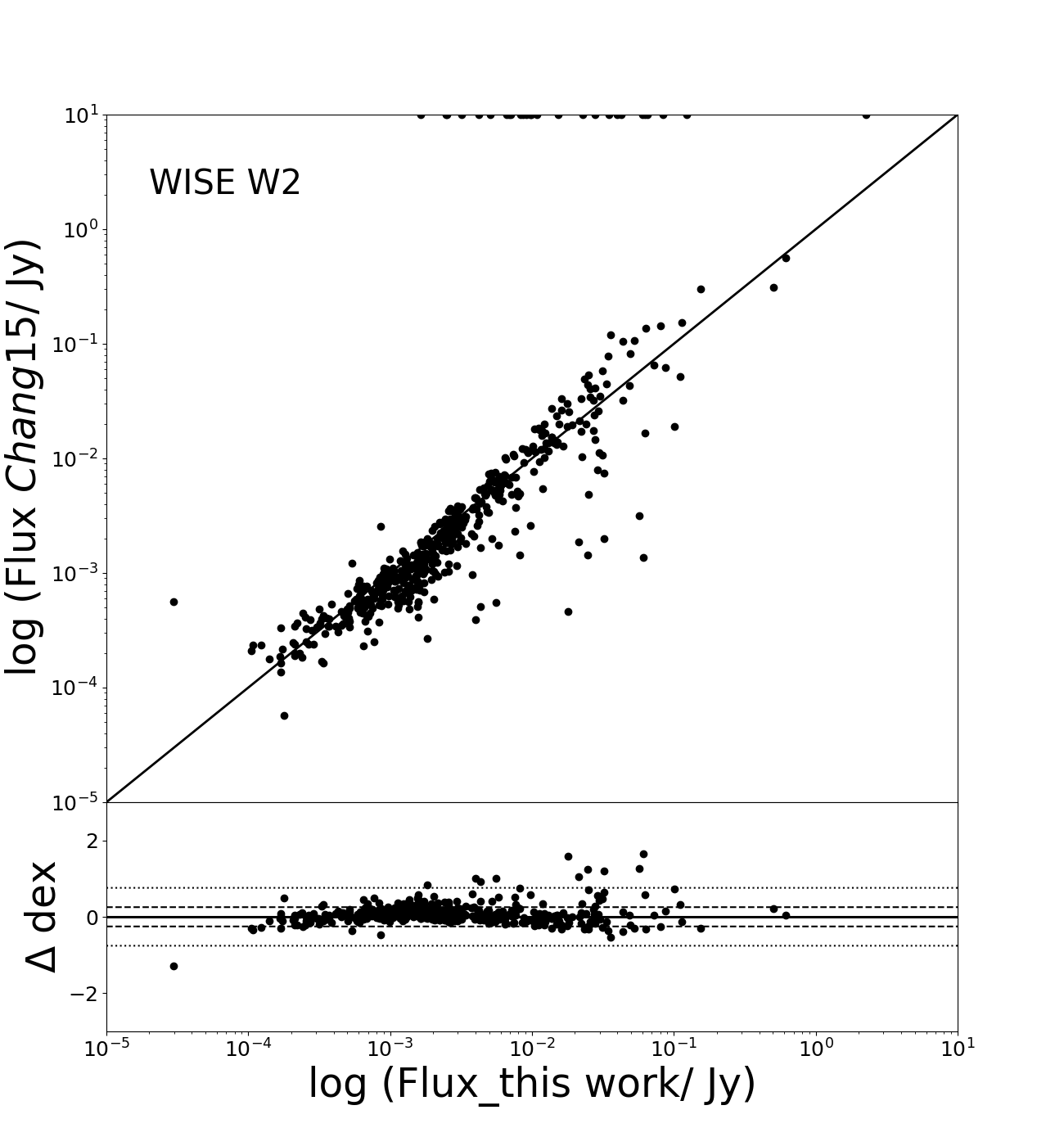}
  \includegraphics[width=0.3\textwidth,clip]{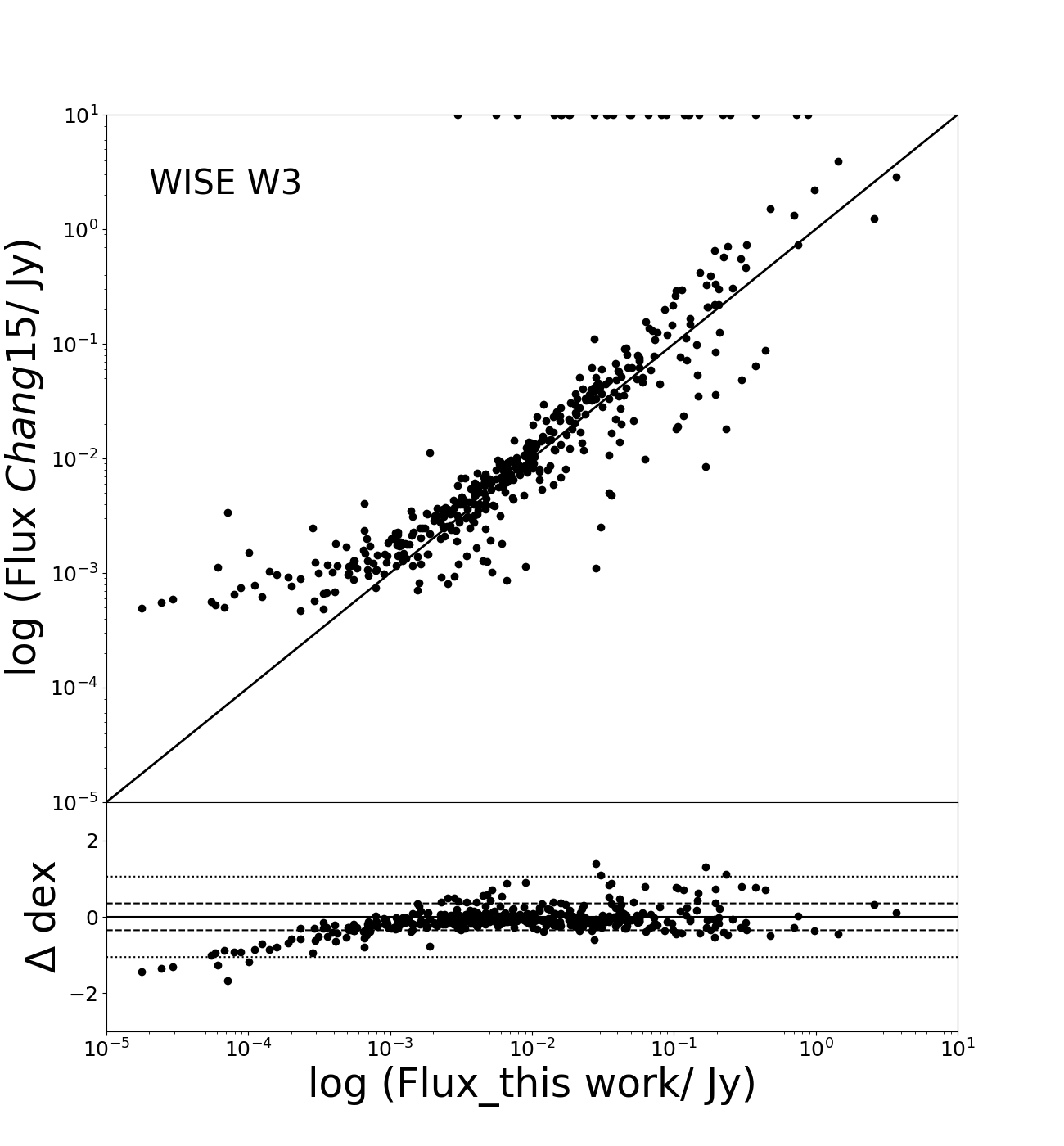}
  \includegraphics[width=0.3\textwidth,clip]{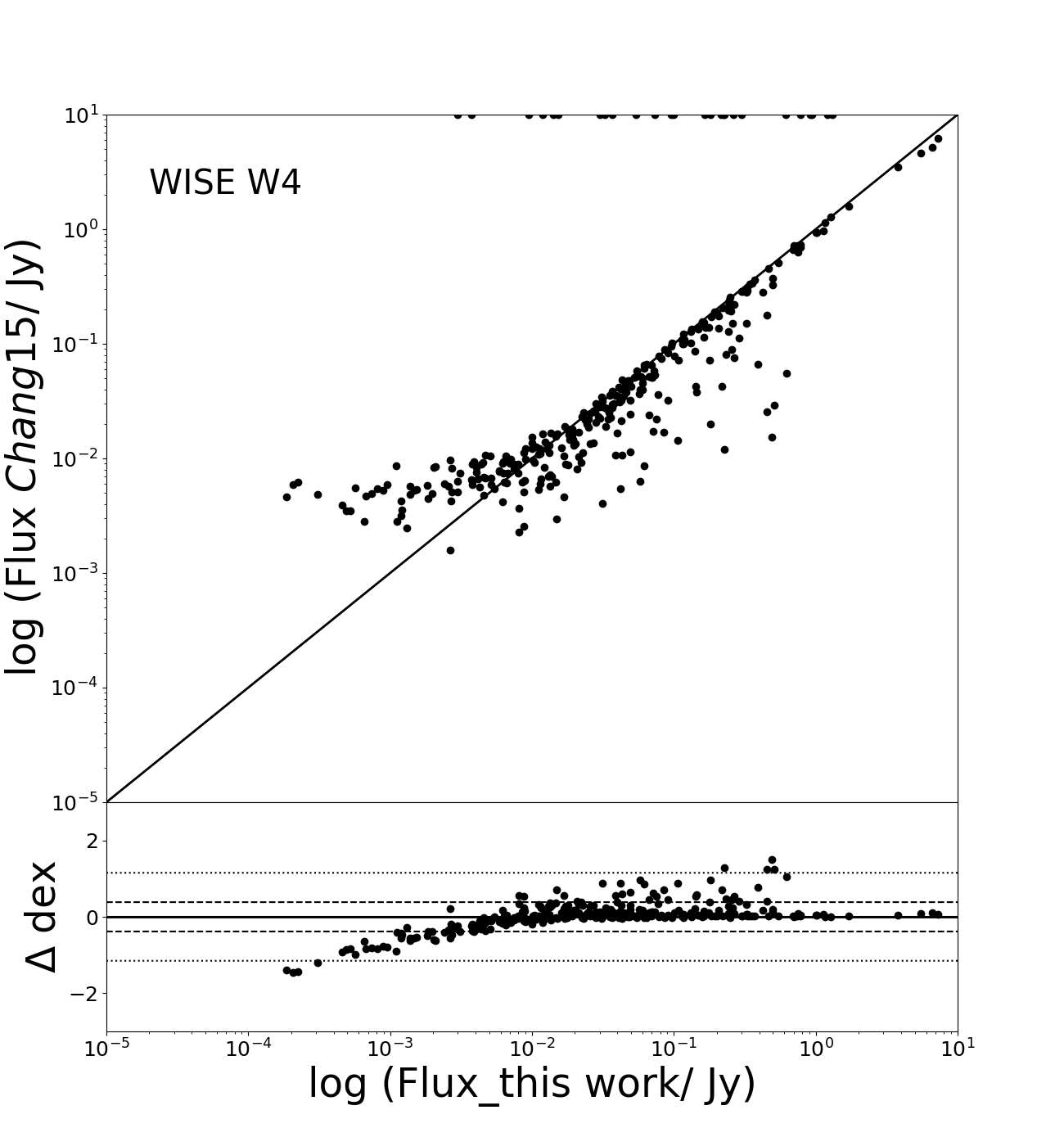}
  \caption{ Comparison  of our aperture flux values  with  those listed by \textsc{Chang15} 
  for all galaxies common to our samples. Specifically, we use the SDSS {\texttt{MODELFLUX}} values 
  and the WISE \textit{mpro} fluxes from \textsc{Chang15}. The solid line in each main panel shows 
  the line of equality and each small panel shows the difference between the two axes. 
  Data points at the highest $y$-axis values in each main panel represent galaxies with fluxes measured by us, 
  but not by \textsc{Chang15}. The dashed and 
  dotted lines show the 1- and 3-$\sigma$ deviation from the one-to-one relation, respectively. 
}  
\label{fig:FiltersFlux_comps}
\end{center}
\end{figure*}

\begin{figure*}[h]
\begin{center}
  \includegraphics[bb=230 0 2650 800,width=\textwidth,clip]{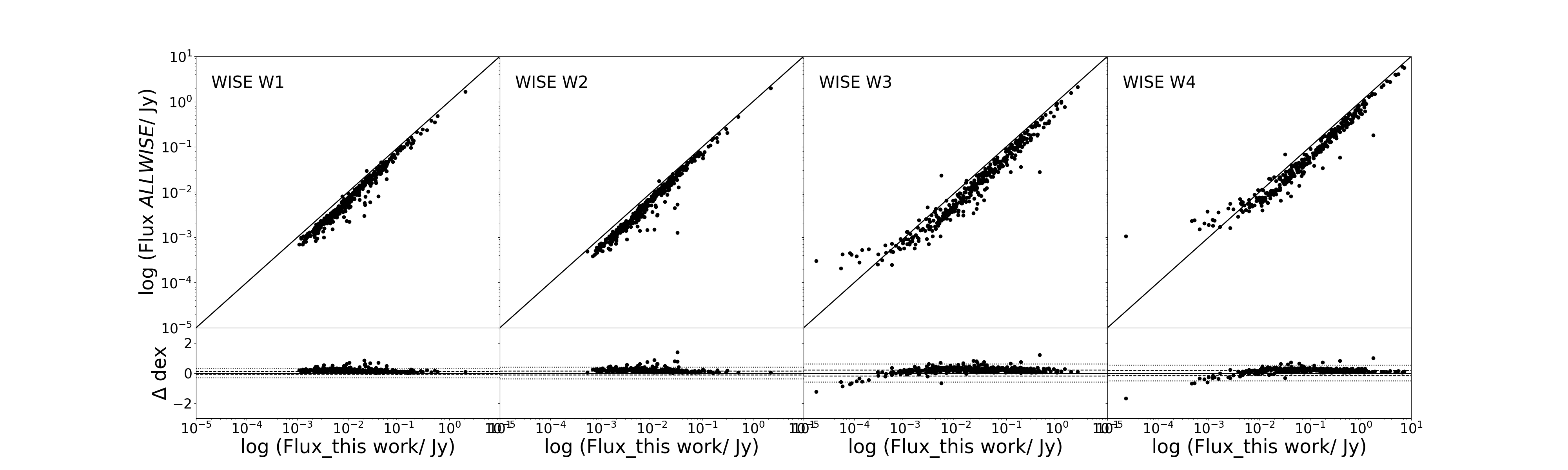}
  \caption{ Flux comparison between our measured photometry and the values in the  WISE table. 
  The black line shows equality. The {\it y}-axes show the WISE \textit{gmag} flux conversion. 
  The dashed  and 
  dotted lines show the 1- and 3-$\sigma$ deviation from the one-to-one relation, respectively.  
}  
\label{fig:FiltersFlux_WISEcomps}
\end{center}
\end{figure*}

Then each image and its resultant SExtractor apertures were checked visually in order to choose the 
best SExtractor parameters for each individual galaxy. The best parameters are those that  show an aperture that encompasses all of the light from the galaxy,  exclude contamination from 
other sources, and  do not exclude any star-forming region that belongs to the galaxy.
In the case when  two individual galaxies were overlapping, we chose the parameters for which 
SExtractor shows an aperture that includes both galaxies within one aperture. 
This could happen in merger stages II and IIIa. 
We also confirmed that the measured flux does not increase when we increase the size of the aperture in 
blank regions of the sky. 

Some examples of how the parameters affect the result for one system can be found in Fig. 
\ref{fig:Photparams}. The top panel shows how SExtractor separates the galaxy into different regions 
for $\rm \sigma$ = 1.5 and n-deblending = 4. The middle panel shows how SExtractor detects the central 
part of the galaxy only, making an aperture that is too small for $\rm \sigma$ = 3.0 and n-deblending = 4. 
Finally, the bottom panel shows how SExtractor successfully detects the galaxy and correctly chooses 
an aperture which is sufficiently large to enclose all of the light from the galaxy for $\rm \sigma$ = 5.0 
and n-deblending = 4. 
It is important to note that the case shown in Fig. \ref{fig:Photparams} is only one
example, and that the most suitable parameters chosen for
this galaxy do not provide acceptable results for other mergers
in our sample. More examples are shown in Appendix \ref{app:ApsExamples}. 
This highlights how redoing the photometry was a necessity for our mergers, 
and demonstrates that performing automated photometry on these types of complex sources is highly challenging.

Our photometric technique can be summarised as follows. We first check which parameters extract all the light 
from each galaxy in the WISE bands W1 and W4, and then  choose the larger aperture 
between W1 and W4 and use that aperture size for all filters. The location of the aperture in the rest of 
the filters is automatically found by Sextractor. In practice, 
we find that once the best set of parameters is found for a particular galaxy from the W1 image, 
running SExtractor with these parameters on the other filters results in similar detections. 
We visually check each filter to ensure that the sources are 
detected entirely and check for possible contamination, and rerun with alternative parameters from the set if 
necessary.

To show an example of how unreliable some catalogued measurements can be, we  selected a merging 
galaxy and show the apertures from the different surveys and our aperture measured by SExtractor using the 
optimal parameters for this galaxy. Figure \ref{fig:Merger_ALLaps} shows 12 images, the first image 
is the SDSS {\it ugriz} image, followed by the 11 filters we use: FUV, NUV, u, g, r, i, z, W1, W2, W3 
and W4 (as described in the bottom right corner of each image). For FUV and NUV we show the 
aperture listed in the GALEX catalogue in cyan. For the {\it ugriz} images, SDSS shows only one aperture size, 
which is from the r-band aperture (in magenta); this value is also used by \textsc{Chang15} to correct their 
W1-3 fluxes. The aperture shown for this galaxy by SDSS is so small that it is barely seen in the figure. 
For the WISE filters, we show the {\it rsemi} values listed in the AllWISE catalogue (in red). Finally, we show our 
measured aperture (green circle) in all the images. In this case, we  selected the aperture size measured 
for W4 since it is larger than that found for W1.
We can see that for all the images the aperture sizes used by 
the different catalogues are much smaller than the galaxy, while our aperture covers the entire 
galaxy in all  11 filters. 

\begin{figure}[h]
\begin{center}
  \includegraphics[bb=0 0 500 400,width=0.48\textwidth,clip]{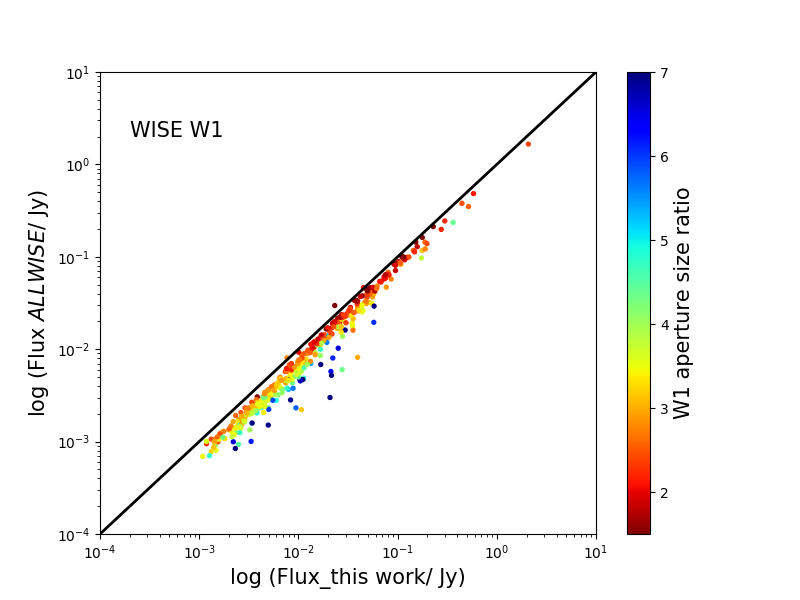}
  \caption{ Comparison of  our total aperture flux (Jy) value  and that listed in the WISE tables for all 
  galaxies in our sample. Symbols are colour-coded according to  the ratio of our measured aperture size to that applied by the 
  WISE team, and the black line shows equality. 
}  
\label{fig:W1mag_comp}
\end{center}
\end{figure}

The final fluxes were corrected for Galactic extinction \citep{SchlaflynFinkbeiner11, Yuanetal13} 
for all filters (except W3 and W4 for which Galactic extinction 
is negligible). They were also  corrected following each survey's specification. The SDSS u  and 
z bands  need a correction of 0.04 and 0.02 mag, respectively, and an extra calibration of 8\% 
for spiral and disc galaxies is needed for W4. 
In order to take possible systematic uncertainties into account, we  added 0.05 in GALEX 
\citep{Marinoetal11}, and 0.02 and 0.1 mag in SDSS and WISE (\textsc{Chang15}), respectively, 
to the statistical uncertainties calculated by SExtractor. 

To check the reliability of our estimated photometric uncertainties, we  performed the 
following test.  
For each galaxy, we compared the flux measured in the aperture using our best set of parameters 
(sky-threshold and n-deblending), as determined by eye. Next we considered the fluxes measured 
within the apertures using the parameters of the eight nearest neighbouring cells of the matrix of 
sky-threshold and n-blend parameters. 
We then calculated the difference between the best flux and the rest of the eight measured fluxes, 
and calculated the standard deviation of these differences. 
We  conducted this test in one representative band for each survey: NUV from GALEX, $r$ from SDSS, 
and W1 from WISE. We find that the standard deviation (0.17, 0.007, and 0.10) for these bands 
is always smaller than the median uncertainty (0.30, 0.234, and 0.15, respectively). 
Thus, from this test we conclude that our photometric uncertainties are 
reasonable estimates. However, we know that if we consider more distant cells  the merger is frequently 
not detected, which confirms  that fully automated photometry for mergers is not trivial. 
We compiled useful information from the different surveys in Table \ref{tab:Filters}  to 
facilitate the use of the different parameters that are required in the photometry process. 

The aperture fluxes measured by us are significantly different from those measured by previous authors. 
Figure \ref{fig:FiltersFluxGALEX_comps} shows a comparison between our aperture flux values ({\it x}-axes) 
for FUV and NUV and the respective values listed in the GALEX catalogue. 
Most of the galaxies show higher fluxes for our measurements. 

Figure \ref{fig:FiltersFlux_comps} shows nine panels comparing our measured fluxes on the {\it x}-axis 
to the  \textsc{Chang15} fluxes. 
The  first five panels (from left to right) show the comparison to the SDSS 
{\texttt{MODELFLUX values}}, the  following three panels show the comparison to WISE W1-3 {\texttt{mpro}} 
fluxes corrected by the radius calibration shown by \textsc{Chang15}, and the last panel shows 
the WISE W4 {\texttt{mpro}} flux which is not corrected by the \textsc{Chang15} calibration. The 
WISE W1-4 panels show that for a fraction of our sample there are no measurements listed 
(see the \textsc{Chang15} fluxes equal to 10 Jy). Thus, 
we  increased the number of useful data for this sample. 
For the first five panels, most of our fluxes are brighter than those listed for {\texttt{MODELFLUX}} and for W4 {\texttt{mpro}}. This is due to the larger apertures we use to measure the flux.  
For the WISE W1-3 filters, we observed that the \textsc{Chang15} values are systematically higher than our fluxes. 
This could be related to their correction, which increases the flux depending on the radius in the r band of the galaxy. 
This correction can be adding more flux than  needed for this type of galaxy, overcorrecting the flux 
for these three filters. The fluxes are affected differently depending 
on the filter, which will  create an offset and will also change the shape of the final SED.

Figure \ref{fig:FiltersFlux_WISEcomps} shows the comparison between the fluxes measured by SExtractor 
($x$-axes) and the 
AllWISE table {\texttt{gmag}} values, which is recommended for extended sources by the WISE team. 
Our fluxes are higher than those listed in the AllWISE tables due to our larger apertures. 
For a small fraction of the galaxies at the faint end of W3 and W4 (6\%\ and 10\%, respectively), we see that 
our measurements are lower those  listed in the AllWISE tables. 

Figure \ref{fig:W1mag_comp} shows the comparison between our W1 photometry measurements on the 
{\it x}-axis and the WISE table W1 {\texttt{gmag}} values (recommended for extended sources), colour-coded according to 
the ratio between our aperture radius and the WISE 
table semi-major axis ({\it rsemi}). This clearly shows the dependence of the measured flux on the aperture 
used during the photometry. For almost all galaxies our measurements give higher fluxes than those 
listed in the WISE table magnitudes, which is primarily due to our (more correct) larger apertures.

The comparisons between our measured fluxes and those listed in GALEX, SDSS, and WISE do not depend 
 on the morphology of the galaxies or on the  stage of the merger. Some of these comparisons can 
be seen in Fig. \ref{fig:UVoptNIR_compscol}.

\subsection{MAGPHYS: SED fitting to obtain M$_*$ and SFR of mergers} \label{sec:MAGPHYS}

We obtained our M$_*$ and SFR values using the publicly available SED fitting code 
MAGPHYS\footnote{\url{http://www.iap.fr/magphys/}} \citep{daCunhaetal2008}.
This program fits photometric data from UV to submillimetre wavelengths. 
We  used the 2003 libraries recommended by their website, which assembles 50000 stellar 
population template spectra \citep{BruzualnCharlot03} for the optical photometric library and other 
50000 polycyclic aromatic hydrocarbon (PAH) plus dust emission template spectra for the infrared photometric library. 
MAGPHYS models galaxy SEDs according to the redshift 
of the given sample and uses a Bayesian approach to interpret the SEDs so as to statistically 
derive different galaxy properties such as M$_*$, SFR, and dust mass, among other quantities. 

In order to obtain more accurate estimations of the M$_*$ and the SFR of our galaxies, our sample was 
chosen such that all mergers have available imaging covering the FUV, NUV, u, g, r, i, z, W1, W2, W3, and 
W4. When a merging galaxy did not show flux in some filter (occasionally FUV, NUV, or W4), 
we set the flux to -99.0 and an uncertainty of 3-$\sigma$ of the average error found for that filter. 
In this manner, MAGPHYS will consider this flux as an upper limit. 

\begin{figure}[!htb]
\begin{center}
  \includegraphics[bb= 20 0 800 500,width=0.48\textwidth,clip]{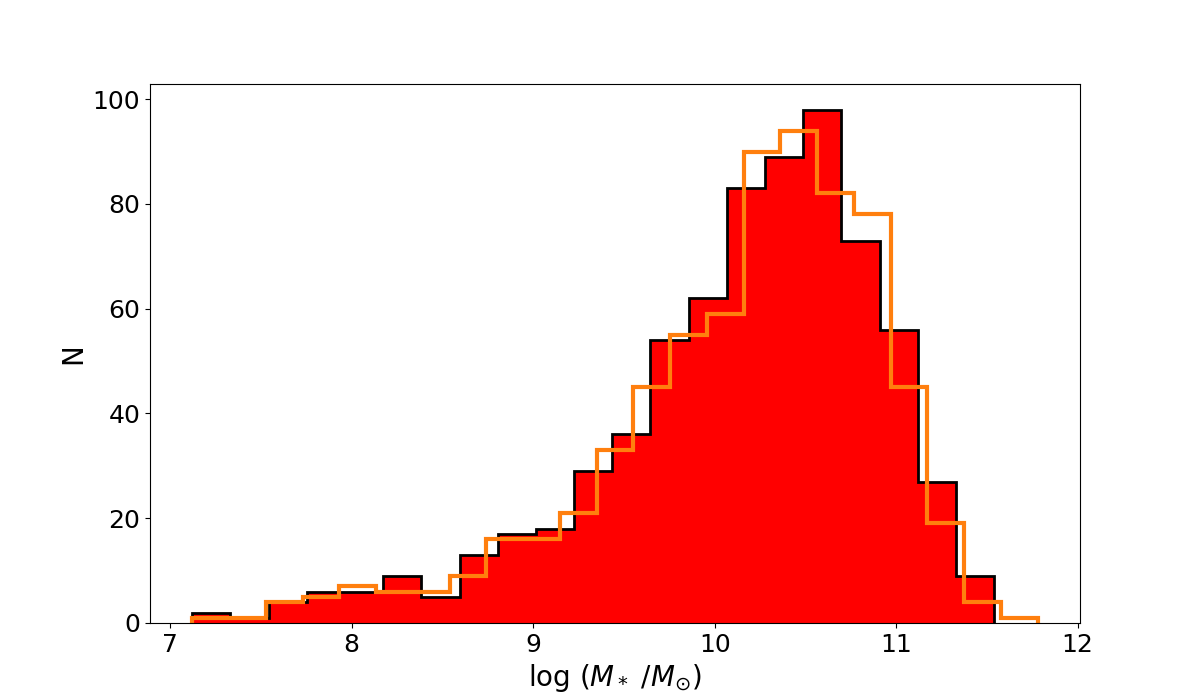}
  
  \includegraphics[bb= 20 0 800 500,width=0.48\textwidth,clip]{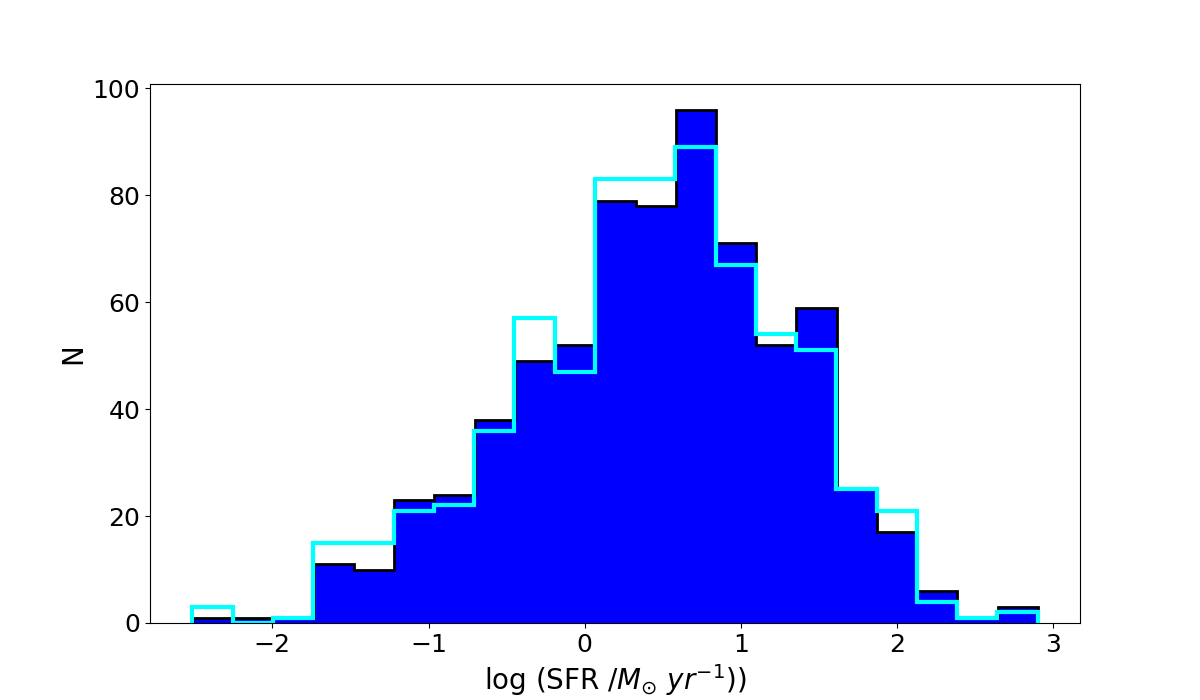}
  \caption{ Distribution of M$_*$ (top) and SFR (bottom) obtained by MAGPHYS 
  for all individual galaxies in our sample. The red and blue histograms show the M$_*$ and SFR 
  distribution using all  11 filters, and the orange and cyan histograms show the results using 
  SDSS+WISE only.
}  
\label{fig:resultshisto}
\end{center}
\end{figure}

The SED fits show a median $\chi^2_r$ = 0.4 and a mean of 1.9. Some SED fitting examples are shown in Fig. \ref{fig:SEDsexamples}.  
The top panel of Fig. \ref{fig:resultshisto} shows the distribution of the estimated M$_*$ (in red); 
for our merger sample the median value is $\rm log(M_{*}/\msun) = 10.28 \pm 0.76$. 
The resulting SFR have a median of $\rm log(SFR/\msun yr^{-1}) = 0.51 \pm 0.86$ with the distribution 
shown in the bottom panel of Figure \ref{fig:resultshisto} in blue. \\ 

We  also estimated M$_*$ and SFR using only the optical and near-infrared (NIR) data in order to compare these results 
to the \textsc{Chang15} catalogue who use the same  limited numbers of filters.
For this sample, we obtain a mean and a median of $\chi^2_r$ = 1.7 and 0.25 for the SED fits. 
A lower $\chi^2_r$ could result because MAGPHYS finds it easier to fit to fewer data points, but  
the fits may also be less accurate since the code is missing information from the young population of the merging galaxy. 
When we include GALEX data, both M$_*$ and SFR estimates are very similar to those derived using 
SDSS+WISE only (orange and cyan histograms in Fig. \ref{fig:resultshisto}, respectively). 

In Figs. \ref{fig:SEDsexamples} and \ref{fig:SEDsexamples2}, we show some examples 
of the SED fits for the sample using all eleven filters and SDSS+WISE only, respectively. 

\subsection{M$_*$ and SFR comparison to the \textsc{Chang15} catalogue} \label{sec:comparisons}

\begin{figure}[ht!]
\begin{center} 
  \includegraphics[width=0.48\textwidth,clip]{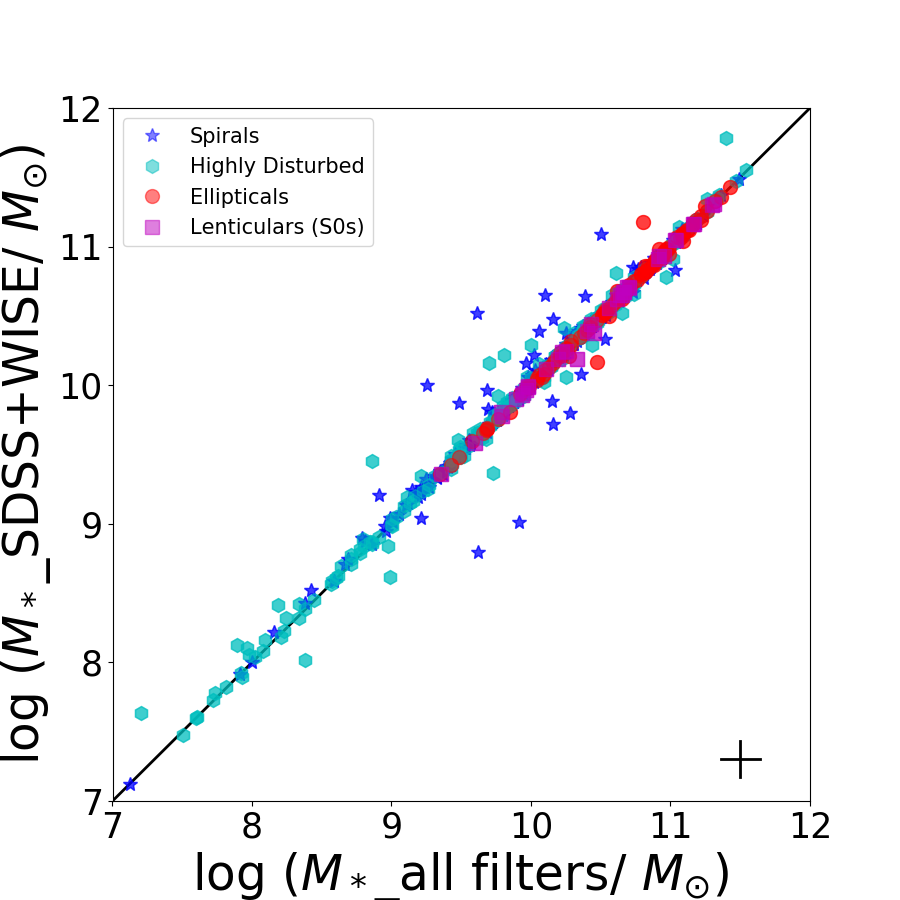}

  \includegraphics[width=0.48\textwidth,clip]{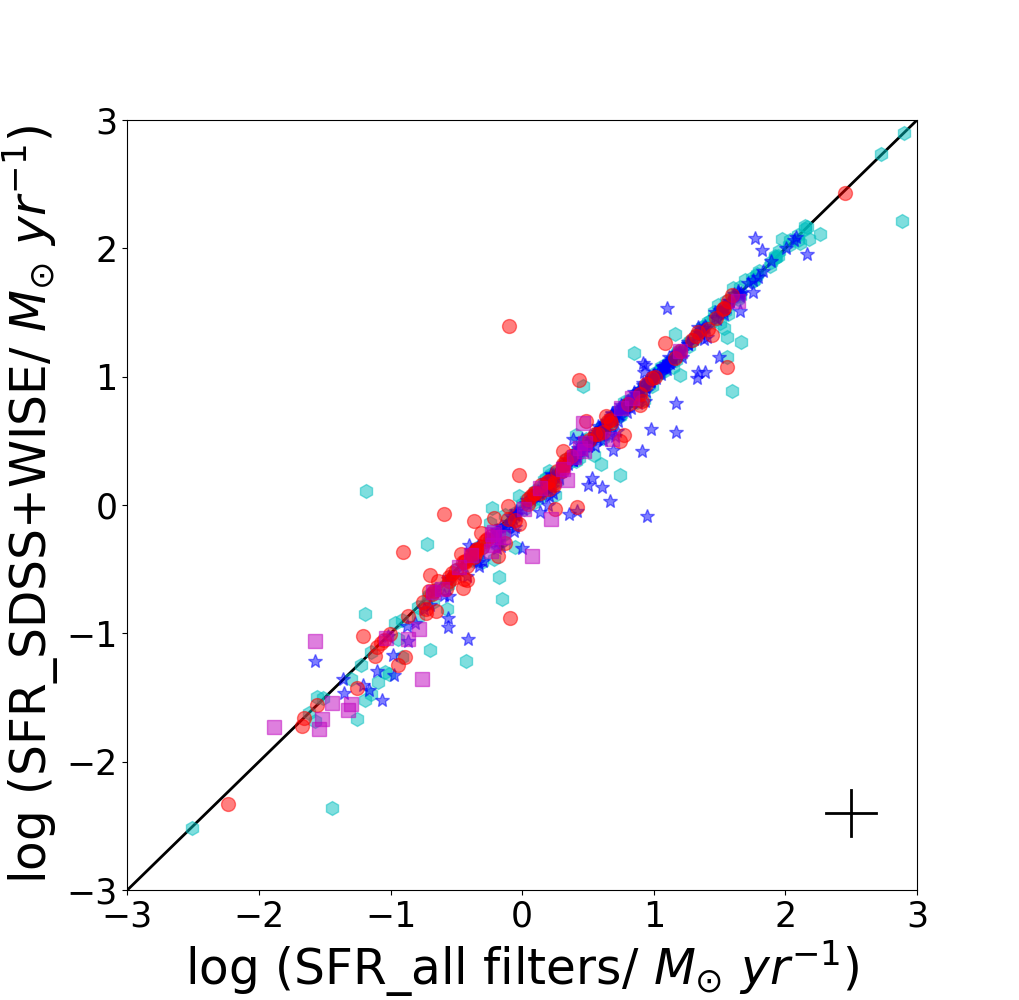}
  \caption{ Comparison of the M$_*$ (top) and SFR (bottom) estimated by MAGPHYS 
  when using all filters (see text) on the $x$-axis and only SDSS+WISE filters on the $y$-axis. 
  The typical error is shown in the bottom right corner of each panel. The sample is colour-coded according to morphology 
  as indicated in the legend. 
}  
\label{fig:comp_ALLfilters_WISESDSS}
\end{center}
\end{figure}

\begin{figure*}[!htb]
  \includegraphics[width=0.5\textwidth,clip]{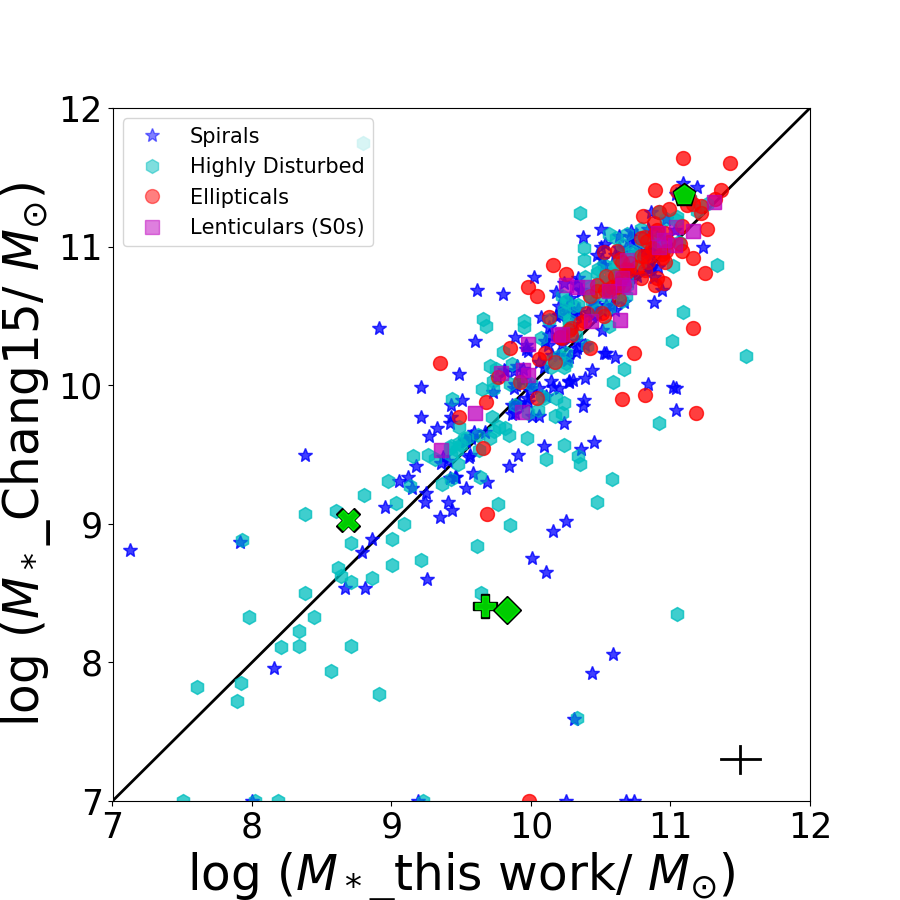}
  \includegraphics[width=0.5\textwidth,clip]{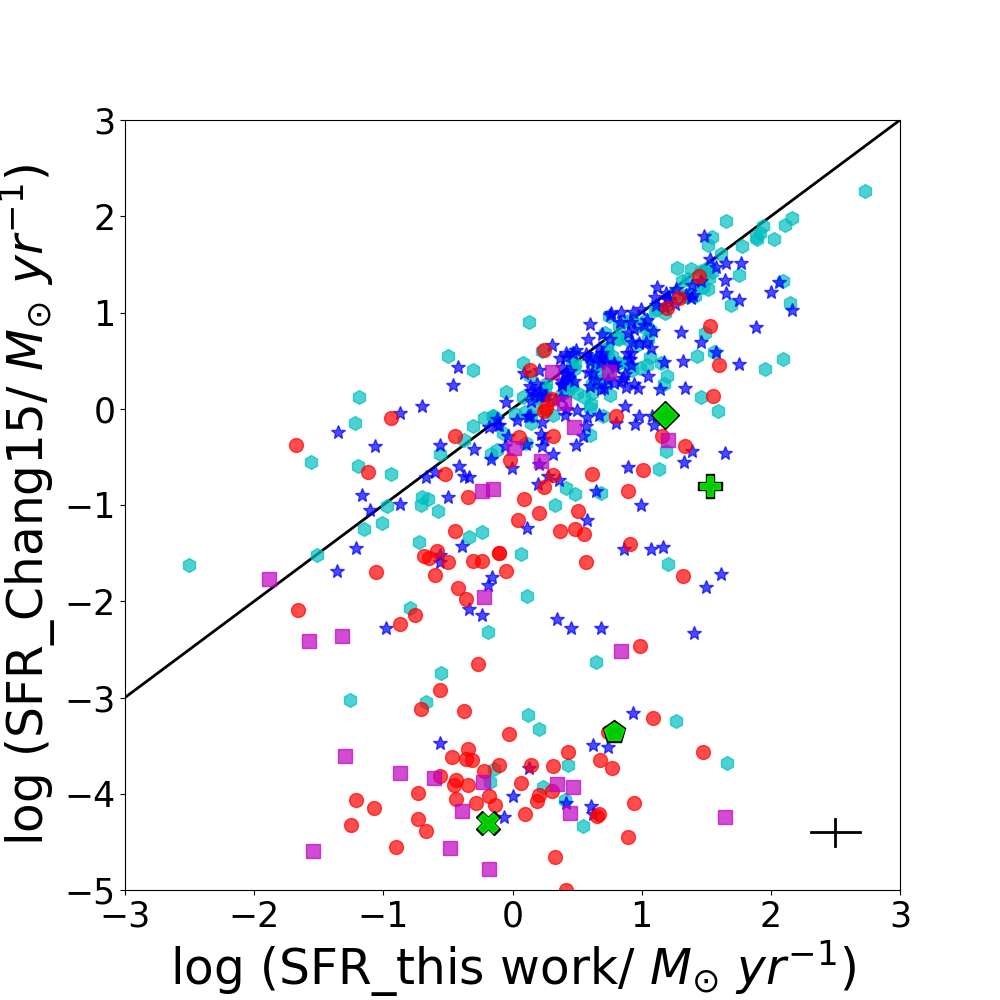}
  \caption{ Comparison between our M$_*$ (left panel) and SFR (right panel) estimates 
  to the estimates listed by \textsc{Chang15}. 
    The typical error is shown in the bottom right corner of each main panel. 
    Data points at the lowest $y$-axis values in the left panel represent galaxies with the minimum value 
    set by \textsc{Chang15}. These data points are not considered for the fits. 
    The merging galaxies are coloured by morphology as shown in the legend. 
}  
\label{fig:Comps}
\end{figure*}

 We compare the results obtained by MAGPHYS using all the filters with those obtained using only SDSS+WISE 
 (see Fig. \ref{fig:comp_ALLfilters_WISESDSS}). We can see that they correlate and have a scatter of 
 0.1 dex for  M$_*$ (top panel) and 0.2 dex for SFR (bottom panel), with no apparent systematic offsets. 
 The scatter may be larger for SFR compared to M$_*$ because UV is a sensitive tracer of 
 recent star formation. This means that including GALEX may not cause large differences in measurements 
 for the majority of the sources, but it can lead to differences as large as 10 and 15 times for M$_*$ and 
 SFR, respectively, in individual galaxies. We colour-coded the merging galaxies by morphology in order to look 
 for dependences. 
 The scatter in M$_*$ is dominated by spirals and highly disturbed galaxies; instead, for the SFR the scatter is similar for all morphologies.

Figure \ref{fig:Comps} shows the comparison between our M$_*$ (left) and SFR (right) results and 
the values listed in the \textsc{Chang15} catalogue. 
We cross-matched our mergers to this catalogue using a distance limit of 5", where all 
coordinates come from SDSS. The mean size of our mergers is 42", thus the distance limit is very small 
compared to the size of the merging galaxies. 
In the case of overlapping galaxies, we only considered the one with the minimum distance. 

Since we also have the SDSS+WISE results, we can directly compare our results to the \textsc{Chang15} 
values. In both cases MAGPHYS is used to estimate M$_*$ and SFR.
The only difference is that \textsc{Chang15} used the measurements shown in SDSS and AllWISE tables 
with an additional correction for W1-3 fluxes based on the aperture size in the SDSS r band, 
whereas we use our own semi-automated photometric approach. 
Thus, any differences that arise are primarily the result of the photometric methodology. 
Stellar masses show good agreement for most of the sample with a scatter of 0.5 dex, considering 
only \textsc{Chang15} sources with $\rm M_* >10^7$\msun, and a scatter of 0.76 dex for SFR, 
considering only \textsc{Chang15} sources with SFR $> 10^{-3} \msun \rm yr^{-1}$. 
The comparisons of M$_*$ and SFR show no dependence on merger stage or separation 
(see Fig. \ref{fig:stellarmass_comps2}). 
The comparison of SFRs shows a large scatter, which indicates that SFRs are more affected by small
differences in photometry compared to M$_*$. 
There are many galaxies with very low SFRs estimated by \textsc{Chang15}, which can be 
underestimations of this property either because  the UV emission is not considered and/or because of  
overcorrections of the W1, W2, and W3 fluxes. This leads to 
changing the SED shape affecting the M$_*$ and SFR estimates, resulting in differences up to 
500 times in M$_*$ and 5000 times in SFR.

The correlation does not clearly depend on morphology, but the scatter of M$_*$ 
is dominated by spirals and highly disturbed galaxies. For the SFR, a large fraction of spirals and 
highly disturbed galaxies are close to the one-to-one relation. However, there is still a large scatter, 
similar to ellipticals and lenticular galaxies. 

We made a linear fit to the distribution of data points in Fig. \ref{fig:Comps}. 
The parameters of the best fit and the scatter about that fit are provided in Table \ref{tab:offsets}.

\begin{table}[!htb]
\begin{center}
\caption{List of the best-fit parameters obtained from the comparison between our M$_*$ and SFR 
results and those listed in the \textsc{Chang15} catalogue.  }
\tiny
\centering
\resizebox{0.4\textwidth}{!}{
\begin{tabular}{c c c c}
\hline
         & slope       & intercept      & scatter   \\ \hline
\hline
M$_*$  & 0.89 & 1.07 & 0.49\\ \hline
SFR & 0.81 & -0.42 & 0.76 \\
\end{tabular}
}
\label{tab:offsets}
\end{center}
\end{table}

\begin{figure*}[!htb]
\begin{center} 
  \includegraphics[width=0.45\textwidth,clip]{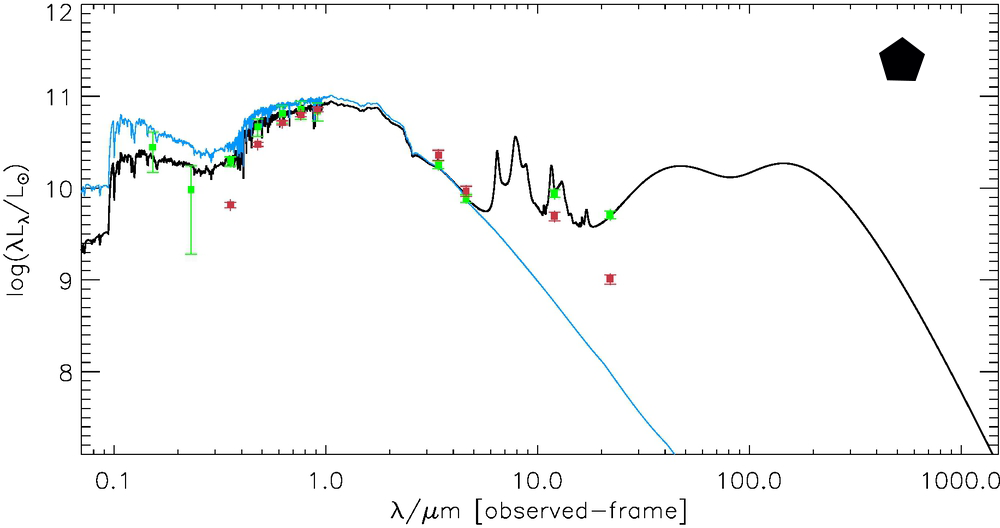}
  \includegraphics[width=0.25\textwidth,clip]{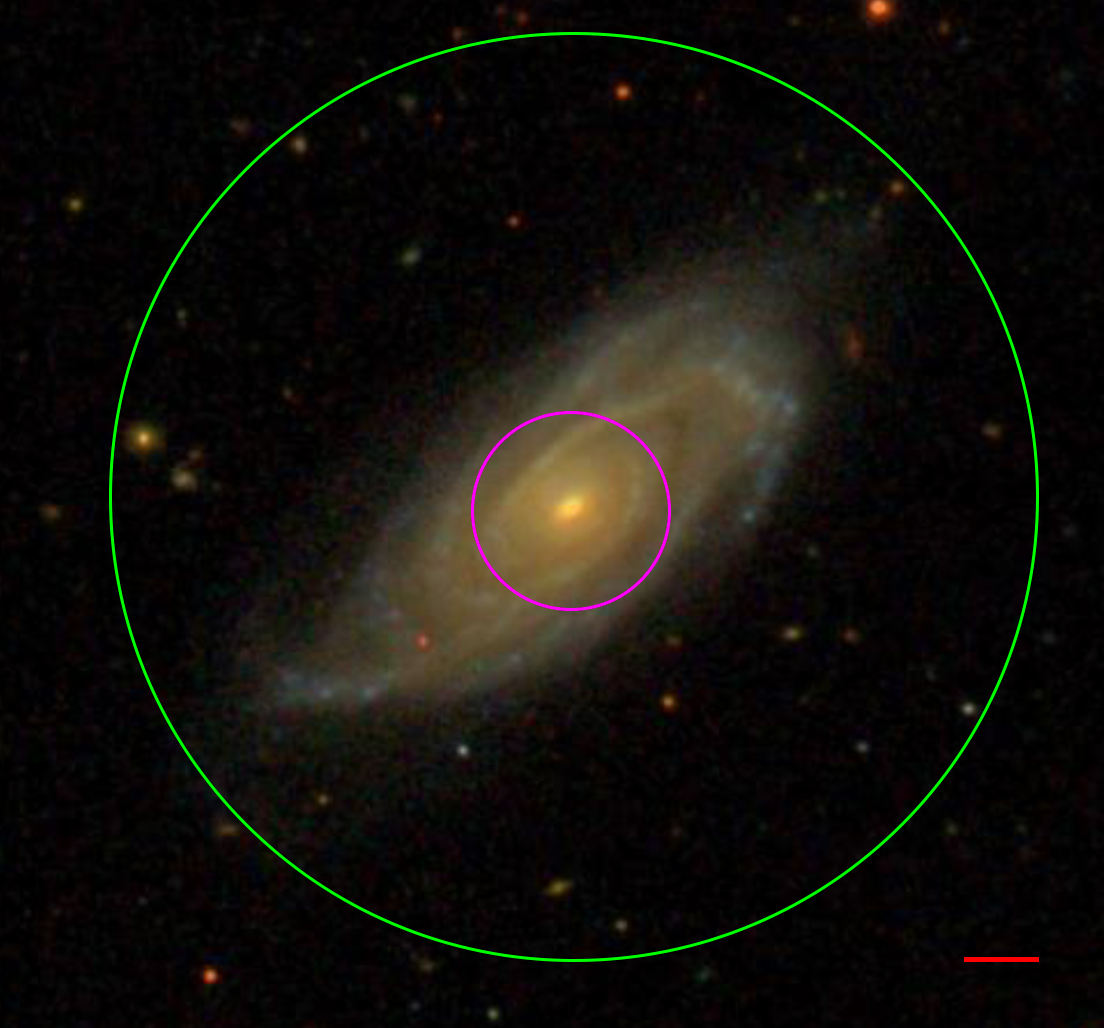}
  
  \includegraphics[width=0.45\textwidth,clip]{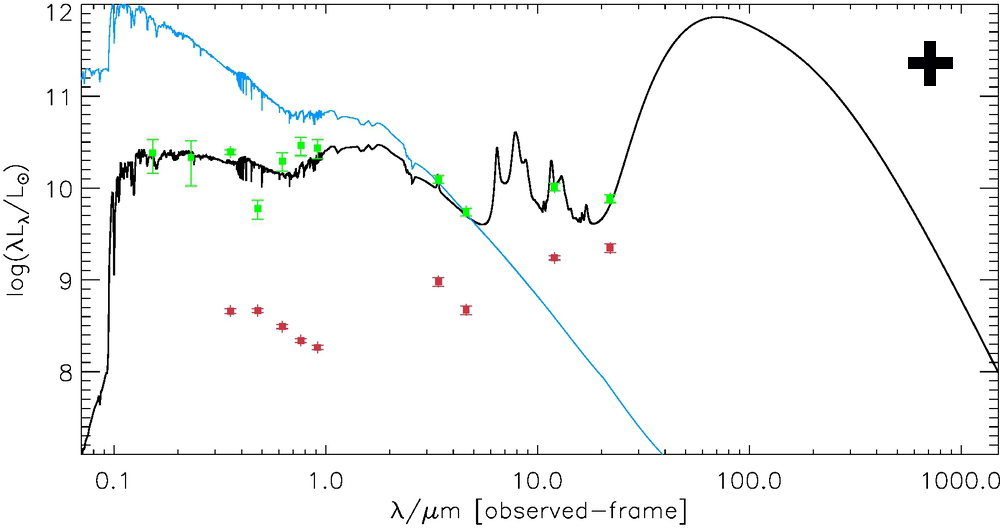}
  \includegraphics[width=0.25\textwidth,clip]{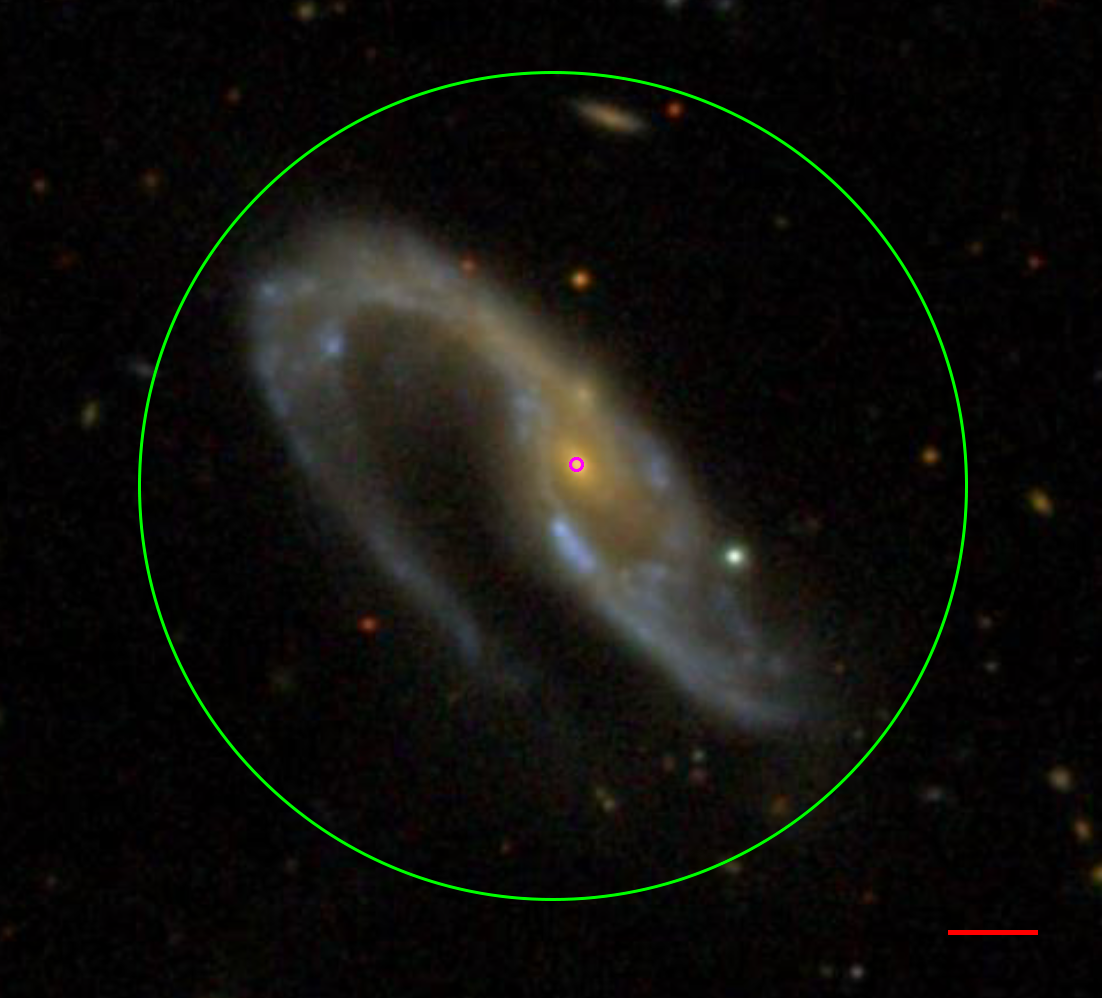}
  
  \includegraphics[width=0.45\textwidth,clip]{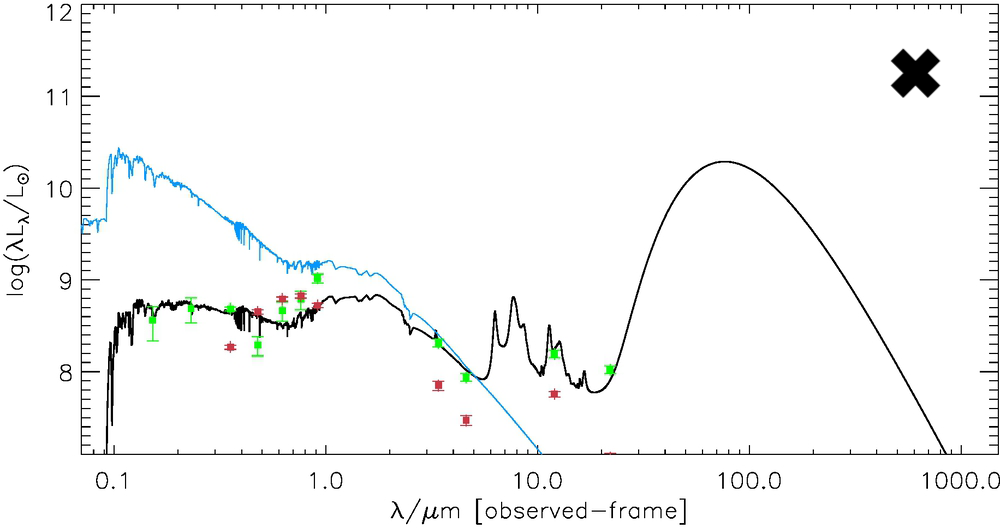}
  \includegraphics[width=0.25\textwidth,clip]{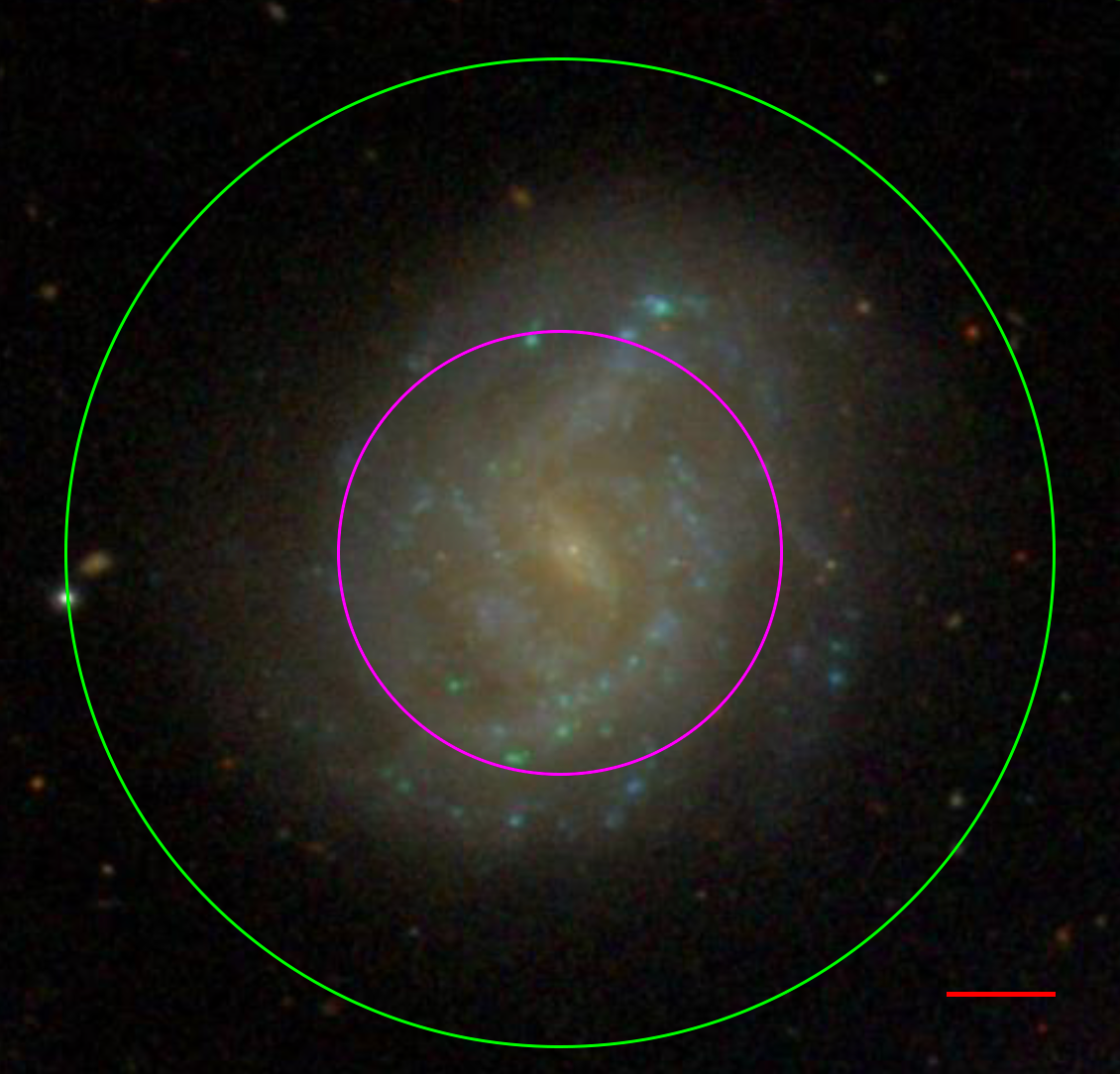}
  
  \includegraphics[width=0.45\textwidth,clip]{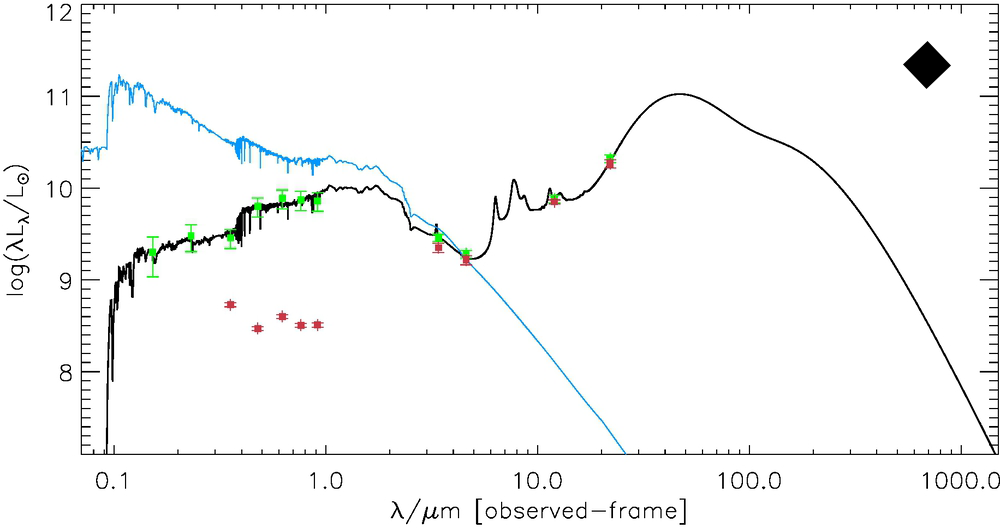}
  \includegraphics[width=0.25\textwidth,clip]{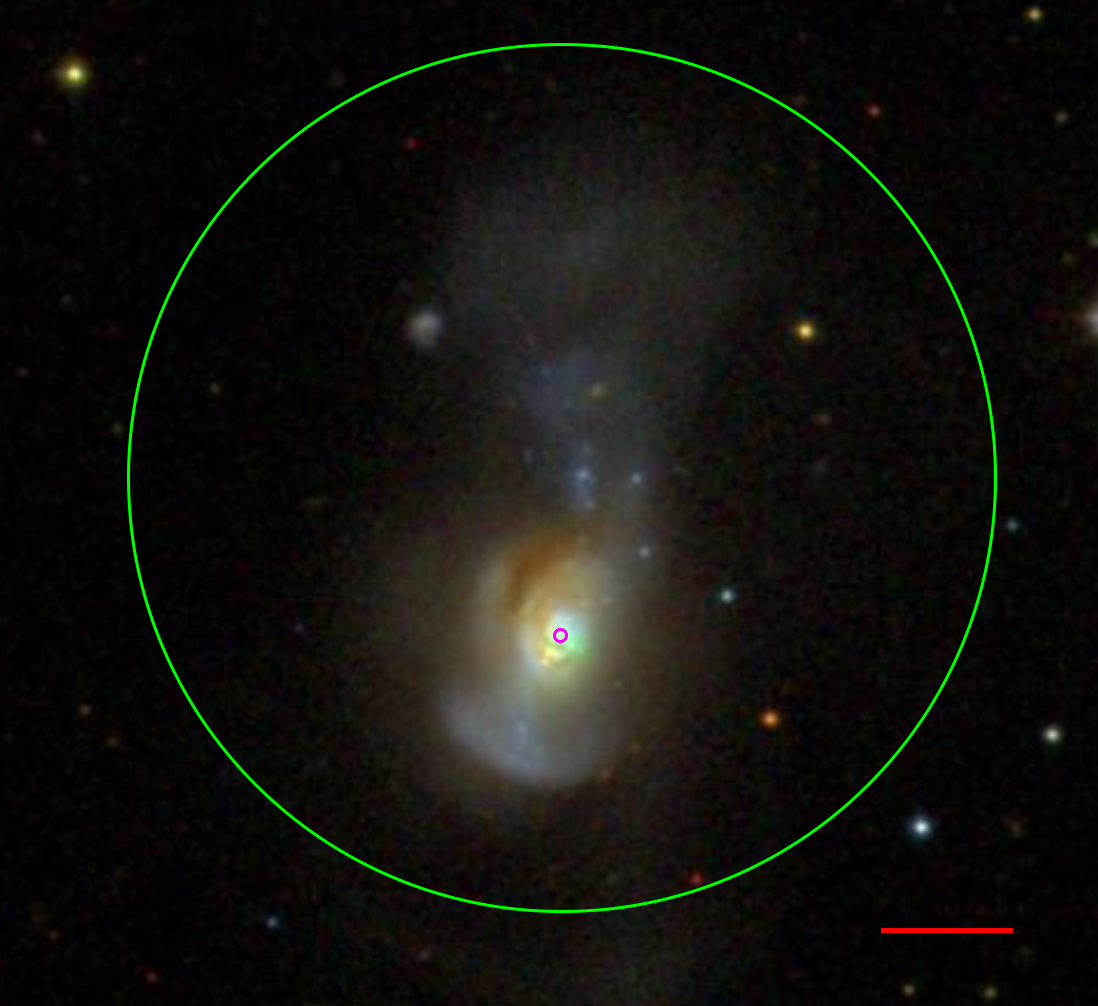}
  \caption{ SED and $ugriz$  images for    galaxies with large differences between 
  our M$_*$ and SFR results and the  \textsc{Chang15} results. The symbol shown in the top right corner of 
  each SED corresponds to the galaxy symbol in Figure \ref{fig:Comps}. The black and blue lines show the 
  attenuated and unattenuated SED fit to our measurements in green. The red dots 
  show the fluxes used by \textsc{Chang15}. The $ugriz$ image show the apertures measured in 
  our study in green and the one measured by SDSS in magenta. The red line represents 20" in each image. 
}  
\label{fig:SEDscomp}
\end{center}
\end{figure*}

In order to better understand the largest differences between our results and \textsc{Chang15}, 
we  plotted the SEDs of some galaxies showing some of the largest differences for M$_*$ and/or SFR. 
Figure \ref{fig:SEDscomp} shows the SEDs of the galaxies shown in Figure \ref{fig:Comps} indicated by the 
same symbol shown in the top right corner of each SED. Next to each SED, we show the $ugriz$ image of each 
galaxy with the aperture we use in green, and the SDSS aperture used by \textsc{Chang15} in magenta.

From top to bottom, as shown in Figure \ref{fig:SEDscomp}:

\pentagon The upper panel shows that most of the fluxes shown by \textsc{Chang15} are very similar 
to our measurements, even when their aperture is much smaller than ours. However, the $u$ and W4 
shows lower values for \textsc{Chang15}. This can be the reason why their SFR value is much lower than 
ours, since these wavelengths show the emission and re-emission of young star formation, respectively. 

\ding{58} The second panel shows a galaxy that had not been measured accurately either by SDSS or 
WISE. The SDSS aperture is very small, barely seen in the $ugriz$ image. Thus, our M$_*$ and SFR values 
are very different to the  \textsc{Chang15} values. 

\ding{54} The third panel shows that the SDSS photometry shows a very different shape for the SED. 
However, the fluxes are not very different at these wavelengths. On the other hand, \textsc{Chang15} 
shows lower fluxes for WISE compared to ours. 
Furthermore, there is no W4 flux shown by \textsc{Chang15}. The similarity in the SDSS fluxes might 
be the origin of the similarity in M$_*$, and the higher WISE fluxes of our measurements can explain our 
higher SFR estimate. 

{\Large $\diamond$} The lower panel shows a very small aperture for SDSS, showing very low fluxes compared to our measurements. 
For WISE, however, the measurements are very similar. This results in very different M$_*$ and SFR 
estimates.

This shows that there are several factors affecting the difference in the results, such as differences in the 
photometric data of one survey compared to the other, low measurements in all of the filters, or 
different measurements made in one or two filters which change the shape of the SED. 
Hence, mergers must be studied with extreme caution if the catalogued values are to  be used.

\subsection{Testing common estimators of M$_*$ and SFR} \label{sec:Inds}

\begin{figure*}[!htb]
\begin{center} 
  \includegraphics[width=0.48\textwidth,clip]{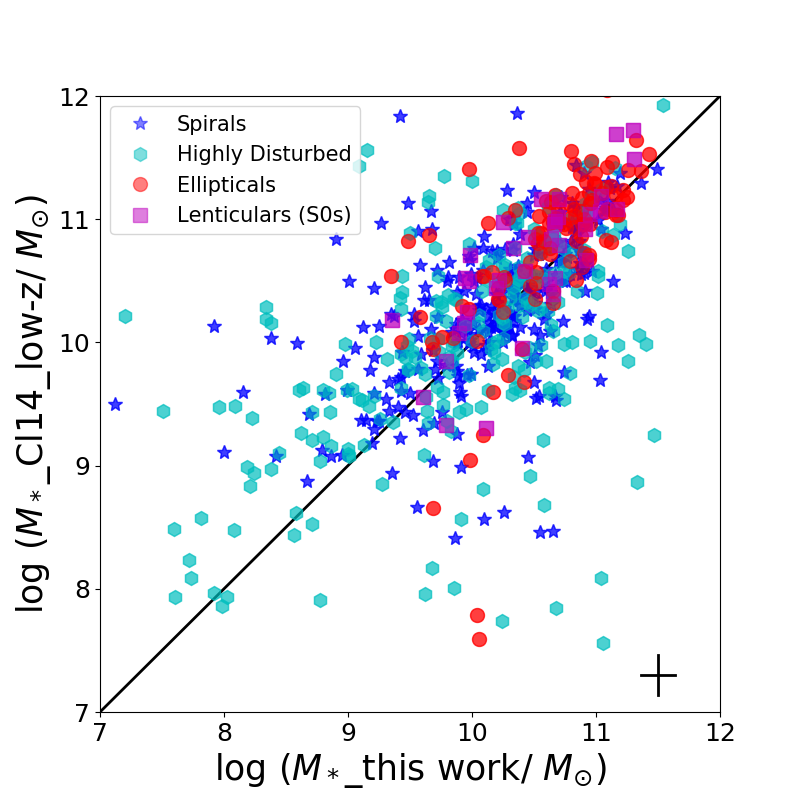}
  \includegraphics[width=0.48\textwidth,clip]{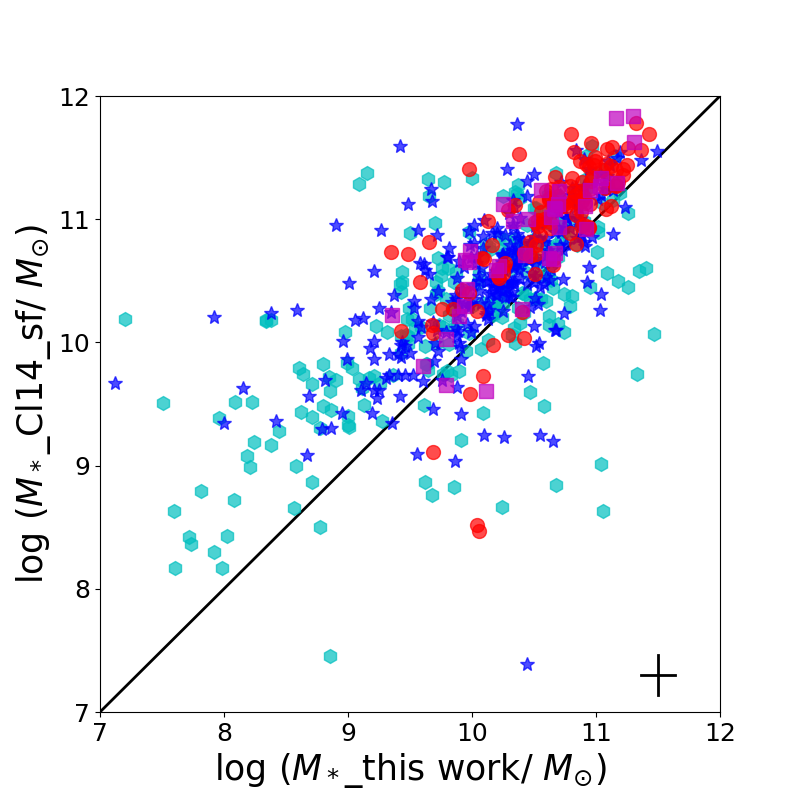}
  \caption{ Comparison of our M$_*$ values (derived from our MAGPHYS fits to data in 
  GALEX, SDSS, and WISE filters) to those estimated using only WISE W1 and W2 photometry combined 
  with the relations provided in \textsc{Cl14}. The left panel shows the comparison to the relation for 
  low-redshift sources, and the right panel shows the comparison to the relation for star-forming galaxies 
  (see text). The typical error is shown in the bottom right corner of each panel. 
  The coloured symbols show the morphology of the merging galaxies as described in the legend. 
}  
\label{fig:CluversComp}
\end{center}
\end{figure*}

  In this section, we consider how various M$_*$ and SFR indicators perform on samples of mergers. 
  The indicators we show in this section were derived using different methods to those used in  this study, but using 
  some of the filters that we use. Thus, we can compare commonly used M$_*$ and SFR indicators 
  from the literature with our results measured using MAGPHYS. The M$_*$ and SFR computed 
  for this section were calculated using our new photometric values and the relations from the literature. 
  We start by comparing M$_*$ indicators.

We also contrast our MAGPHYS-derived M$_*$ values with those estimated from one- or two-band 
photometry by \citet[hereafter \textsc{Cl14}]{Cluveretal14}, \citet[hereafter B03]{Belletal03}, and 
\citet[hereafter T11]{Tayloretal11}. 
\textsc{Cl14} have studied two of their equatorial fields in the 
Galaxy and Mass Assembly (GAMA) Survey. 
They note that `the typical W1 1-$\sigma$ isophotal radius is more than a factor of $\sim$2 in scale 
compared to the equivalent 2MASS $K_s$-band isophotal radius', and that WISE {\texttt{gmags}} should be 
used with caution since no deblending or star subtraction has been made in WISE tables, which is 
further reason to apply our semi-automatic approach.
They also show empirical relations between the M$_*$ derived from 
synthetic stellar population models and W1 and W2 colours and W1 luminosity, 
which they separate into three equations following the form 
$ log M_{stellar} / L_{W1} = a(W1 - W2) - b$, with 
$ L_{W1} (\lsun) = 10^{-0.4(M_{W1}-3.24)}$, where $\rm M_{W1}$ is the absolute magnitude in W1. 
For low-redshift sources {\it a} and {\it b} are $-$2.54 and 0.17, respectively; 
for star-forming galaxies {\it a} and {\it b} are 0.04 and $-$1.93, repectively;  and  {\it a} = $-$1.96 and 
{\it b} = 0.03 are the best-fit values for the entire sample.

Figure \ref{fig:CluversComp} shows the comparison between our M$_*$ and 
the estimates using the \textsc{Cl14} relations. The left panel shows the M$_*$ estimated using the 
\textsc{Cl14} relation for low-redshift galaxies, and the right panel shows the M$_*$ estimations 
using the \textsc{Cl14} relation for their entire sample, and using the star-forming relation for 
star-forming galaxies. Both methods tend to provide higher values of M$_*$ compared to our 
results, and with large scatter  (1.1 dex and 1.0 dex, respectively), leading to 
differences of up to a factor of 1000. 
The correlations of these comparisons do not show a clear dependence on the morphology of the 
merging galaxies, but the scatter is dominated by spirals and highly disturbed galaxies.

\begin{figure*}[!htb]
\begin{center} 
  \includegraphics[bb=0 10 550 550,width=0.45\textwidth,clip]{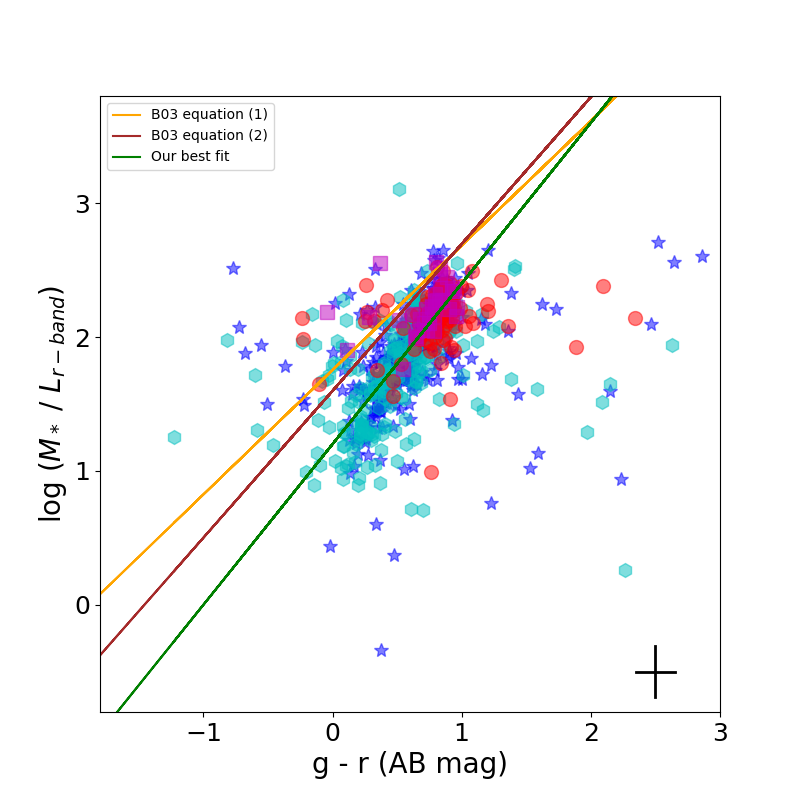}
    \includegraphics[bb=0 10 550 550,width=0.45\textwidth,clip]{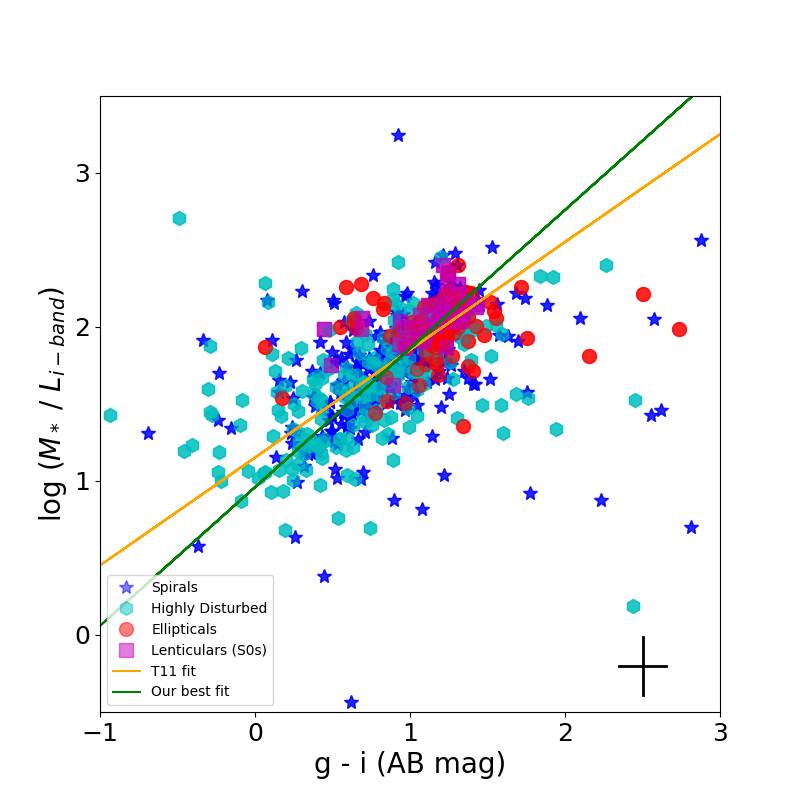}
       \caption{ Left: Ratio of  M$_*$ to light  versus the colour relation from B03. 
  The orange line shows equation (1) and the brown line shows equation (2) from B03. 
  Our best fit is shown in green. 
   Right: Ratio of  M$_*$ to light  versus colour relation from T11. 
  The orange line shows M$_*$-colour relation from T11. 
  Our best fit is shown in green. The typical error is shown in the bottom right corner of each panel. 
  The merging galaxies are colour-coded according to their morphology as indicated in the legend.  
}  
\label{fig:MstellarL_colour}
\end{center}
\end{figure*}  
  
  The  relation shown by B03 is frequently used, this  
  relates the M$_*$ to optical colours following 
  $\rm log(M_*/L_r) $ $=$ $-0.15+0.93(g-r)$(1) and $\rm log(M_*/L_r) $ $=$ $-0.306+1.097(g-r)$(2) for 
  galaxies in the range $0.3 < (g-r) < 1$, with $L_r$ being the luminosity ($\lsun$) in the r band. 
Figure \ref{fig:MstellarL_colour} (left panel) shows the relation between 
  $\rm log(M_*/L_r)$ and the ($g-r$) colour. Equations (1) and (2) from B03 are shown by the 
  orange and brown lines, respectively. Our best fit (in green) shows a steeper slope than for B03 sample:
  $\rm log(M_*/L_r) $ $=$ $0.9+1.69(g-r)$   
  and redder colours for 
  a large fraction of our sample. This could be related to the higher dust mass of merging galaxies 
  compared to unperturbed galaxies.

Figure \ref{fig:MstellarL_colour} (right panel) shows the relation between the $\rm M_*/L_i$ and 
the ($g-i$) colour. 
The orange line shows the relation determined by T11 for the GAMA sample: 
$\rm log(M_*/L_i) $ $=$ $-0.68+0.70(g-i)$, with $\rm L_i$ being the luminosity in the i band, in 
$\rm \lsun$. Our best fit (green line) shows a steeper slope, 
$\rm log(M_*/L_i) $ $=$ $0.96+0.90(g-i)$, but no offset with colour. 
This suggests that T11 can properly correct for dust, as they also use near-infrared filters (WISE), and 
they additionally include Herschel. 
The large scatter could be due to different photometry used for their and our calculations.  

These two relations do not show a clear dependence on morphology and the scatter is large for all 
morphologies. 
However, spirals and highly disturbed galaxies show larger scatter compared to elliptical and lenticular galaxies.

\begin{figure*}[!htb]
\begin{center} 
  \includegraphics[width=0.48\textwidth,clip]{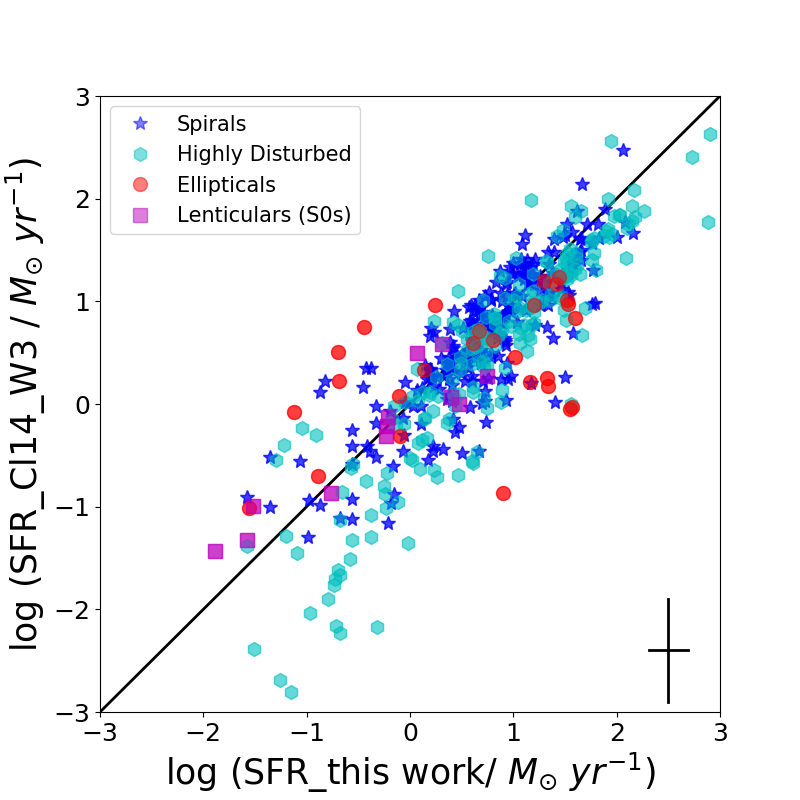}
  \includegraphics[width=0.48\textwidth,clip]{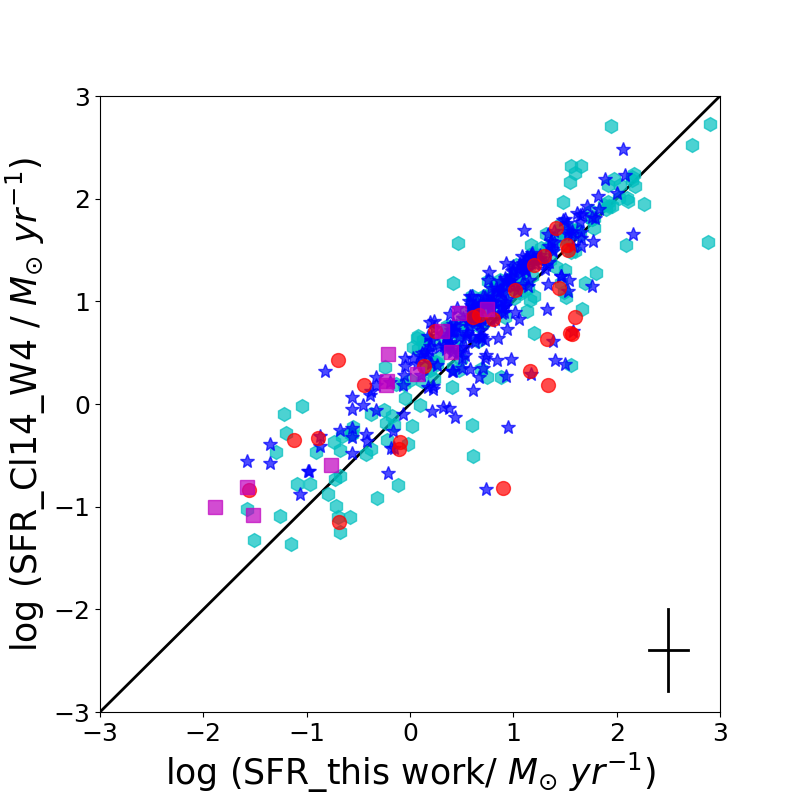}
  \caption{ Comparison of our SFR values (derived from our MAGPHYS fits to data in GALEX, SDSS, and WISE filters) to those estimated using only WISE W3 and W4 photometry combined with the relations provided in \textsc{Cl14}. 
  The left panel shows the comparison to the relation using W3, and the right panel shows the 
  comparison to the relation using W4. 
   The typical error is shown in the bottom right corner of each panel. 
   The sample is colour-coded according to morphology, as shown in the legend. 
}  
\label{fig:CluversComp2}
\end{center}
\end{figure*}

\begin{figure*}[!htb]
\begin{center} 
  \includegraphics[width=0.48\textwidth,clip]{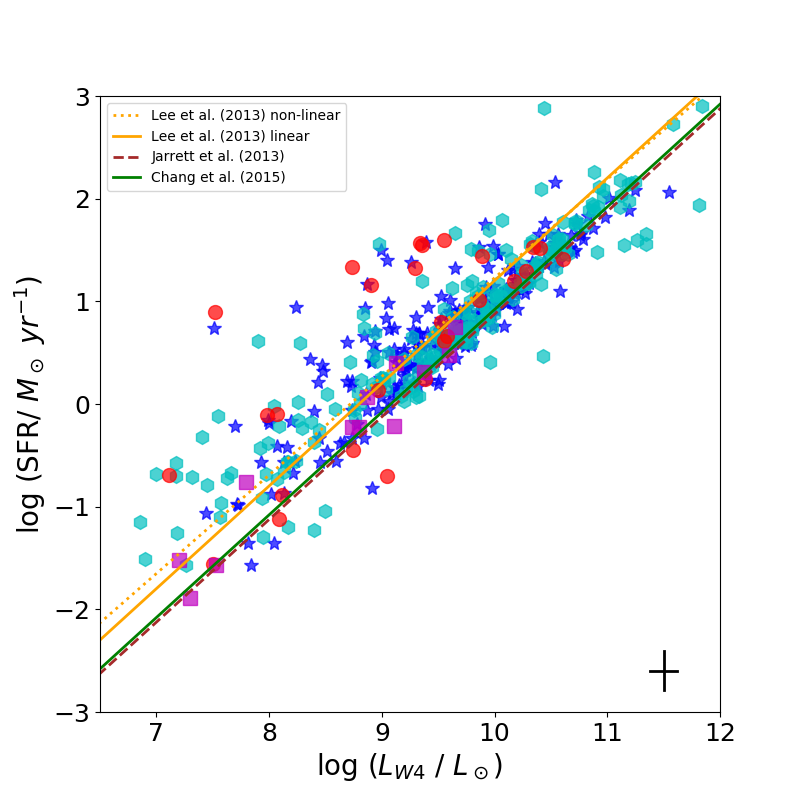}
  \includegraphics[width=0.48\textwidth,clip]{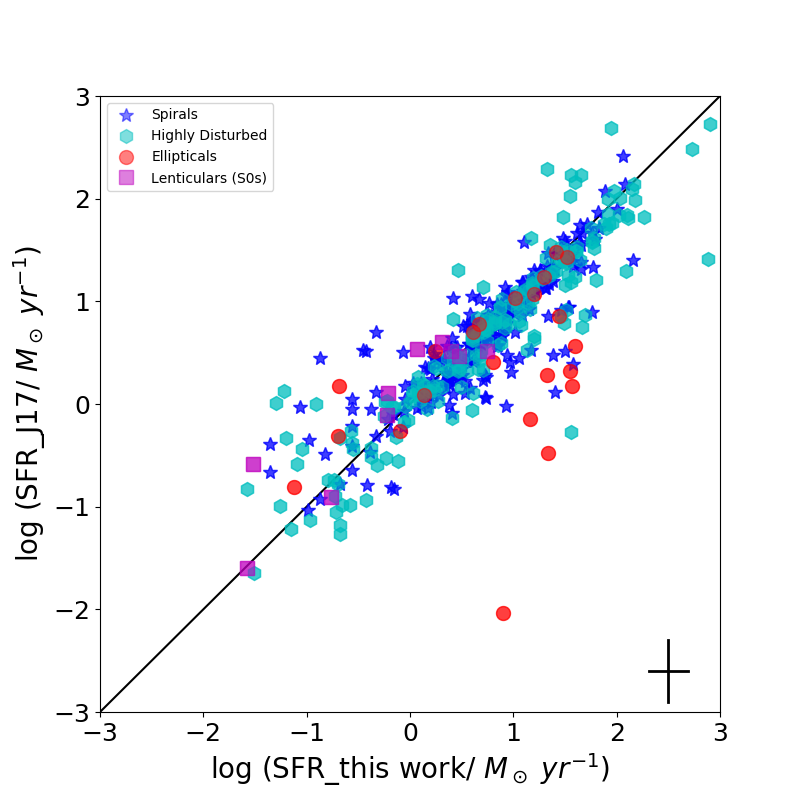}
  \caption{ Left panel: Relation between the SFR and the W4 luminosity ($L_{W4}/L_\odot$). 
  The lines show different relations from the literature. The orange dotted line shows the non-linear 
  relation found by \citet{Leeetal13}, while the orange solid line shows the linear relation from the same study. 
  The brown dashed line shows the relation by \citet{Jarrettetal13} and the green line shows the relation 
  found by \textsc{Chang15}.
  Right panel: Comparison between our MAGPHYS SFR results and the calculation using the SFR relation 
  defined by J17. The black line shows the equality line. 
  The typical error is shown in the bottom right corner of each panel. 
  The coloured symbols represent the morphology of the merging galaxies, as described in the legend. 
}  
\label{fig:SFR_Ind}
\end{center}
\end{figure*}  

We now compare our MAGPHYS-derived SFR results to those estimated by \textsc{Cl14}, 
\citet{Leeetal13}, \citet{Jarrettetal13}, \textsc{Chang15}, and \citet[hereafter J17]{Janowieckietal17}. 
\textsc{Cl14} also present a relation between a dust-corrected $H_{\alpha}$-derived SFR and W3 and W4 
luminosities separately: 

$\rm log~ SFR_{H_{\alpha}}(\msun~yr^{-1}) = 1.13~ log~ \nu L_{W3}(\lsun)-10.24,$

 $\rm log~ SFR_{H_{\alpha}}(\msun~yr^{-1}) = 0.82~ log~ \nu L_{W4}(\lsun)-7.3.$

Figure \ref{fig:CluversComp2} shows the comparison between our SFR and the estimates from the 
\textsc{Cl14} relations for W3 (left panel) and for W4 (right panel). \textsc{Cl14} SFR correlate closely to 
our results, showing larger scatters for SFR estimated from W3 compared to estimations from W4 
(0.5 compared to 0.4, respectively). 
This could occur because W3 is more affected by PAH 
emission, as shown in previous studies \citep[\textsc{Cl14}, \textsc{Chang15}]{Jarrettetal13}. 
It is important to note that even if they seem to relate closely, the estimations can lead to differences of 
up to a factor of 50 for the W4 relation and a factor of 500 for the W3 relation.

The scatter in the results using the W3 relation is large for all morphologies. Nevertheless, a large fraction of 
spirals and highly disturbed galaxies seem to be closer to the one-to-one relation. The comparison to the SFR 
using the W4 filter show no dependence on mophology.

 The left panel of Fig. \ref{fig:SFR_Ind} shows the relation between the SFR and the luminosity in W4. 
We show different relations found in the literature. \citet{Leeetal13} show two relations, a non-linear 
(orange dotted line) and a linear relation (orange solid line). The brown dashed line shows the relation found by 
\citet{Jarrettetal13} and the green line shows the relation found by \textsc{Chang15}. 
Our relation seems to be  between the \citet{Leeetal13}  and \textsc{Chang15} relations. 
There is a large scatter ($\sim$0.4 for the \citealt{Leeetal13} equations, and $\sim$0.5 for \citealt{Jarrettetal13} and \textsc{Chang15}), showing higher SFR for the same $\rm L_{W4}$, which might indicate that 
$\rm L_{W4}$ does not fully trace all the SFR of the galaxy. This result does not depend on the morphology of 
the merging galaxy. 

Figure \ref{fig:SFR_Ind} right panel shows a comparison of the SFR we derive from MAGPHYS and that 
derived using the SFR estimator of J17. The  J17 approach combines 
the SFR from the NUV light and the attenuated light in W4 following 
$\rm SFR = SFR_{NUV} + SFR_{W4}$, with \\
$\rm SFR_{NUV}(\msun~yr^{-1}) = 10^{-28.165} * L_{NUV} (erg/s/Hz)$ \\
from \citet{Schiminovichetal07} and \\
$\rm SFR_{W4}(\msun~yr^{-1}) = 7.50x10^{-10} (L_{W4} - 0.04 L_{W1})(\lsun)$ \\ 
from \citet{Jarrettetal13} with an extra correction for stellar contamination. 
Our results and those of   J17 both show a tight correlation for a large fraction of the sample, except for galaxies 
with low SFR. The correlation and scatter are very similar for all morphological classifications. 

\subsection{Colour-morphology relation} \label{sec:col-morph}

\begin{figure}[!htb]
\begin{center} 
  \includegraphics[width=0.5\textwidth,clip]{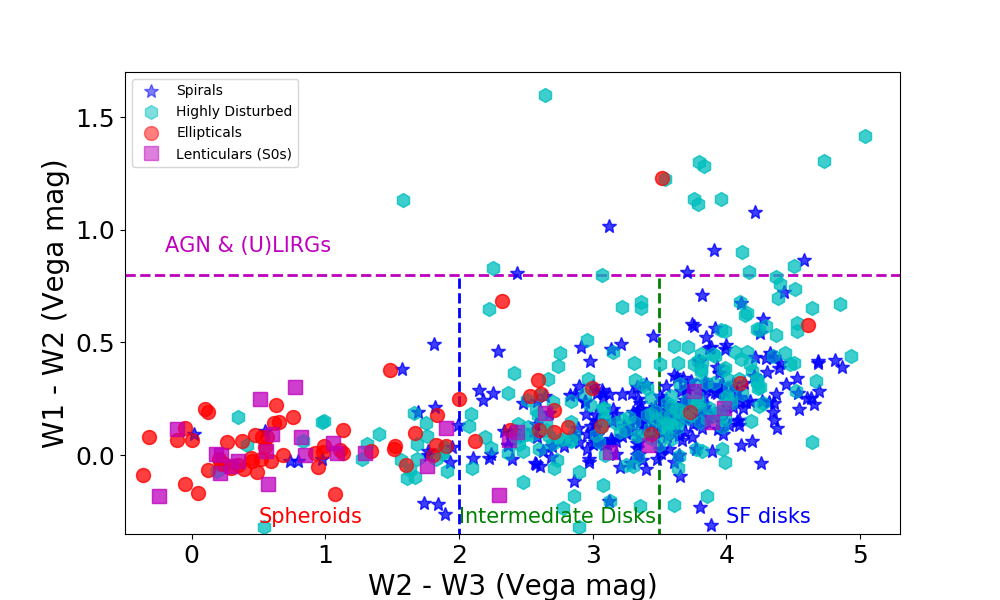}
  
  \caption{ WISE colour-colour diagram showing morphology separation, as in Fig. 5 of 
  \textsc{Cl14}, where galaxies can be classified as Spheroids, Intermediate discs, 
  star-forming discs, and AGNs and (U)LIRGs. Our merging galaxies are colour-coded according to morphology, as indicated 
  in the legend.
}  
\label{fig:JarrettCCDs}
\end{center}
\end{figure}

The mid-infrared colour-colour relation distinguishes galaxies by morphology, luminosity, and AGN content. 
Figure \ref{fig:JarrettCCDs} shows the colour-colour relation \citep{Jarrettetal11, Cluveretal14, Jarrettetal17}, 
which allows us to identify galaxies as   AGNs and  ULIRGs, and 
also classifies galaxies by morphology, for example spheroids (ellipticals and lenticulars) and discs 
(intermediate and star-forming). The lines shown in the figure follow the separations shown by \textsc{Cl14}.

We can see that, overall, merging galaxy morphologies differentiate similarly to unperturbed galaxies 
on the WISE colour-colour diagram. Elliptical (red) and lenticular (magenta) galaxies are mostly in the 
spheroid region, and spiral (blue) and highly disturbed (cyan) galaxies are in the disc region. 
We can see a high number density of data points in the star-forming disc region where we expect to find 
spirals and starburst galaxies. 
It is also important to note that, as expected, merging galaxies can be found spread over this plot; 
we can see blue ellipticals and red spirals as they change colour due to the merging process.

\subsection{Specific star formation rate} \label{sec:SFR-Mplane}

\begin{figure}[!htb]
\begin{center} 
  \includegraphics[bb=10 20 590 580, width=0.4\textwidth,clip]{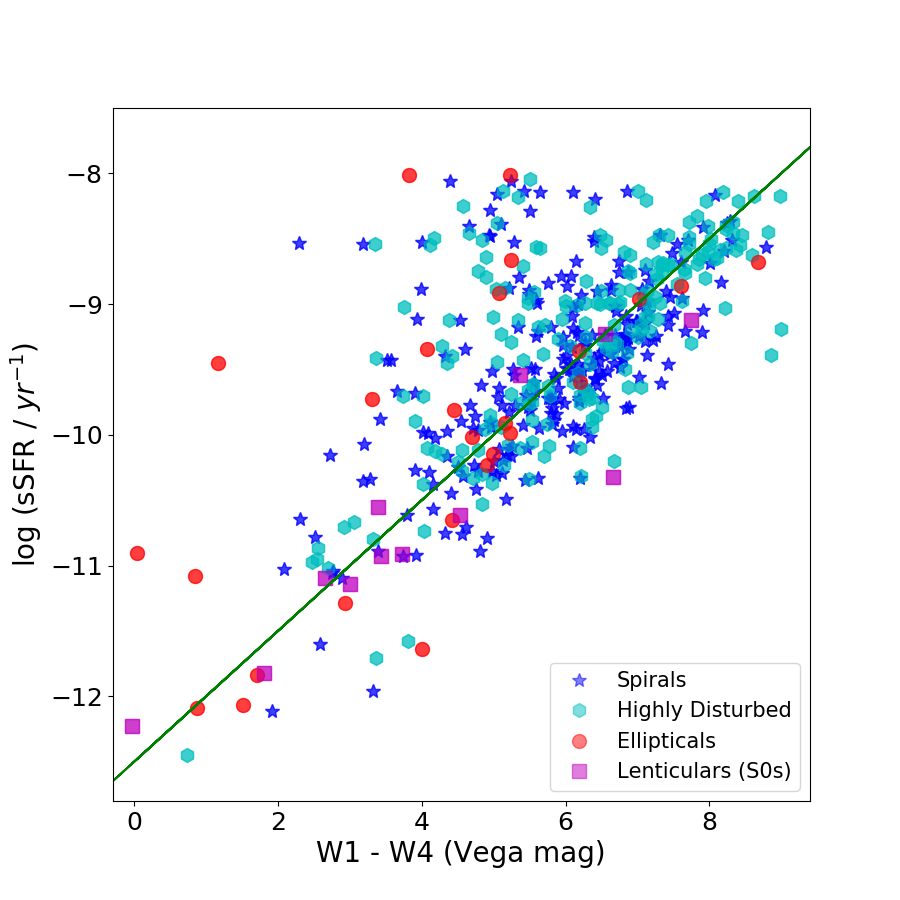}
  \caption{ sSFR - WISE colour relation for mergers. The green line shows the best fit to our sample 
  (see text). The coloured symbols show the morphology of our mergers (see legend).
}  
\label{fig:sSFRWISEi_colour}
\end{center}
\end{figure}  

In order to look for a specific star formation rate (sSFR) indicator using one or two photometric bands, 
we look for a relation between 
this parameter and a combination of the stellar component and the dust/obscured star-forming 
component, which can be traced by the mid-infrared colour W1 - W4 (Vega mag) from the WISE filters. 
Figure \ref{fig:sSFRWISEi_colour} shows the relation found for our merger sample, which 
is described by Equation \ref{eq:sSFR_W1W4}:
\begin{equation}
\rm log (sSFR/ yr^{-1}) = 0.34 ~ (W1 - W4) - 11.42 
\label{eq:sSFR_W1W4}
\end{equation}

The scatter of this relation is similar for all morphologies except for lenticular galaxies, which show a 
smaller (0.2 dex) scatter compared to that shown by the other morphologies (0.5 dex).

\section{Public catalogue} \label{sec:catalogue}

The public  catalogue shows all the information we have used for our results. We list all the information of the  merging galaxies, 
showing the same name for pairs and adding a `b' to the name for the companion. Coordinates are 
centred in the individual merging galaxy, coordinates of overlapping galaxies  are centred in each of their nuclei. 
We present the morphology and merger stage as classified in this study. 
We also provide the fluxes and their errors as measured and explained in Sect. \ref{sec:photo}. 
Stellar masses (M$_*$) and star formation rates (SFR) are listed in log-scale, and we also provide the 
$\chi^2$ values of each fit made using MAGPHYS.  
This catalogue will be publicly available on the ViZiER website\footnote{\url{http://vizier.u-strasbg.fr}}; 
a direct link to the catalogue can be found in the online data from ASO/NASA Astrophysics Data System.
See Table \ref{tab:Cat} for the description of the catalogue's columns. 

\begin{table*}[!htb]
\begin{center}
\caption{Output Catalogue}
\centering
\begin{tabular}{c l c l}
\hline
 \# & Name of Column          & Units       & Description      \\ \hline
\hline
  1 & Name                  & String              & Name is in the format PCC\_\# (for companion: PCC\_\# b) \\
  2 & RA                    & degrees             & Right ascension in decimal degrees  \\
  3 & Dec                  & degrees              & Declination in decimal degrees  \\
  4 & z                      & float                   & Redshift  \\
  5 & Morph               & Integer               & Morphology (0=Spiral, 1=Elliptical, 2=S0, 3=HD) \\
  6 & MrgStg              & Integer               & Merger Stage (1,2,3,4,5,6,7,8 = I,II,IIIa,IIIb,IIIc,IVa,IVb,V) \\
  7 & FUV                  & Jy                       & FUV flux  \\
  8 & FUV\_err           & Jy                       & FUV flux error  \\
  9 & NUV                  & Jy                       & NUV flux   \\
  10 & NUV\_err           & Jy                       & NUV flux error  \\
11 & u                       & Jy                       & u flux   \\
12 & u\_err                & Jy                       & u flux error  \\
13 & g                       & Jy                       & g flux   \\
14 & g\_err                & Jy                       & g flux error  \\
15 & r                       & Jy                       & r flux   \\
16 & r\_err                & Jy                       & r flux error  \\
17 & i                        & Jy                       & i flux   \\
18 & i\_err                 & Jy                       & i flux error  \\
19 & z                       & Jy                       & z flux   \\
20 & z\_err                & Jy                       & z flux error  \\
21 & W1                    & Jy                       & W1 flux   \\
22 & W1\_err             & Jy                       & W1 flux error  \\
23 & W2                    & Jy                       & W2 flux   \\
24 & W2\_err             & Jy                       & W2 flux error  \\
25 & W3                    & Jy                       & W3 flux   \\
26 & W3\_err             & Jy                       & W3 flux error  \\
27 & W4                    & Jy                       & W4 flux   \\
28 & W4\_err             & Jy                       & W4 flux error  \\
29 & W\_flag              & Integer               & Photometry flag (0=ok, 1=two components within same aperture)  \\
30 & chi2                    & float                  & $\chi^2$ of the fit performed by MAGPHYS \\
31 & logMst\_Per2p5        & log(M$_*$)        & Percentile 2.5th of the log(M$_*$) \\
32 & logMst\_Per16          & log(M$_*$)        & Percentile 16th of the log(M$_*$) \\
33 & logMst\_Per50          & log(M$_*$)        & Percentile 50th of the log(M$_*$) \\
34 & logMst\_Per84          & log(M$_*$)        & Percentile 84th of the log(M$_*$) \\
35 & logMst\_Per97p5      & log(M$_*$)        & Percentile 97.5th of the log(M$_*$) \\
36 & logSFR\_Per2p5   & log(M$_*$yr$^{-1}$)        & Percentile 2.5th of the log(SFR) \\
37 & logSFR\_Per16     & log(M$_*$yr$^{-1}$)        & Percentile 16th of the log(SFR) \\
38 & logSFR\_Per50     & log(M$_*$yr$^{-1}$)        & Percentile 50th of the log(SFR) \\
39 & logSFR\_Per84     & log(M$_*$yr$^{-1}$)        & Percentile 84th of the log(SFR) \\
40 & logSFR\_Per97p5 & log(M$_*$yr$^{-1}$)        & Percentile 97.5th of the log(SFR) \\

\end{tabular}
\label{tab:Cat}
\end{center}
\end{table*}

\section{Discussion and conclusions} \label{sec:discussion}

We have assembled a sample of 540 mergers in isolated environments. The galaxies forming part of the merger  
were constrained to have similar redshift, which ensures that we are including only galaxies that 
could merge (i.e. we exclude clear cases of fly-bys). 
The mergers have been classified by morphology and  we 
have also classified the mergers according to their current phase in the merging process.

We  performed photometry in a semi-automated manner in order to extract the flux of the entire 
galaxy. We find this is necessary as the automated photometry from publicly available catalogues is  
unreliable for this type of galaxy. Our semi-automated photometry was performed in 11 bands, 
including ultraviolet 
(FUV and NUV from GALEX), optical (u, g, r, i, and z from SDSS), and the near-infrared 
(W1, W2, W3, and W4 from WISE). 

We find that most of the galaxies show higher fluxes using our semi-automated photometry compared 
to the GALEX, SDSS, and AllWISE catalogues. This is a result of the larger apertures that we find are required 
in order to capture all of the galaxy's light in comparison to the fluxes from these catalogues in the literature.
This demonstrates that automated methods are not efficient in extracting all of the light in merging galaxies, 
often missing the light in the outskirts where tidal features may be found, and breaking up the galaxy into 
individual objects instead of recognising they are from the same galaxy (e.g. star-forming regions). 
Also, we find that radial the corrections  performed by \textsc{Chang15}, which were designed to correct for 
aperture effects, seem to result in overestimating  the flux for W1-3. This shows that corrections of this 
type may have poor accuracy in merging galaxies.
These combined results have convinced us to pursue a less automated approach to perform the 
photometry in order to make sure that we are measuring the light of the entire galaxy in all the filters 
mentioned above.

Comparing the results for M$_*$ and SFR using the same method (MAGPHYS) but varying the number 
of filters used as an input (GALEX+SDSS+WISE versus SDSS+WISE), the one-to-one relations are tight for 
most of the sample, while it  can lead to differences of a factor of 10 for M$_*$ and 15 for SFR for 
individual galaxies. This indicates that UV is clearly important to the SFR estimation, as might be expected, 
and that  it is significant for M$_*$ estimations. When we compare our measured fluxes to the 
\textsc{Chang15} fluxes, 
we find that our measurements are higher than theirs, except for W1-3. 
This might be a result  of the aperture correction, which tends to overcorrect the fluxes on these filters. 
This can cause MAGPHYS to fit an altered SED resulting in our measurements having lower 
M$_*$ and higher SFR.

We note that the scatter in M$_*$ and SFR is smaller than the scatter seen when comparing fluxes. 
This suggests that the SED fitting is smoothing out some of the scatter seen in the fluxes, which could be 
partially related to the resolution of some of the filters that contribute the most to M$_*$ and SFR 
estimations. 
However, it is important to note that even when the scatter appears to be small, the difference in photometry 
for some of the galaxies can lead to differences in M$_*$ of a factor of 1000 and a factor of 10000 
for SFR. This suggests that the previous values in the literature should be used with caution, and that  
adequate photometry must be conducted for these types of galaxies before estimating M$_*$ and SFR. 

The differences in photometry results show no clear dependence on either morphology or merger stage. 
The only dependence is on the aperture size used when extracting the light from each source. 
There is also no clear dependence on morphology or merger stage in the M$_*$ and SFR estimates, which leads us to conclude that the major factor affecting our results is the photometry performed. 

The M$_*$ and SFR indicators based on optical colours and NIR and/or NUV fluxes, 
respectively, also show large scatter. 
For the M$_*$ indicators, our results show lower $\rm log(M_*/L_{r})$ for the same colours compared to the 
B03 relations. On the other hand, our M$_*$  shows a closer correlation to the T11 M$_*$ 
indicator, which is also based on SED fitting of photometric data spanning from the UV to the far-infrared, 
but using different photometry. 
For the SFR indicators, the scatter seems to be larger mainly for highly disturbed and elliptical galaxies. 
This can also be related to the photometry used at these wavelengths;  the fluxes in the NIR filters 
can be underestimated (for highly disturbed galaxies) or the  NUV fluxes not considered for some 
of the indicators (for both highly disturbed and elliptical galaxies).
It is important to note that these catalogues and indicators have been optimised for statistical studies. 
Thus, it is not surprising that they show issues for mergers' estimations. 

The near-infrared colour-colour diagram separates the morphologies of our sample as it is designed to do. 
However, we can see different morphologies in the various regions of the colour-colour diagram, 
suggesting that these galaxies go through colour changes when they are involved in a merging process. 
Also, we observe that mergers in this diagram are not mainly located in the AGN region (above 
the magenta line). This suggests either that not many mergers host an AGN or that the AGN in mergers 
are not luminous enough to outshine the brightness of the hosting merging galaxy. This will be  
discussed further in Paper II, as will the SF enhancement of mergers and their location in the M$_*$-SFR plane 
separated by merger stage.

Finally, the estimation of sSFR from W1-W4 colours is only recommended for lenticular galaxies 
as other morphologies show larger scatter. Instead, sSFR should be estimated using M$_*$ from W1 
(\textsc{Cl14}) and SFR from W4 (\textsc{Cl14}). 

\begin{acknowledgements}
We would like to thank the referee for a constructive report which helped us improve this manuscript. 
P.C-C. was supported by CONICYT (Chile) through Programa Nacional de Becas de Doctorado 
2014 folio 21140882. 
NN acknowledges support from Conicyt (PIA ACT172033, Fondecyt 1171506, and BASAL AFB-170002).
S.K.Y. acknowledges support from the Korean National Research Foundation 
(NRF-2017R1A2A1A05001116). This study was performed under the umbrella of the joint collaboration 
between Yonsei University Observatory and the Korean Astronomy and Space Science Institute. 
G.O. acknowledges the support provided by CONICYT (Chile) through FONDECYT postdoctoral research grant 
no 3170942. 
R.L. acknowledges support from Comité Mixto ESO-GOBIERNO DE CHILE, GEMINI-CONICYT FUND 32130024, 
and FONDECYT Grant 3130558. 
T.M.H. acknowledges the support from the Chinese Academy of Sciences(CAS) and the 
National Commission for Scientific and Technological Research of Chile (CONICYT) through a 
CAS-CONICYT Joint Postdoctoral Fellowship administered by the CAS South America Center for Astronomy 
(CASSACA) in Santiago, Chile. 
We acknowledge the use of the following databases and codes: 
ARP (\url{http://arpgalaxy.com}), 
GZ (\url{http://data.galaxyzoo.org}), 
VV (\url{www.sai.msu.su/sn/vv}), 
GOALS (\url{http://goals.ipac.caltech.edu}), 
GALEX (\url{http://galex.stsci.edu/data/}), 
SDSS (\url{https://dr13.sdss.org/sas/dr13/}), 
WISE (\url{http://irsa.ipac.caltech.edu/applications/wise/}), 
Chang et al. 2015 catalogue (\url{http://www.asiaa.sinica.edu.tw/~yychang/sw.html}), 
MPA-JHU (\url{https://www.sdss3.org/dr10/spectro/galaxy_mpajhu.php}),
NSA (\url{http://www.nsatlas.org/}), 
SExtractor (\url{https://www.astromatic.net/software/sextractor}), 
and MAGPHYS (\url{http://www.iap.fr/magphys/}). 

\end{acknowledgements}

\bibliographystyle{aa} 
\bibliography{Merging_galaxies_in_isolated_environments} 

\appendix

\section{Examples of morphology}

\begin{figure}[!hb]
\begin{center}
  \includegraphics[width=0.3\textwidth,clip]{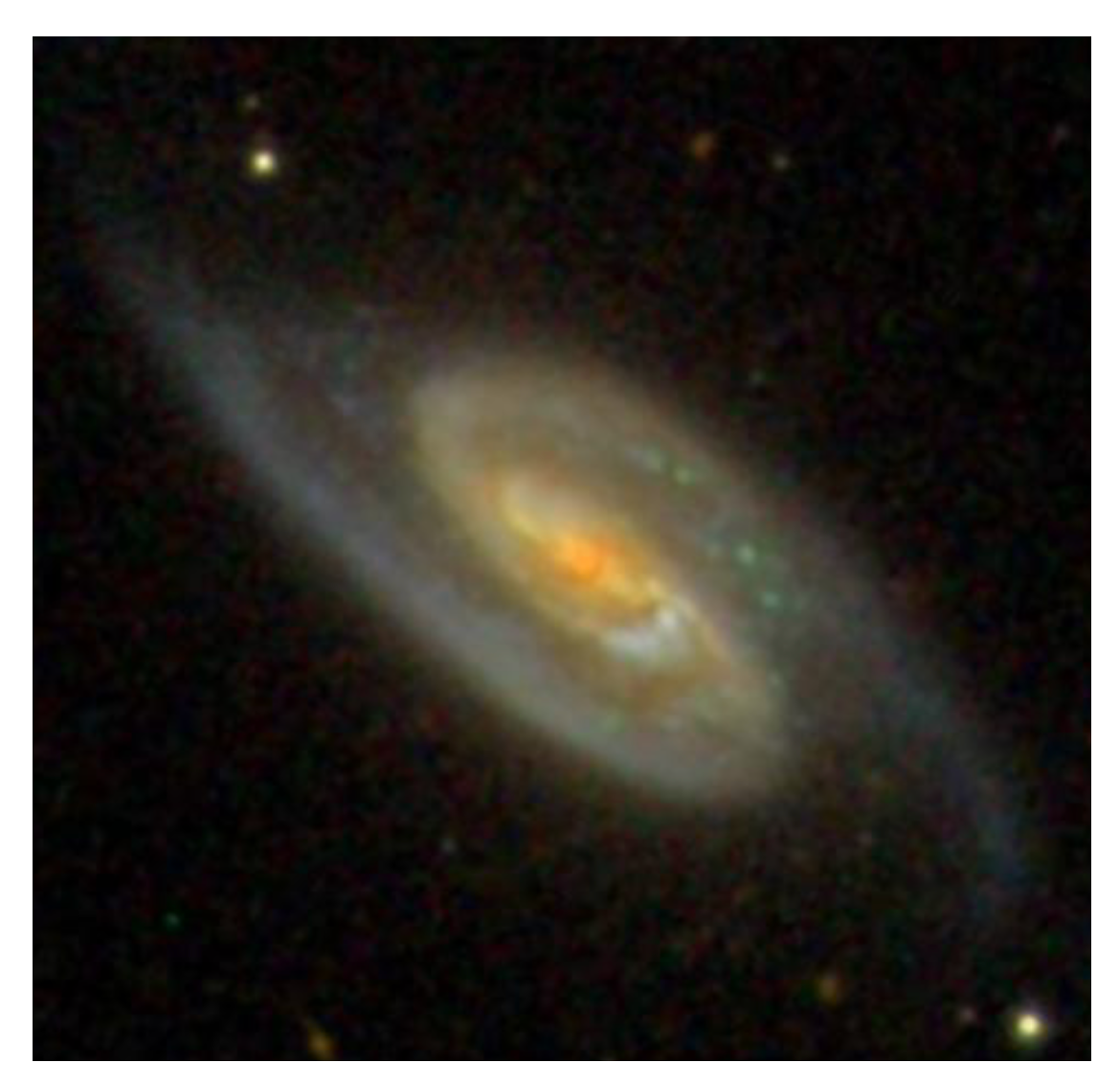}
  \includegraphics[width=0.3\textwidth,clip]{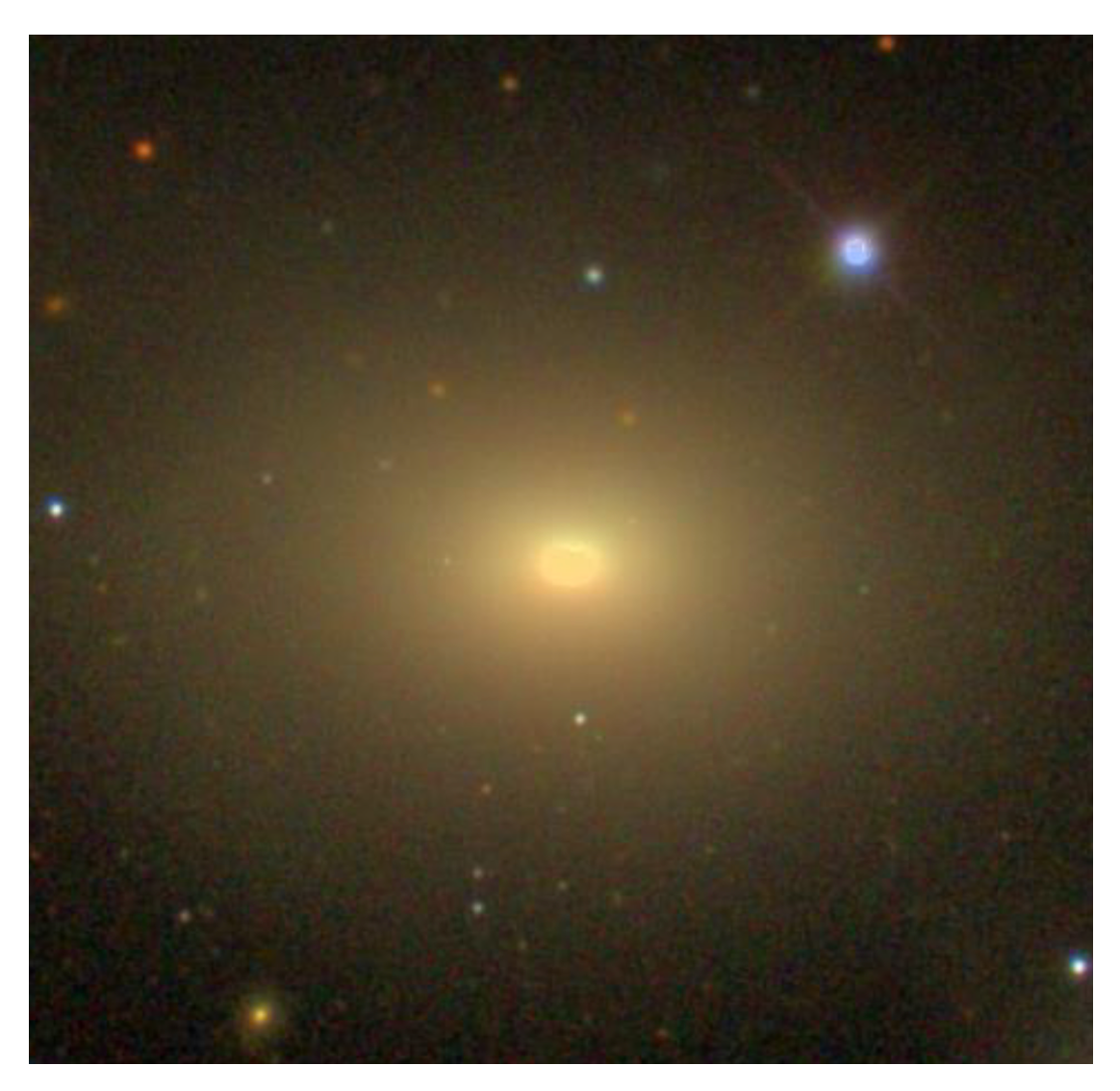}
  \includegraphics[width=0.3\textwidth,clip]{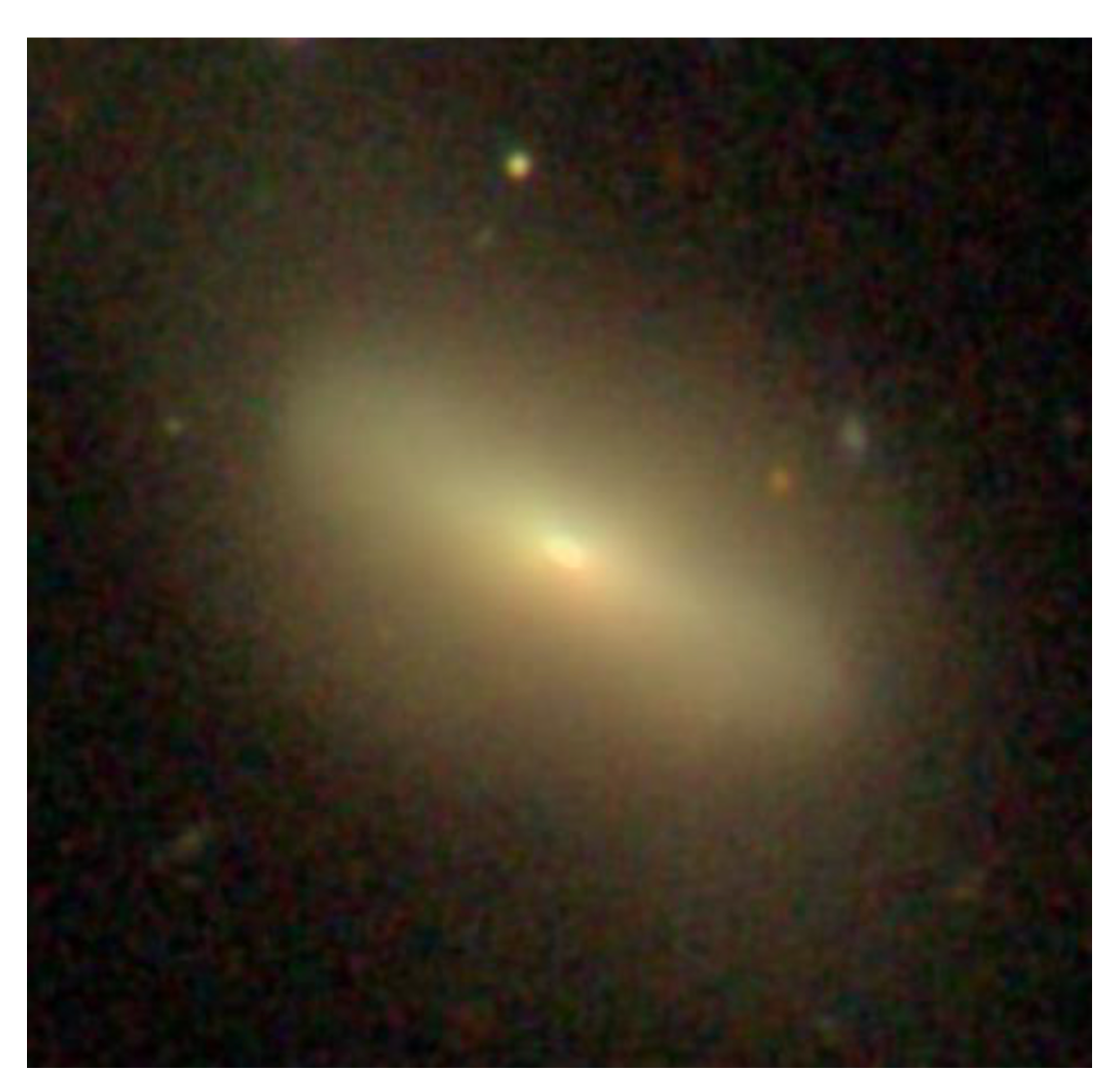}
  \includegraphics[width=0.3\textwidth,clip]{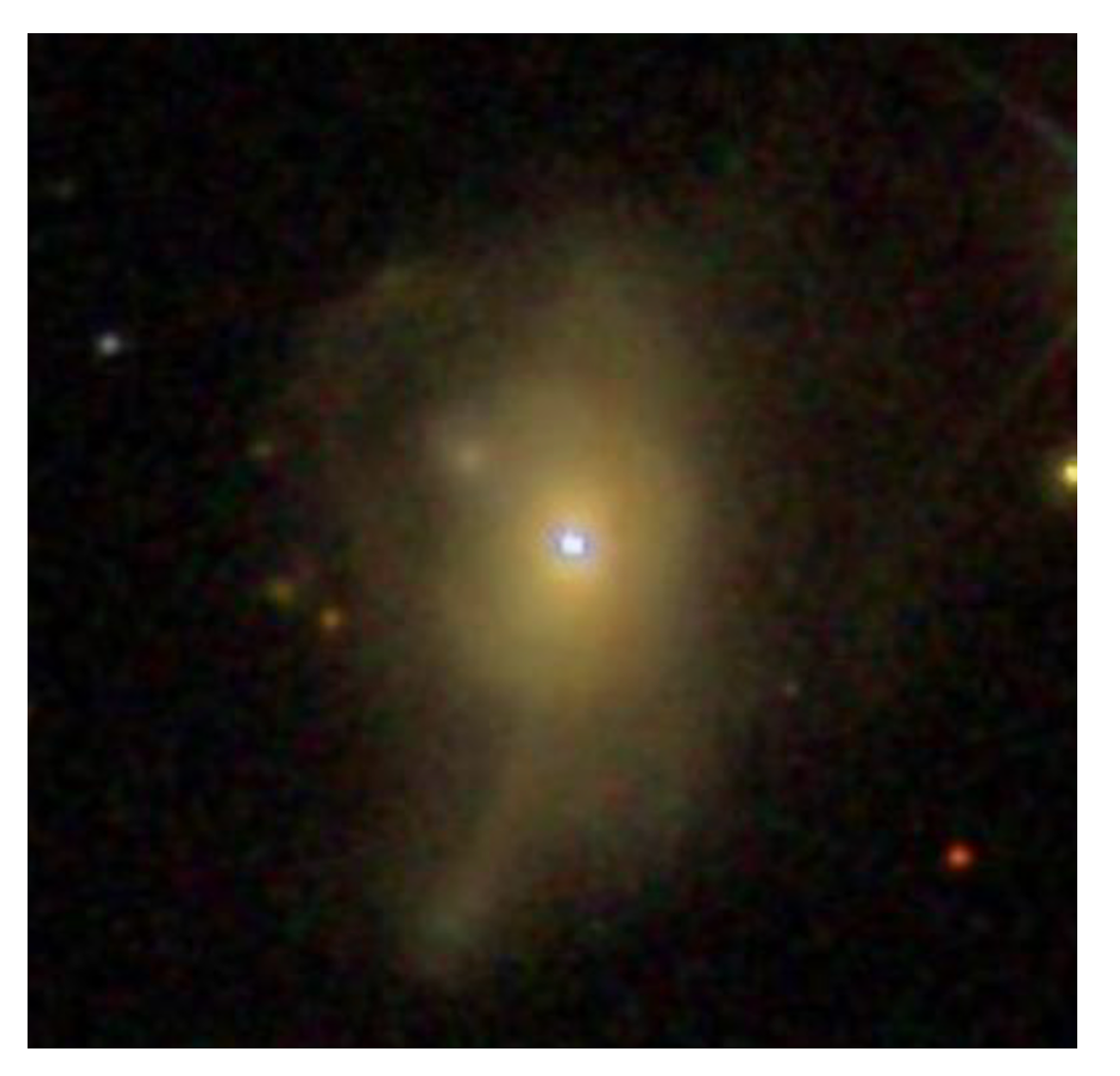}
  \caption{Examples of the morphology classification. From top to bottom: spiral, elliptical, lenticular, and 
  highly disturbed galaxies.  
}  
\label{fig:Morphs}
\end{center}
\end{figure}

In Sect. \ref{sec:class} we show the distribution of the galaxy morphologies as classified in this study. 
Here we present an example of each morphology classification. Figure \ref{fig:Morphs} shows, 
from top to bottom, a spiral, an elliptical, a lenticular (S0), and a highly disturbed galaxy from our merger 
sample. Specifically, to be classified as a spiral, galaxies  have to show clear spiral arms not heavily 
perturbed. Ellipticals show red colours and round shapes with a bright nucleus. Lenticular galaxies show 
similar morphologies to  ellipticals, but they all show a clear disc. Finally, highly disturbed galaxies are those that cannot be classified in any of the previous classes. These galaxies could show highly perturbed 
spiral arms, various shells, tidal features, among others.

\section{Survey parameters}

To perform the photometry, we  searched for the relevant parameters from each survey. 
We  assembled Table \ref{tab:Filters} with the reference information  used in these surveys.

\begin{table*}[!htb]
\begin{center}
\caption{Relevant parameters of the filters and images used from the different surveys. 
$^1$All-sky $^2$Deep, and $^3$Medium Imaging Surveys from GALEX. }
\centering
\resizebox{1\textwidth}{!}{
\begin{tabular}{|c|c|c|c|c|c|c|c|c|}
\hline
Survey          & Band       & Effective      & Zero point     & Resolution & Sensitivity   & Pixel Scale        &  $\Delta$ m \\
                     &                & wavelength & magnitude                     & (arcsec)     &                    & (arcsec/pixel)     & ($m_{AB}=m_{Vega} + \Delta m$) \\
\hline
\hline
  GALEX        & FUV        & 1528 \AA{}  &   18.82                    & 4.2            & 20 (AIS$^1$) / 22.7 (MIS$^2$) / 24.8 (DIS$^3$) & 1.5             & 2.22 \\
                      & NUV        & 2271 \AA{}  &   20.08                       & 5.3            & 21 / 22.7 / 24.4 (ABmag)              & 1.5              &  1.69 \\
\hline
\hline
  SDSS          & u             & 3551 \AA{}  &      22.5          &                  &  22.0 (ABmag)                               & 0.396                      & 0.91\\
                      & g             & 4686 \AA{}  &    22.5            &                  &  22.2  (ABmag)                              & 0.396                      & -0.08 \\
                      & r              & 6165 \AA{}  &   22.5             &      1.3       &  22.2   (ABmag)                             & 0.396        & 0.16\\
                      & i              & 7481 \AA{}  &    22.5         &                  &  21.3    (ABmag)                            & 0.396                     & 0.37\\
                      & z             & 8931 \AA{}  &    22.5           &                  &  20.5     (ABmag)                           & 0.396                     & 0.54 \\
\hline
\hline
  WISE          & W1           & 3.4 \mum   &   20.73           & 6.1            & 0.08 mJy (16.5 Vegamag)                      & 1.375          & 2.699 \\
                     & W2           & 4.6 \mum   &   19.56          & 6.4            & 0.11 mJy  (15.5 Vegamag)                     & 1.375        & 3.339 \\
                     & W3           & 12 \mum    &   17.60         &  6.5           & 1mJy       (11.2 Vegamag)                      & 1.375          &  5.174 \\
                     & W4           & 22 \mum    &   12.98         & 12.0          & 6 mJy       (7.9 Vegamag)                       & 1.375       & 6.620 \\
\hline
\end{tabular}}
\label{tab:Filters}
\end{center}
\end{table*}

\section{Examples of apertures}
\label{app:ApsExamples}

Here, we show more examples of the different apertures measured by SExtractor depending on the 
parameters $\sigma$ and n-deblending (see Sect. \ref{sec:photo}). Each figure shows three examples for 
each galaxy using different $\sigma$ and n-deblending. 
We show the best aperture at the bottom of each figure. 

\begin{figure}[h]
\begin{center}
  \includegraphics[width=0.4\textwidth,clip]{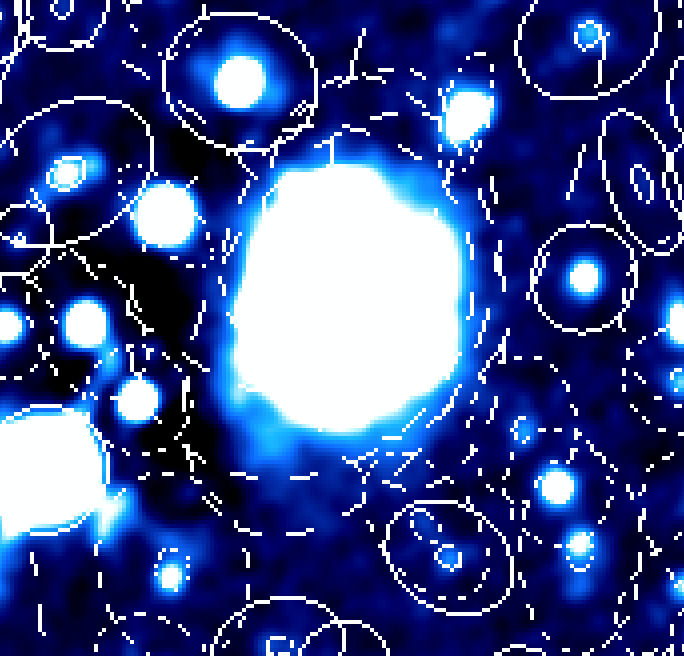}
  \includegraphics[width=0.4\textwidth,clip]{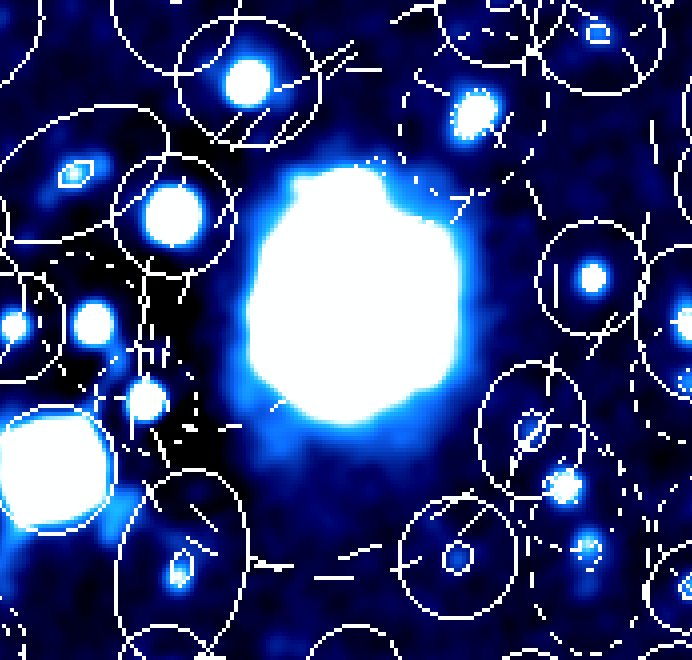}
  \includegraphics[width=0.4\textwidth,clip]{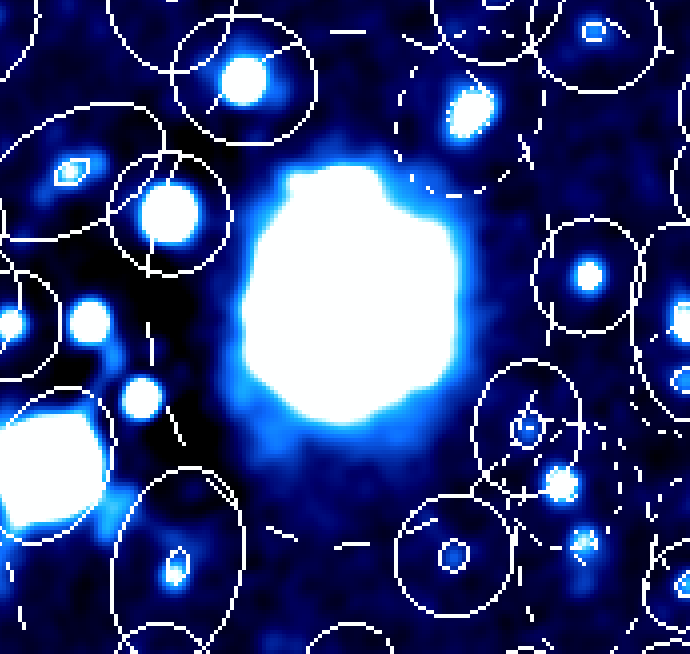}     
    \caption{ From top to bottom: $\sigma$ = 1.5, n-deblending=32; 
    $\sigma$ = 5.0, n-deblending=32;
    $\sigma$ = 3.0, n-deblending=2. 
}  
\end{center}
\end{figure}

\begin{figure}[h]
\begin{center}
  \includegraphics[width=0.365\textwidth,clip]{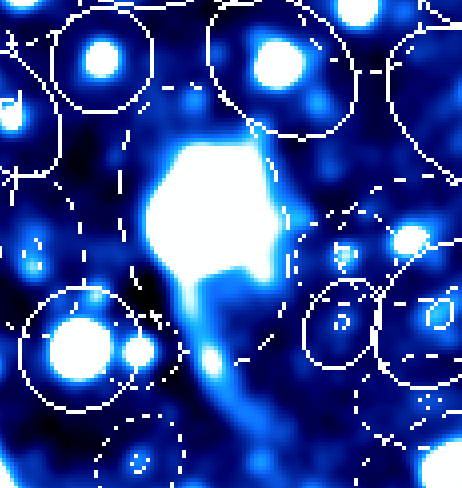}
  \includegraphics[width=0.365\textwidth,clip]{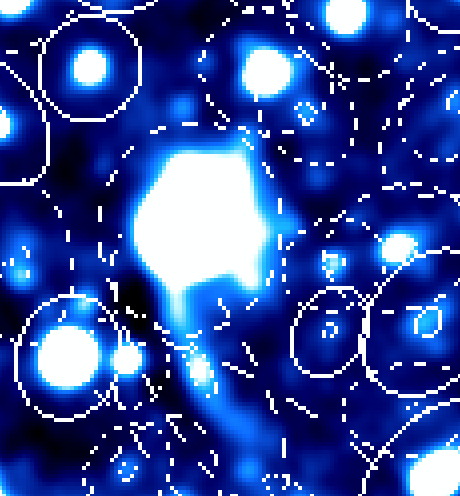}
  \includegraphics[width=0.365\textwidth,clip]{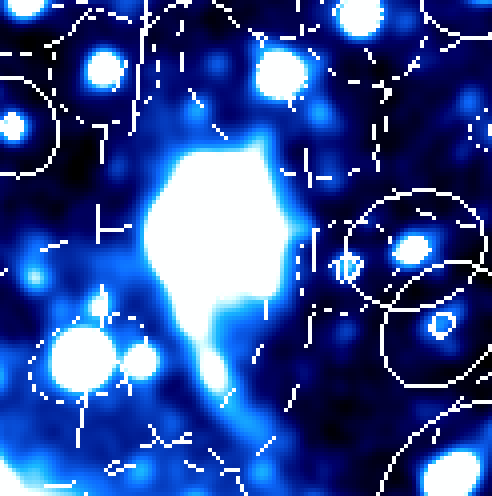}     
    \caption{ From top to bottom: $\sigma$ = 1.5, n-deblending=2; 
    $\sigma$ = 1.5, n-deblending=32;
    $\sigma$ = 1.5, n-deblending=4. 
}  
\end{center}
\end{figure}

\begin{figure}[h]
\begin{center}
  \includegraphics[width=0.385\textwidth,clip]{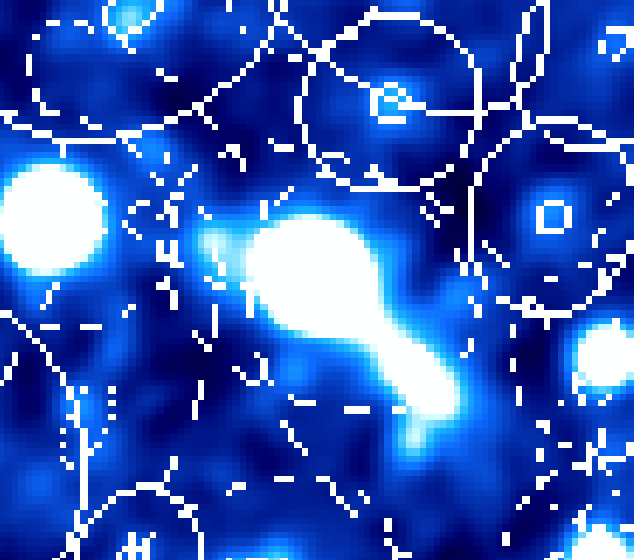}
  \includegraphics[width=0.385\textwidth,clip]{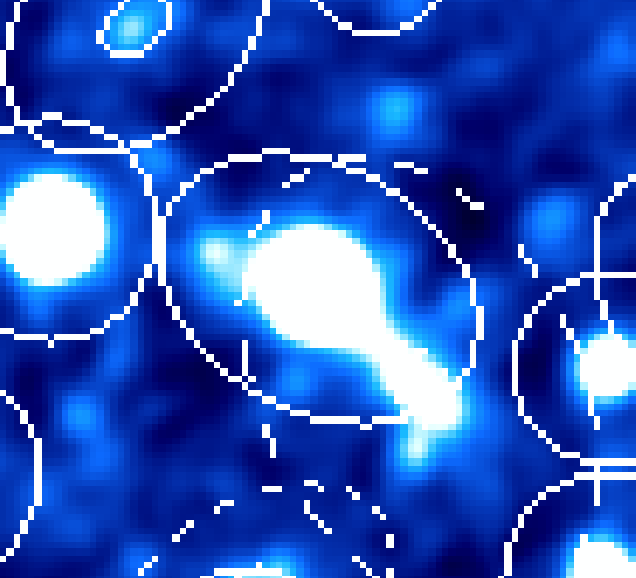}
  \includegraphics[width=0.385\textwidth,clip]{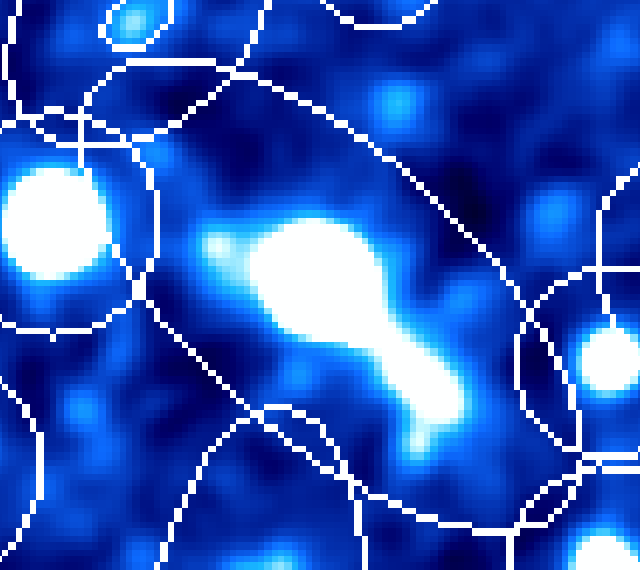}     
    \caption{ From top to bottom: $\sigma$ = 1.5, n-deblending=32; 
    $\sigma$ = 5.0, n-deblending=32;
    $\sigma$ = 5.0, n-deblending=2. 
}  
\end{center}
\end{figure}

\begin{figure}[h]
\begin{center}
  \includegraphics[width=0.4\textwidth,clip]{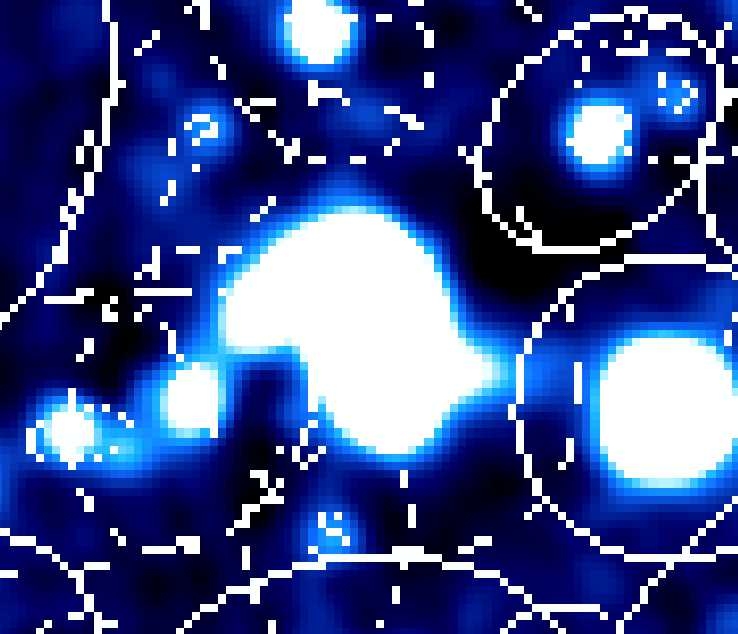}
  \includegraphics[width=0.4\textwidth,clip]{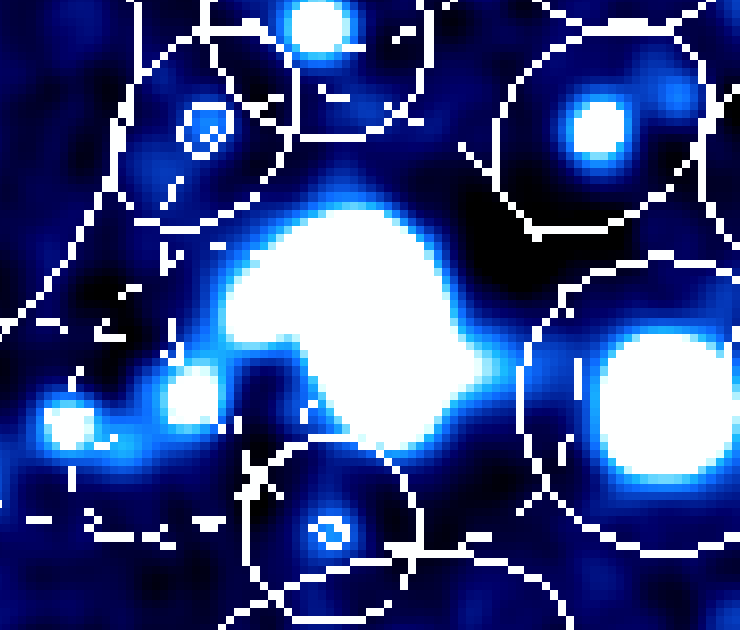}
  \includegraphics[width=0.4\textwidth,clip]{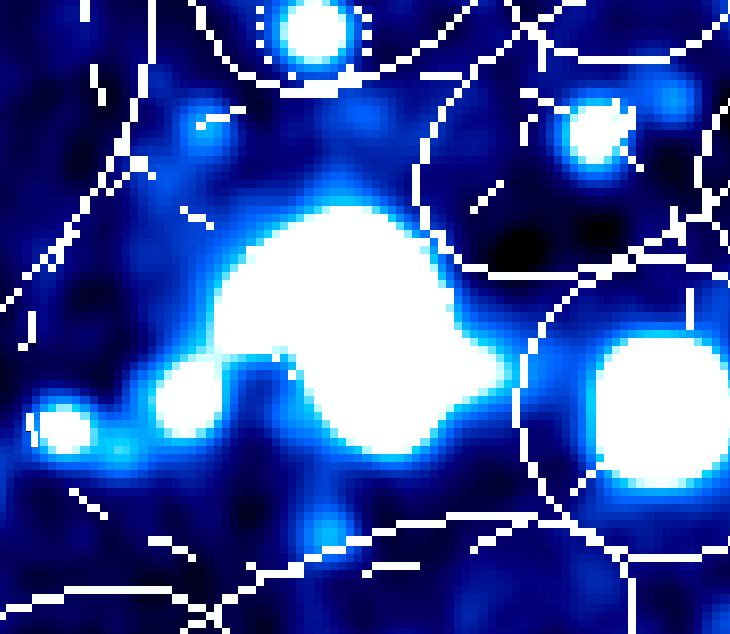}     
    \caption{ From top to bottom: $\sigma$ = 1.5, n-deblending=32; 
    $\sigma$ = 3.0, n-deblending=32;
    $\sigma$ = 3.0, n-deblending=4. 
}  
\end{center}
\end{figure}

\section{Fluxes comparisons according to different parameters}

In Sect. \ref{sec:photo}, we showed the comparison between our measured GALEX NUV fluxes and the 
GALEX GR6/GR7 catalogue, the SDSS DR13 r-band fluxes of our measurements and the values listed by 
\textsc{Chang15}, and our measured WISE W1 fluxes and the catalogued values shown in AllWISE 
tabulated as {\texttt{gmag}}. Here we colour-coded the comparisons in order to look for any dependences in 
morphology, merger stage, or photometry flag. The photometry flag shows whether the apertures, 
measured by our semi-automated method, are 
completely separated or joined (both galaxies are within the same aperture). 

\begin{figure*}[!hb]
\begin{center}
  \includegraphics[width=0.3\textwidth,clip]{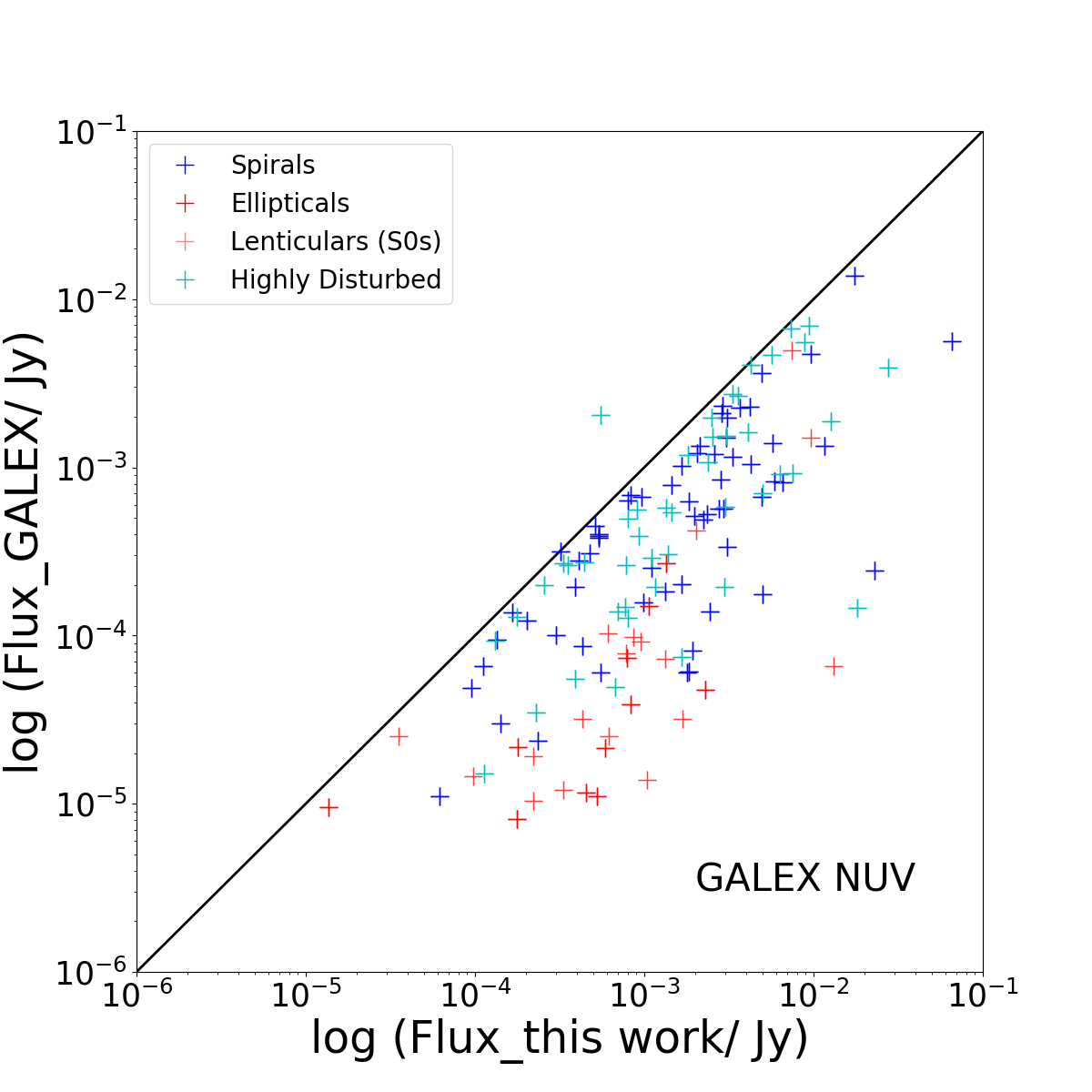}
  \includegraphics[width=0.3\textwidth,clip]{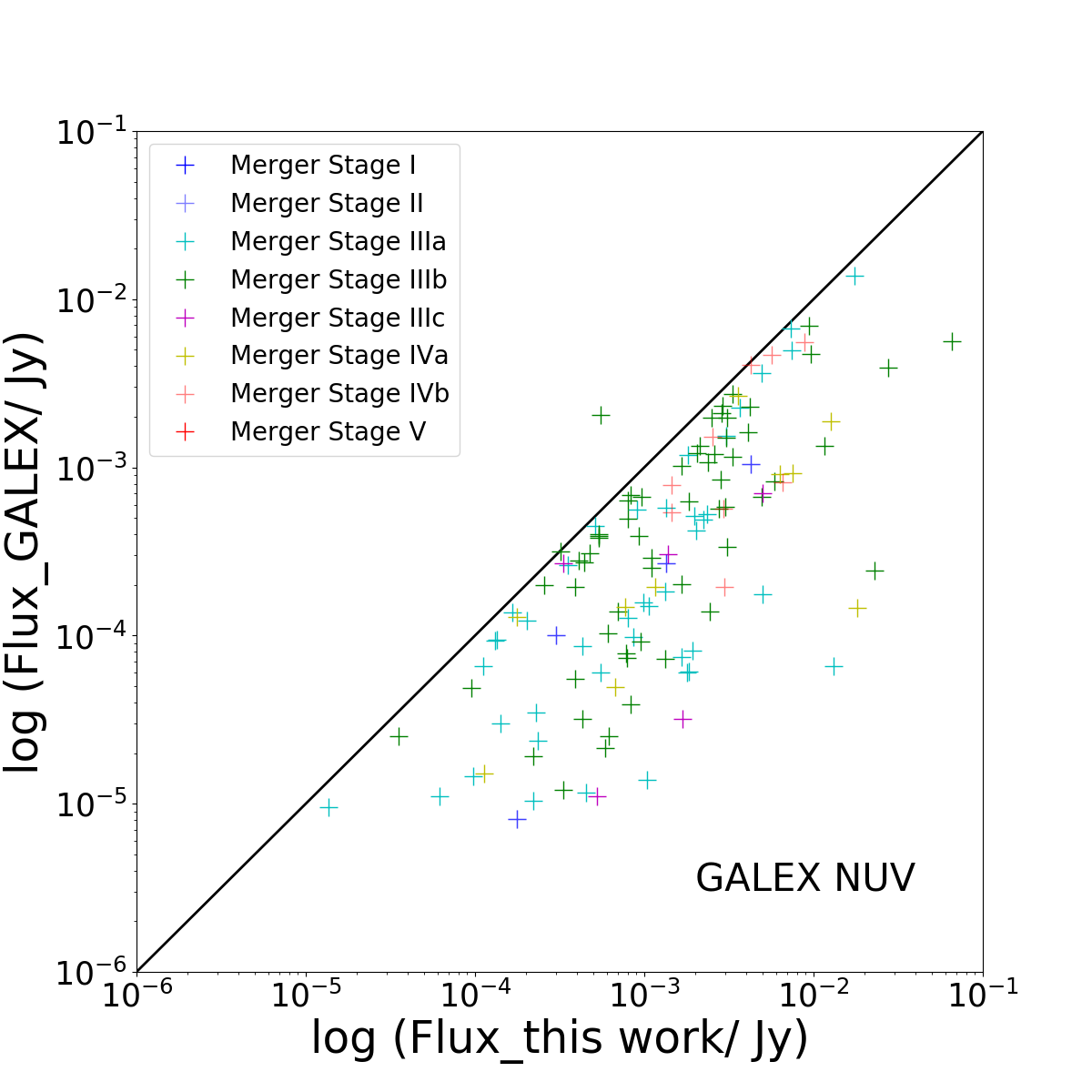}
  \includegraphics[width=0.3\textwidth,clip]{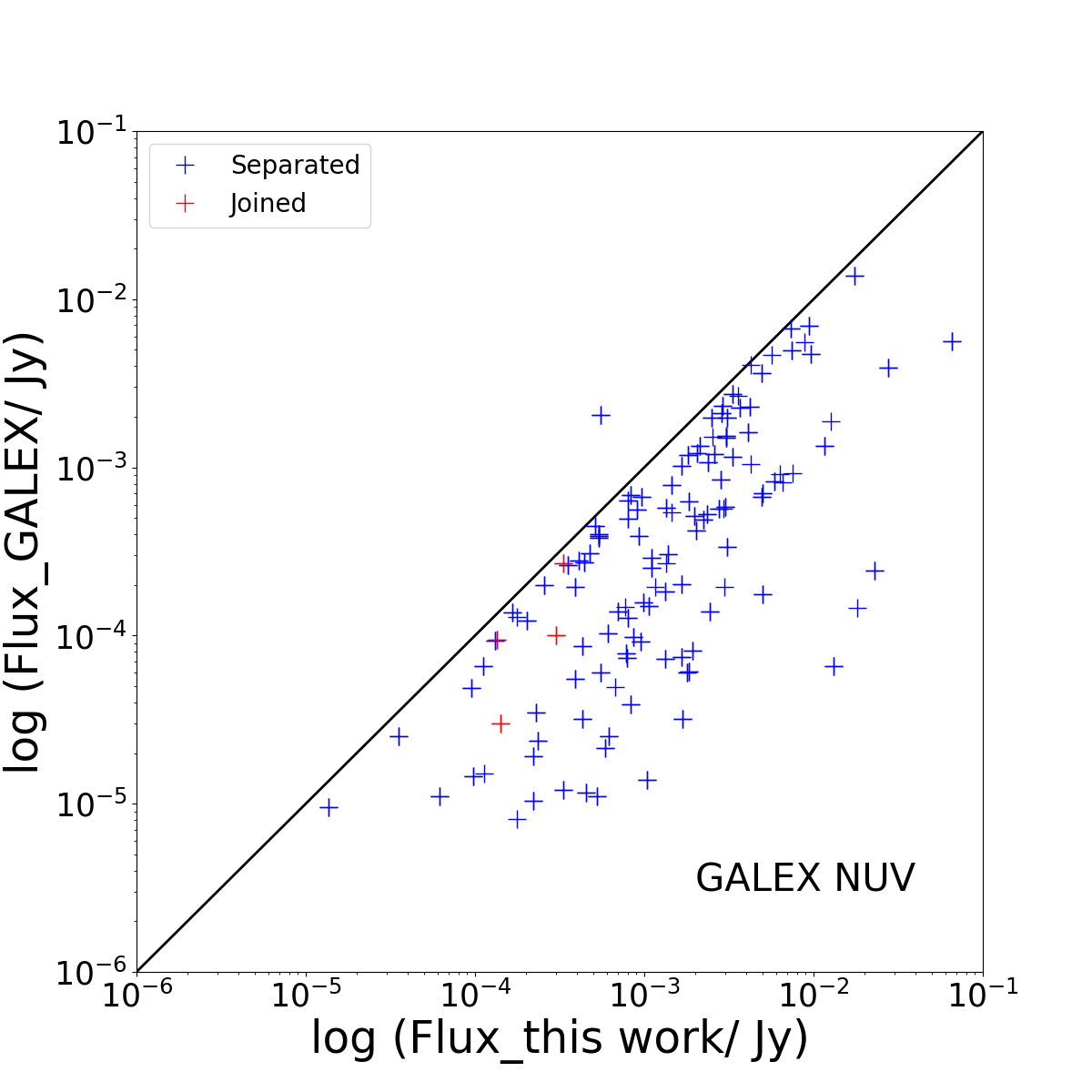}
  \includegraphics[width=0.3\textwidth,clip]{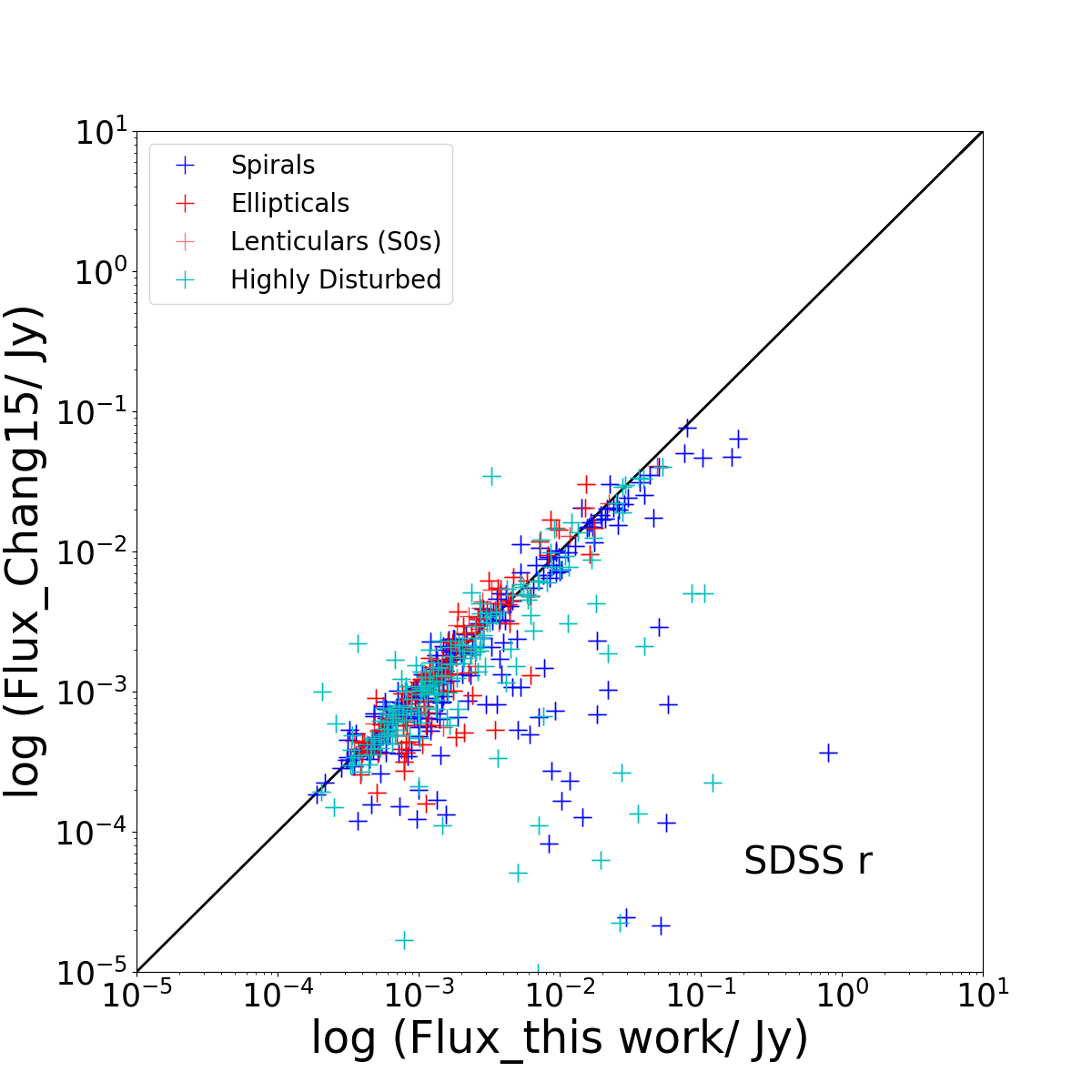}
  \includegraphics[width=0.3\textwidth,clip]{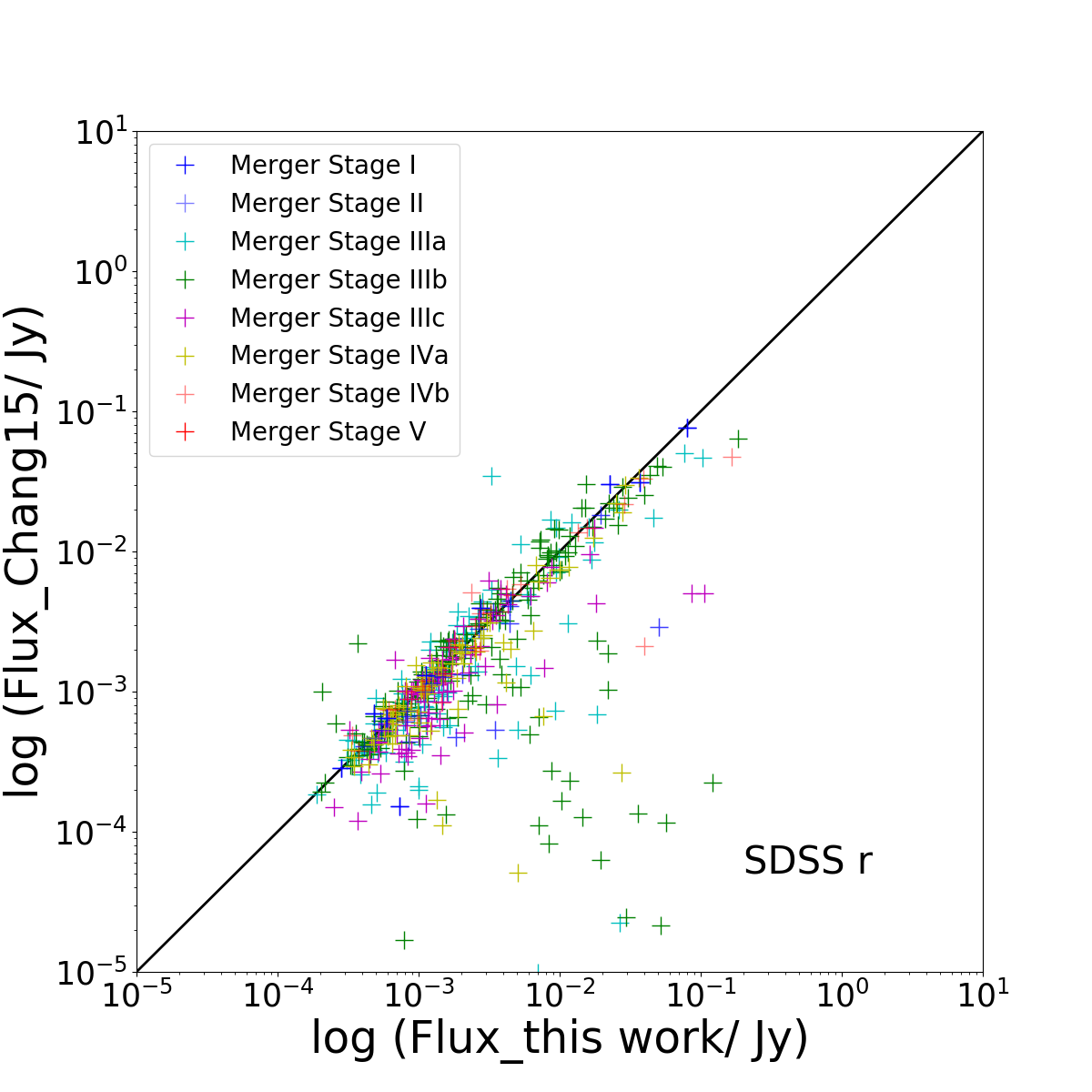}
  \includegraphics[width=0.3\textwidth,clip]{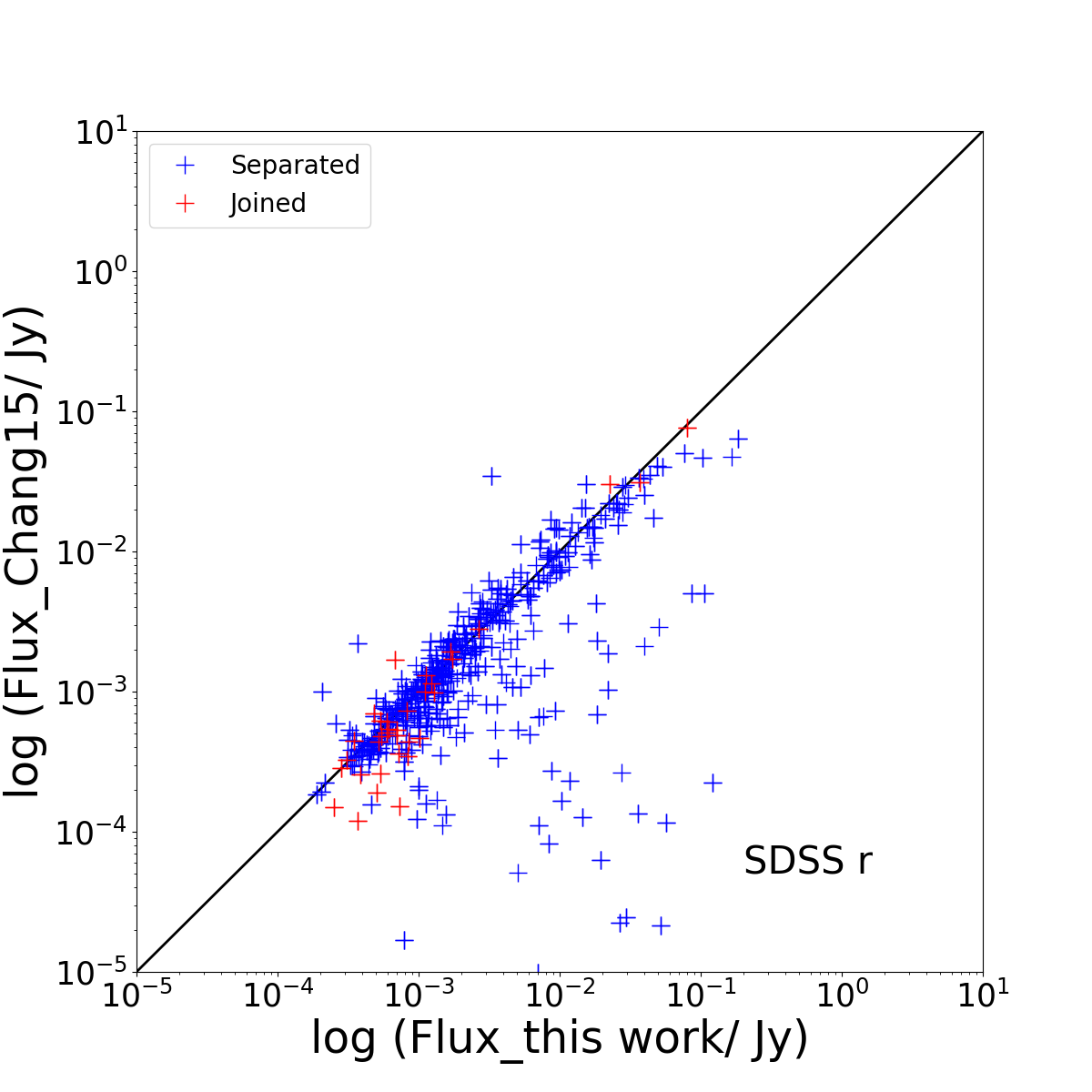}
  \includegraphics[width=0.3\textwidth,clip]{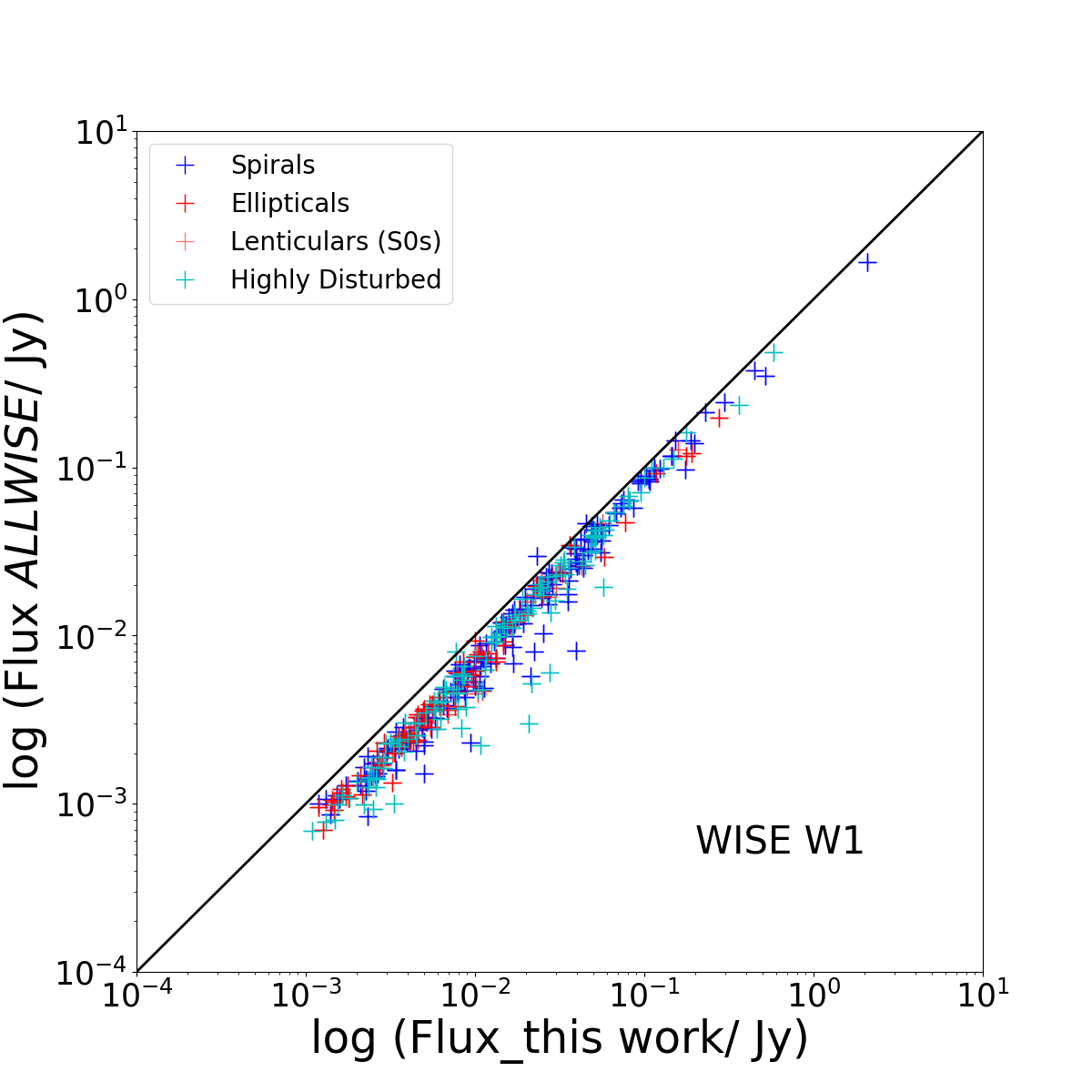}
  \includegraphics[width=0.3\textwidth,clip]{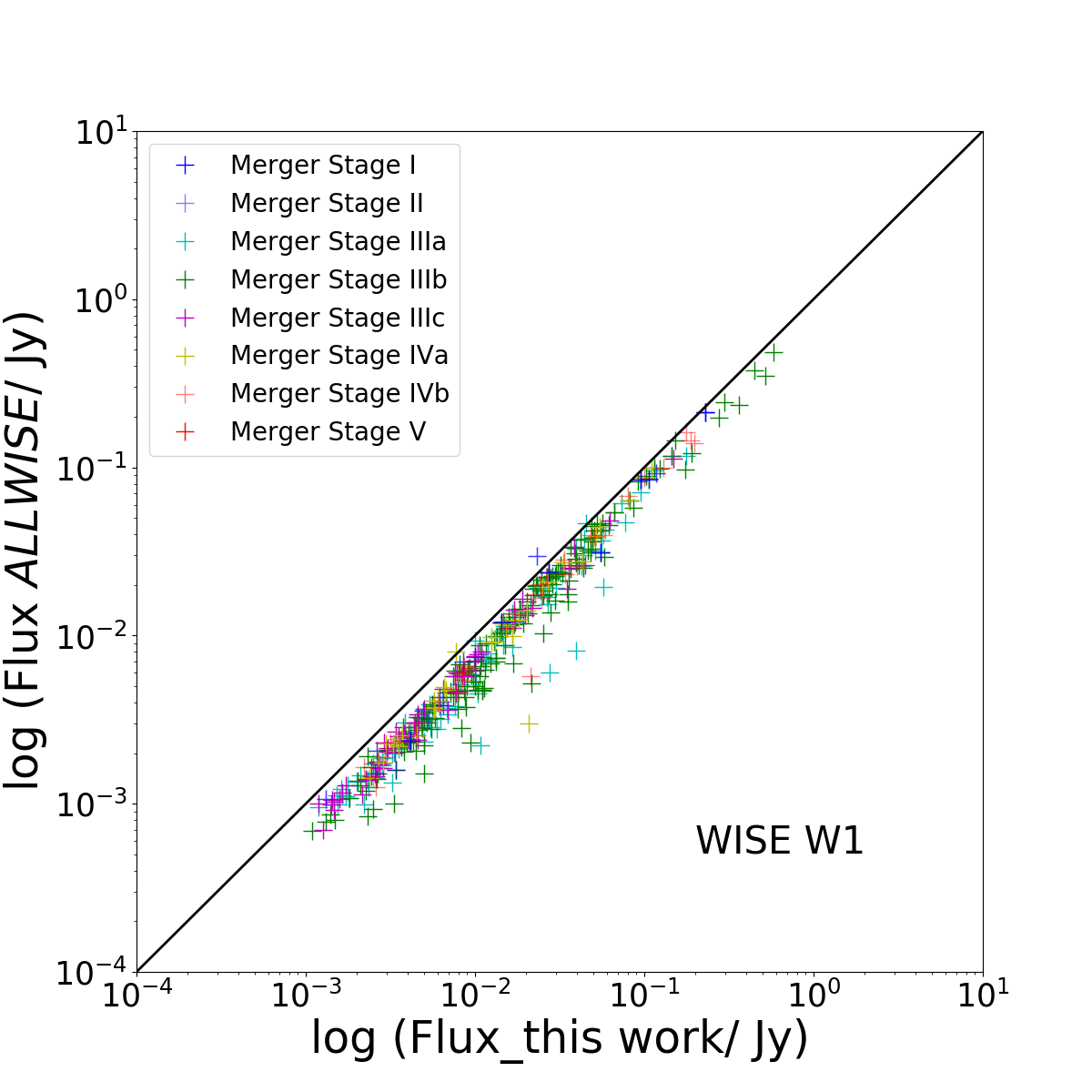}
  \includegraphics[width=0.3\textwidth,clip]{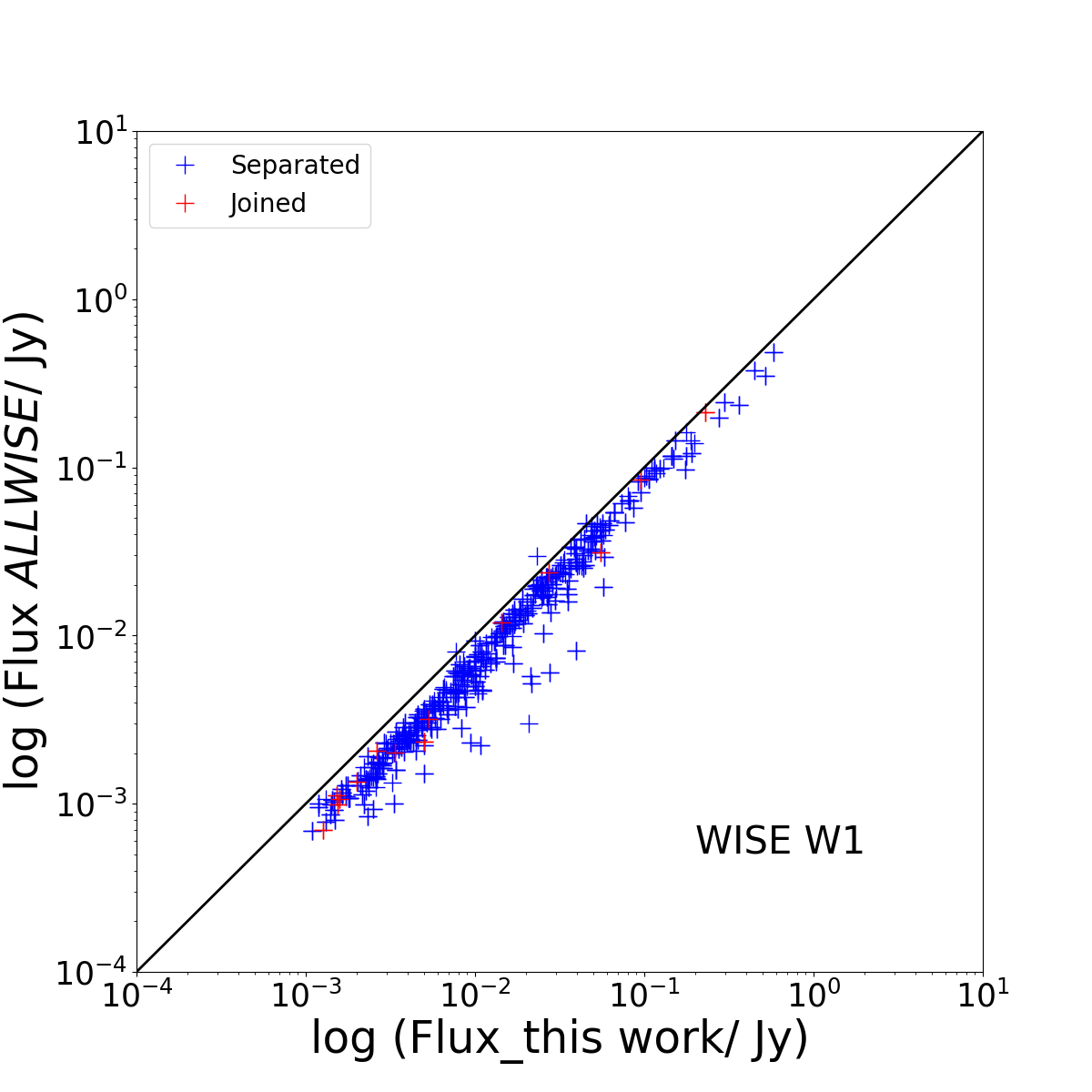}
\caption{Comparison between our GALEX NUV (top panels), SDSS r band (middle panels), 
and WISE W1 (bottom panels) measured fluxes and GALEX, SDSS, and AllWISE catalogued values. 
Colour-coded according to morphology (left panels), merger stage (middle panels), and photometry flag (right panels).
}  
\label{fig:UVoptNIR_compscol}
\end{center}
\end{figure*}

Figure \ref{fig:UVoptNIR_compscol} shows the fluxes we measured from 
GALEX NUV (top row), SDSS r-band (middle row), and WISE W1 images (bottom row) compared 
to the values catalogued by GALEX, SDSS, and AllWISE. They are colour-coded according to morphology (left panels), 
merger stage (centre panels), and photometry flag (right panels). 
The correlations do not show a clear dependence on any of these parameters. However, spirals and 
highly disturbed galaxies show a larger scatter. The scatter in the SDSS (centre panel, middle row) is dominated 
by merging galaxies at merger stage IIIb, where a large fraction of the merging galaxies are either spiral or 
highly disturbed. 
This leads to the conclusion that differences in these values are related mainly to the photometry performed.

\section{MAGPHYS SED fits}

\begin{figure*}[h]
\begin{multicols}{2}
  \begin{center}
   \includegraphics[width=0.42\textwidth,clip]{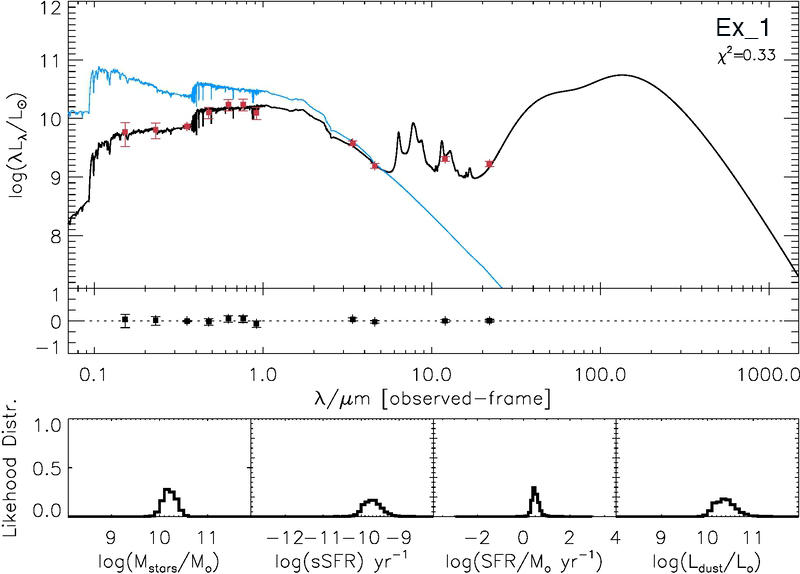}\par 
   \includegraphics[width=0.42\textwidth,clip]{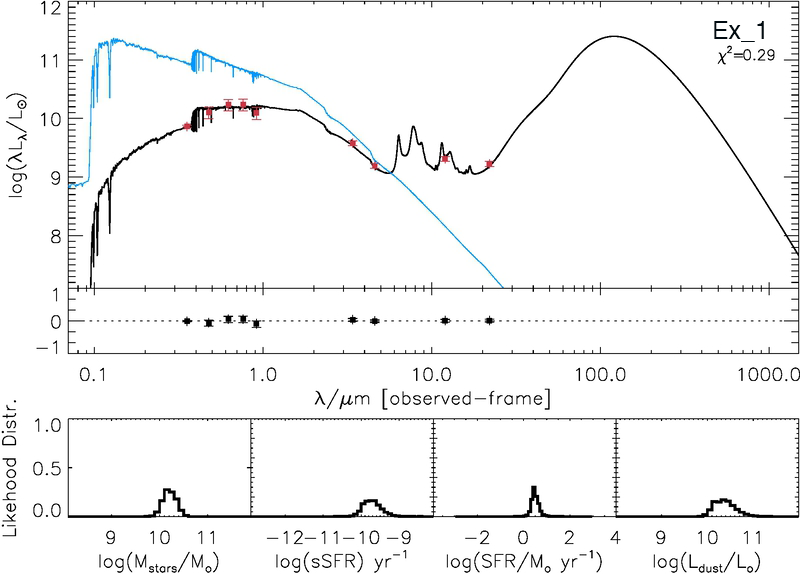}\par 
  \end{center}
    \end{multicols}
\begin{multicols}{2}
  \begin{center}
    \includegraphics[width=0.42\textwidth,clip]{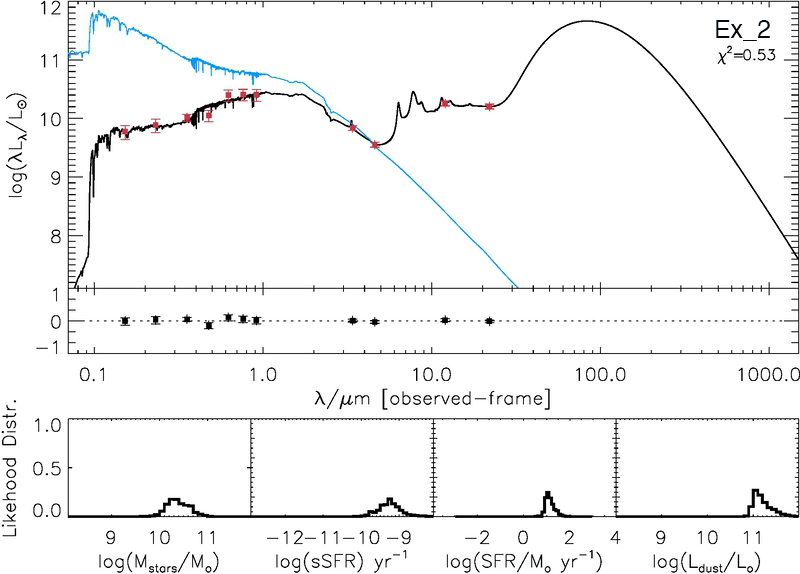}\par
    \includegraphics[width=0.42\textwidth,clip]{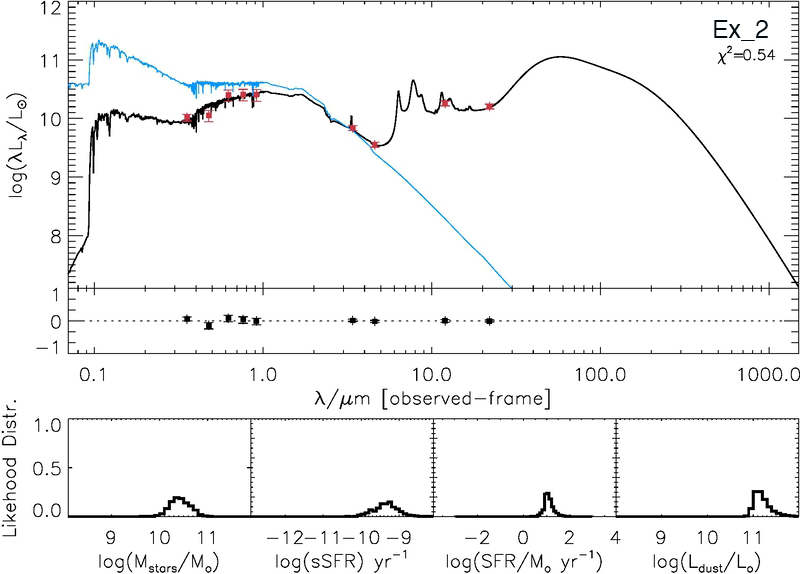}\par
  \end{center}
\end{multicols}
\begin{multicols}{2}
  \begin{center}
    \includegraphics[width=0.42\textwidth,clip]{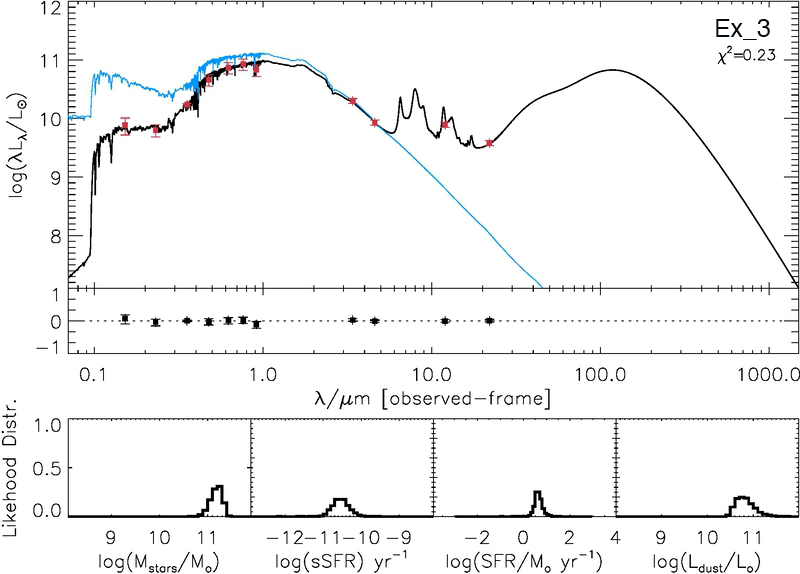}\par
    \includegraphics[width=0.42\textwidth,clip]{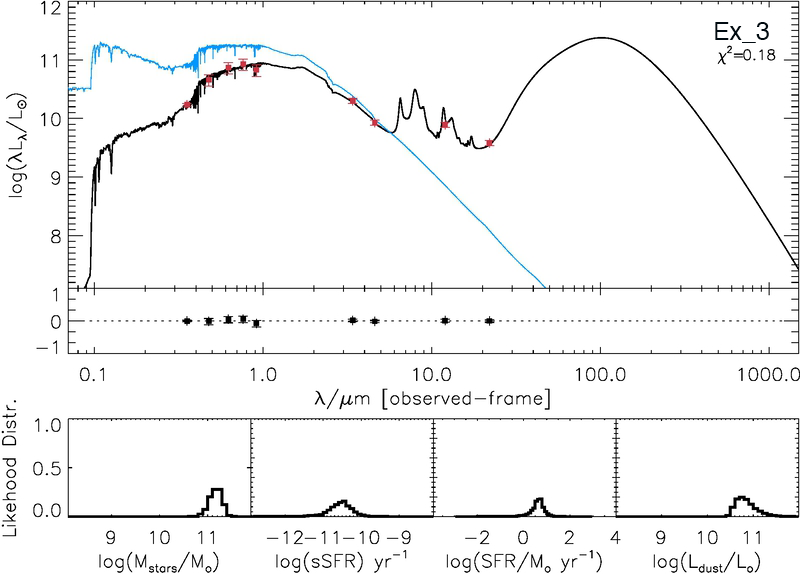}\par
  \end{center}
\end{multicols}
  \caption{Examples of MAGPHYS SED fits for some of our mergers 
  using GALEX+SDSS+WISE filters (left) and  SDSS+WISE filters (right). We show the SED fits for the same 
  object in each row. 
  In the upper panel, the red points show our photometric data, the black curve shows the best-fit 
  SED  to the photometry, and the blue curve shows the unattenuated SED. 
  In the top right corner of each panel the $\chi^2_r$ of the fit is shown.
  The middle panel shows the residuals of the SED fit.
  In the lower part, there are four small panels showing the PDFs of the different parameters estimated by MAGPHYS. In this case, we show the PDFs of the M$_*$, sSFR, SFR, and $\rm M_{dust}$ of each fit.  
}  
\label{fig:SEDsexamples}
\end{figure*}

\begin{figure*}[h]

\begin{multicols}{2}
  \begin{center}
    \includegraphics[width=0.42\textwidth,clip]{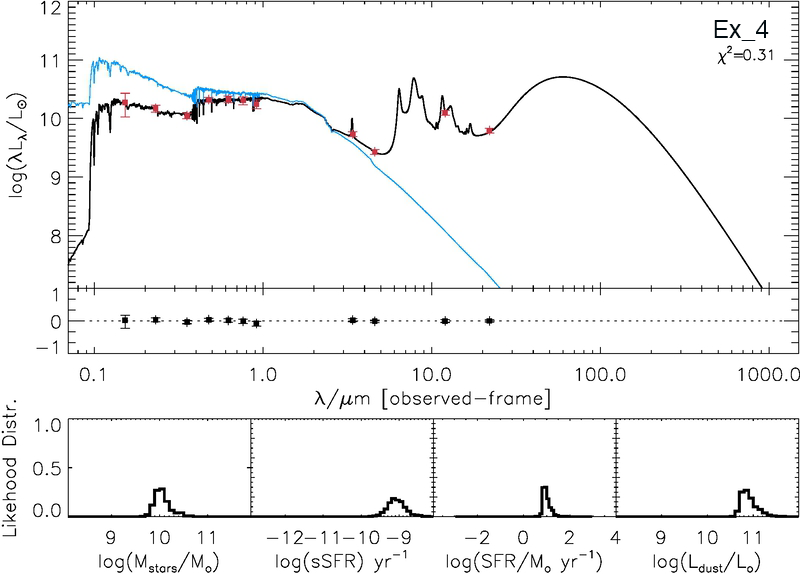}\par
    \includegraphics[width=0.42\textwidth,clip]{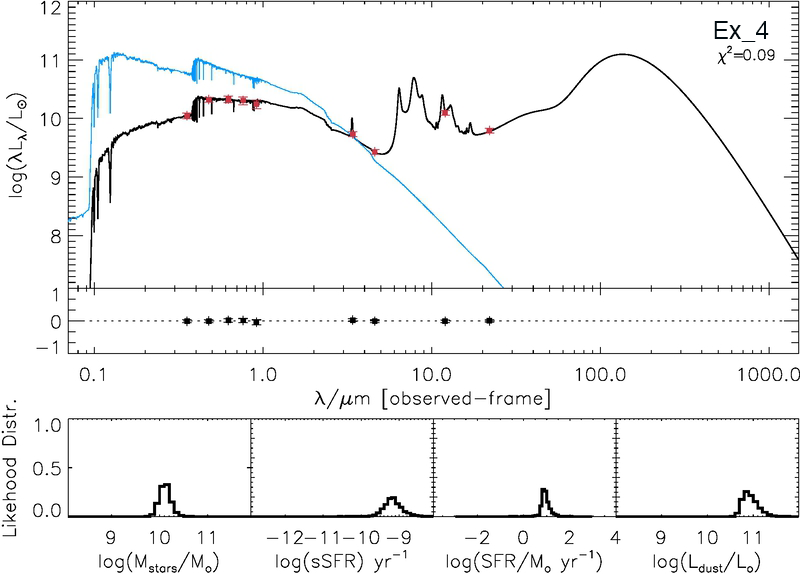}\par
  \end{center}
\end{multicols}
\begin{multicols}{2}
  \begin{center}
      \includegraphics[width=0.42\textwidth,clip]{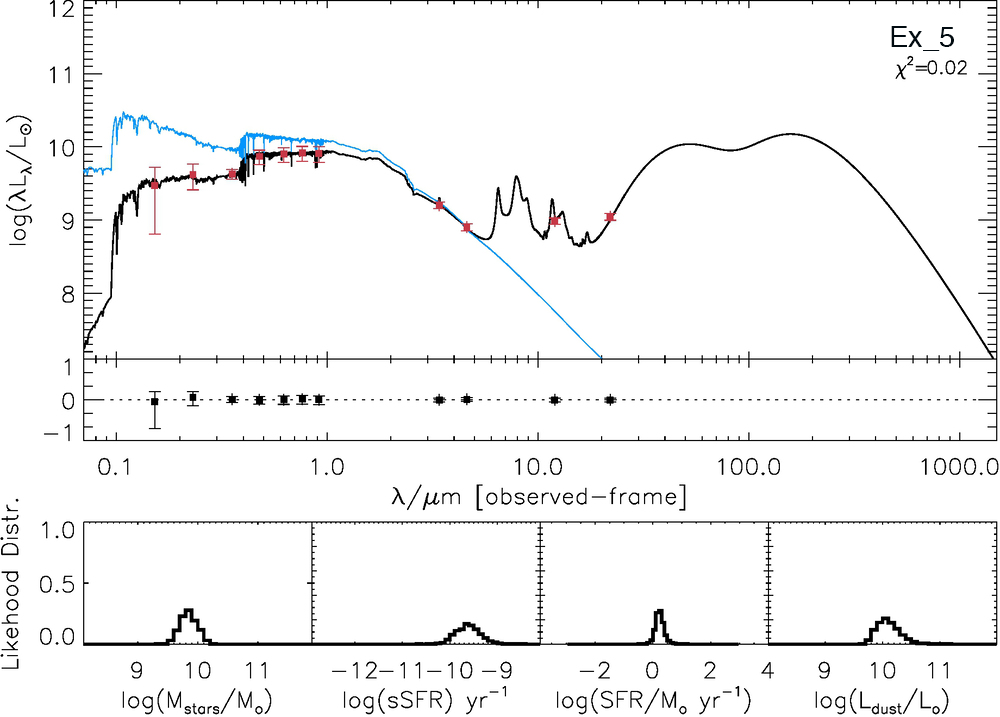}\par
      \includegraphics[width=0.42\textwidth,clip]{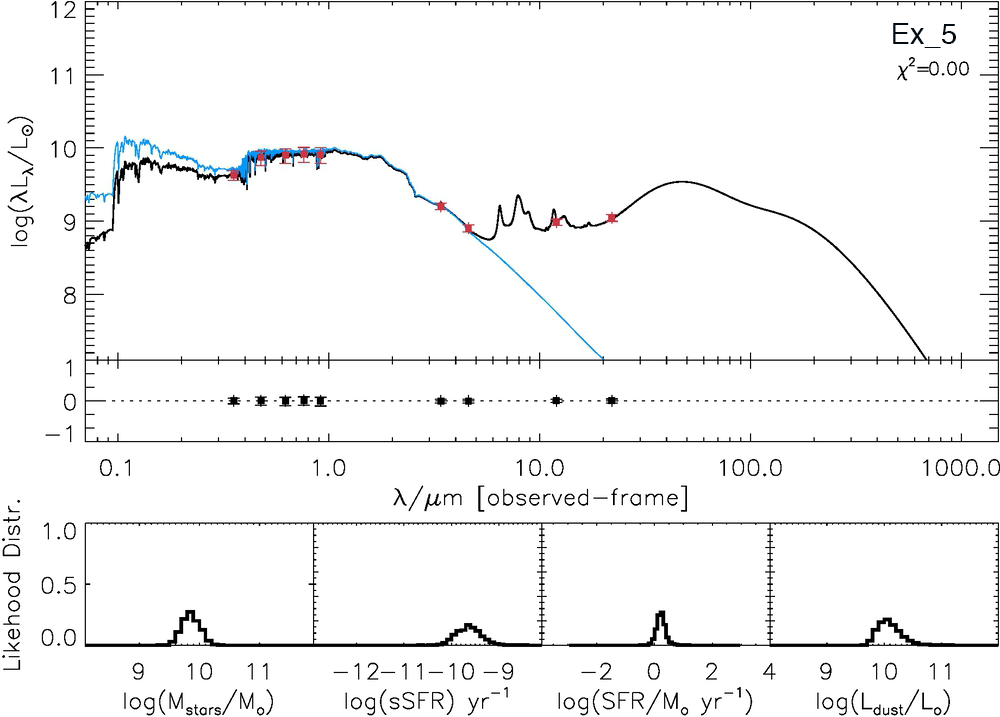}\par
  \end{center}
\end{multicols}
\begin{multicols}{2}
  \begin{center}
   \includegraphics[width=0.42\textwidth,clip]{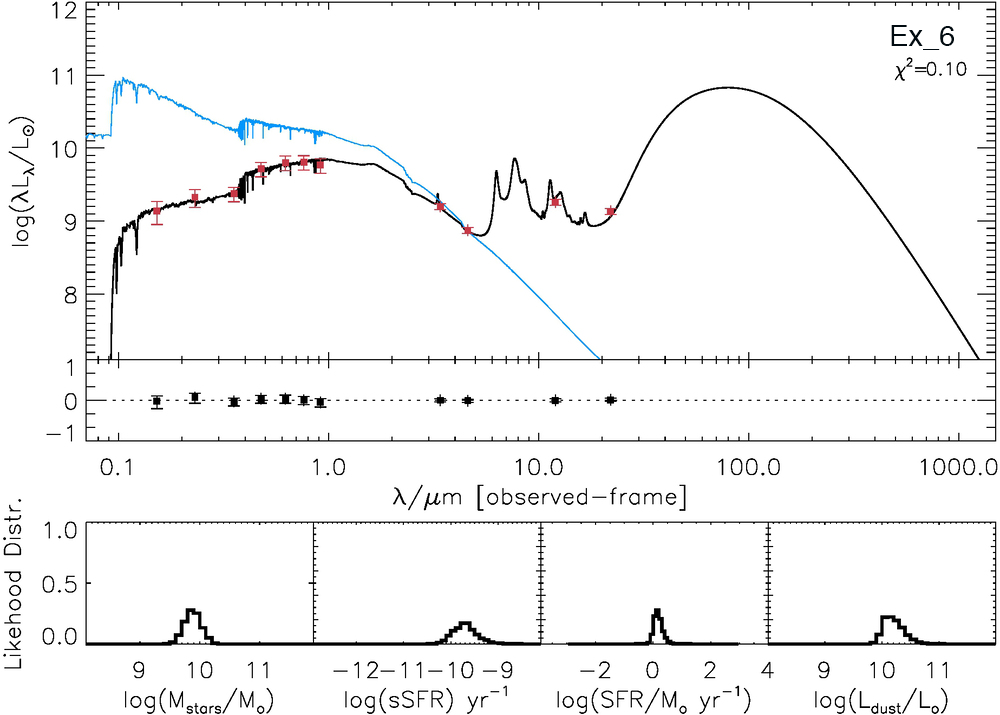}\par 
   \includegraphics[width=0.42\textwidth,clip]{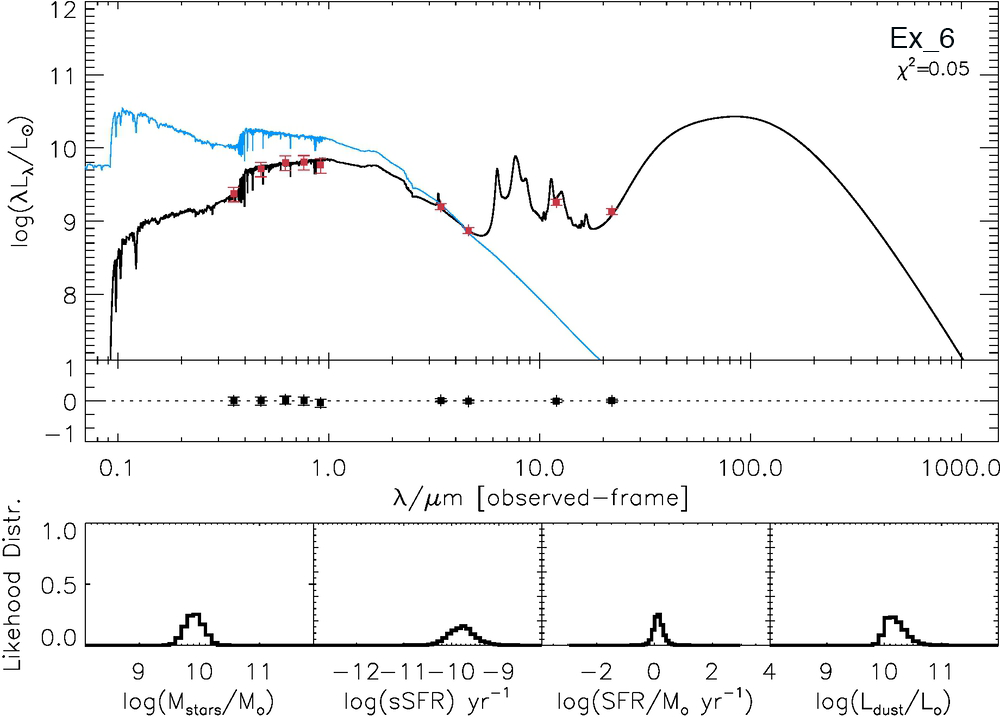}\par 
  \end{center}
    \end{multicols}
  \caption{More examples of the MAGPHYS SED fits, as explained in Fig. \ref{fig:SEDsexamples}.
  }  
\label{fig:SEDsexamples2}
\end{figure*}

Some examples of fitted SEDs are shown in Figures \ref{fig:SEDsexamples} and \ref{fig:SEDsexamples2}. 
The figures show six examples of MAGPHYS SED fit results. Each row shows the SED fits of the same 
object, using all the filters (GALEX, SDSS, and WISE) on the left and the SED fits only using SDSS and WISE  
on the right. The top panel of each fit shows the 
photometric points in red, the best-fit SED  in black, and the unattenuated SED in blue. The reduced chi-squared 
($\chi^2$) of the fit is shown in the top right corner of each main panel. The panel 
below shows the residual between the fit and the photometric points. In the lower panel of each fit, 
we show four probability distribution functions (PDFs) of some of the parameters estimated by 
MAGPHYS. From left to right, we show the PDFs of the M$_*$, the sSFR, the SFR, and the dust mass 
($\rm M_{dust}$).

\section{Stellar Masses comparison according to different parameters}

\begin{figure*}[!htb]
\begin{center}
  \includegraphics[width=0.35\textwidth,clip]{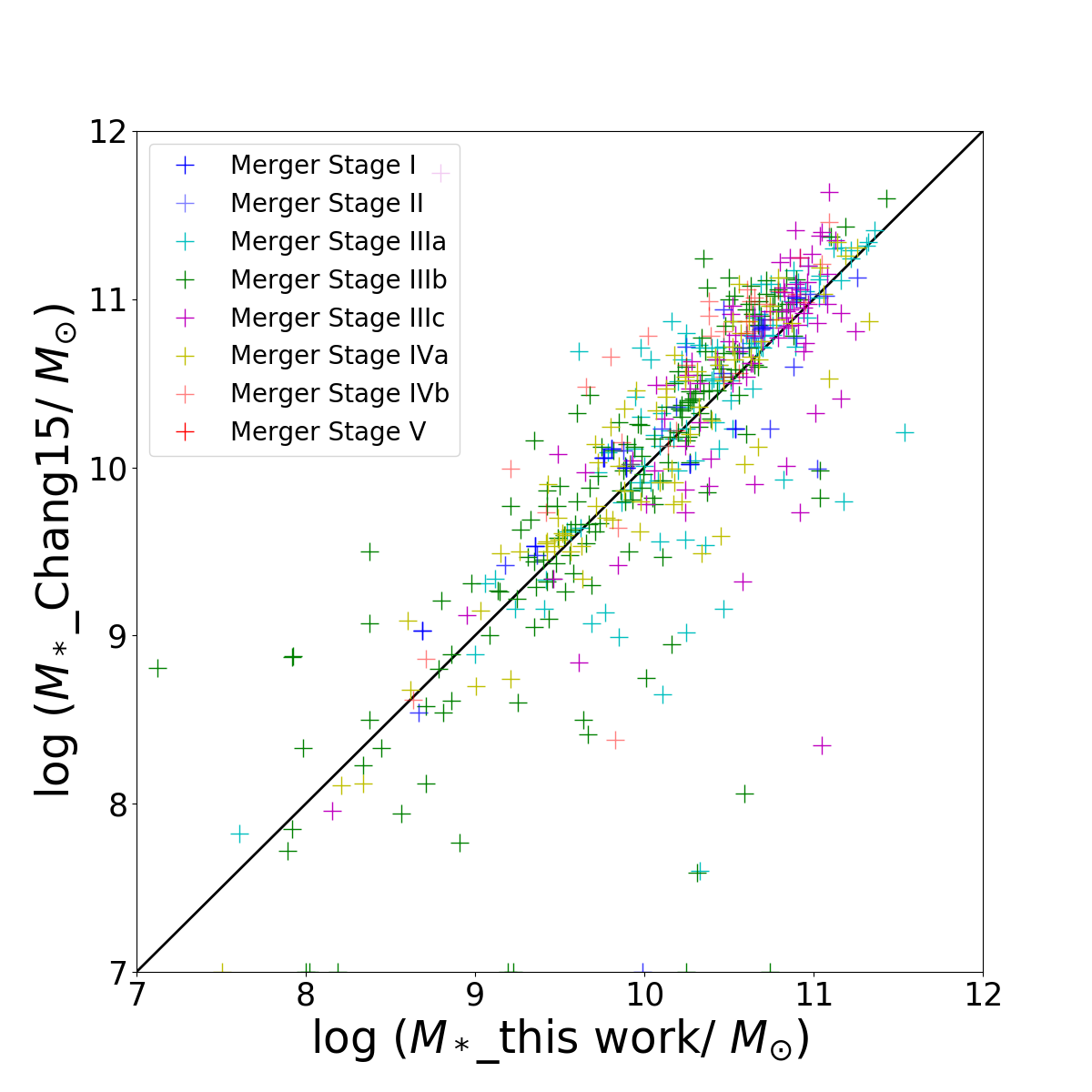}
  \includegraphics[width=0.35\textwidth,clip]{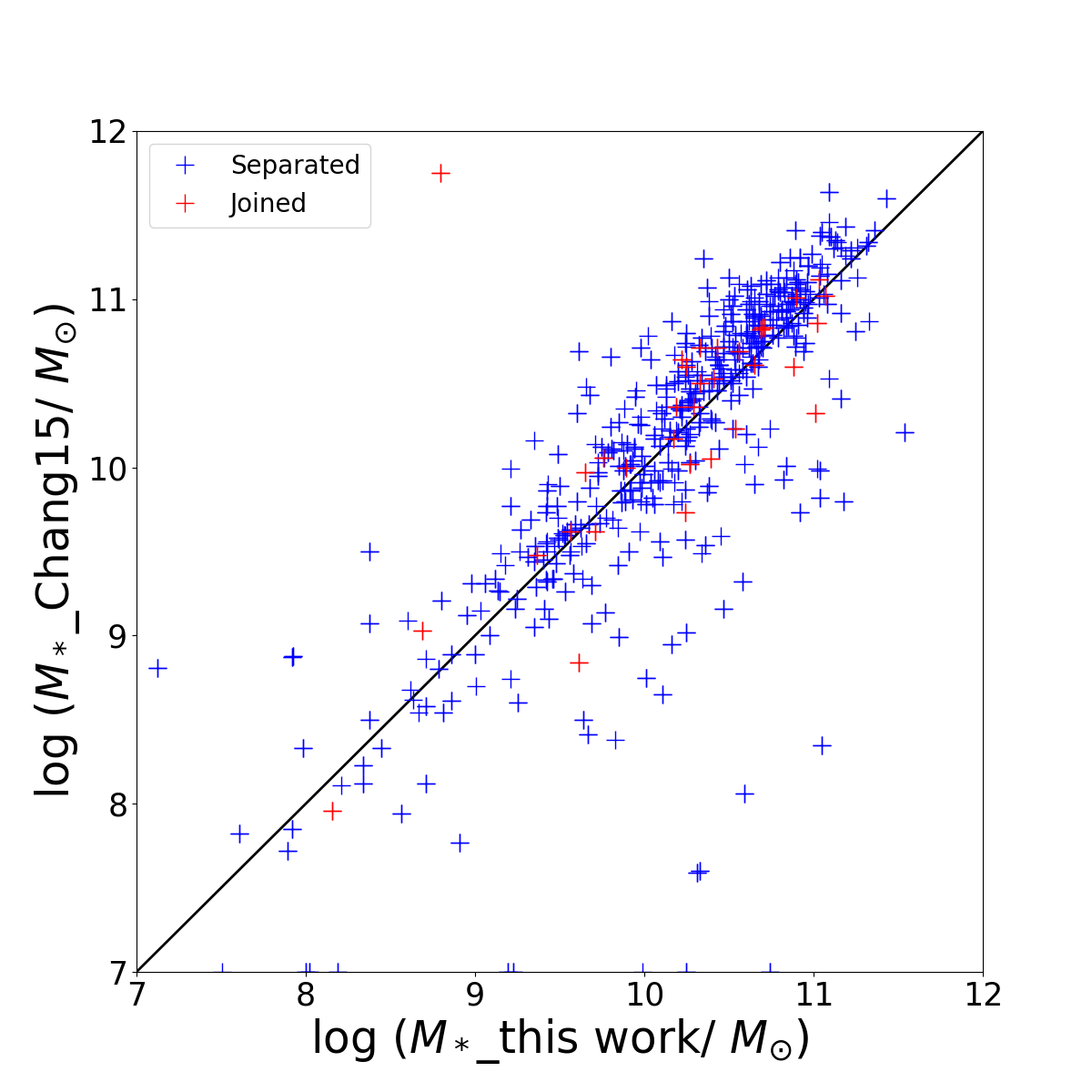}
  \includegraphics[width=0.35\textwidth,clip]{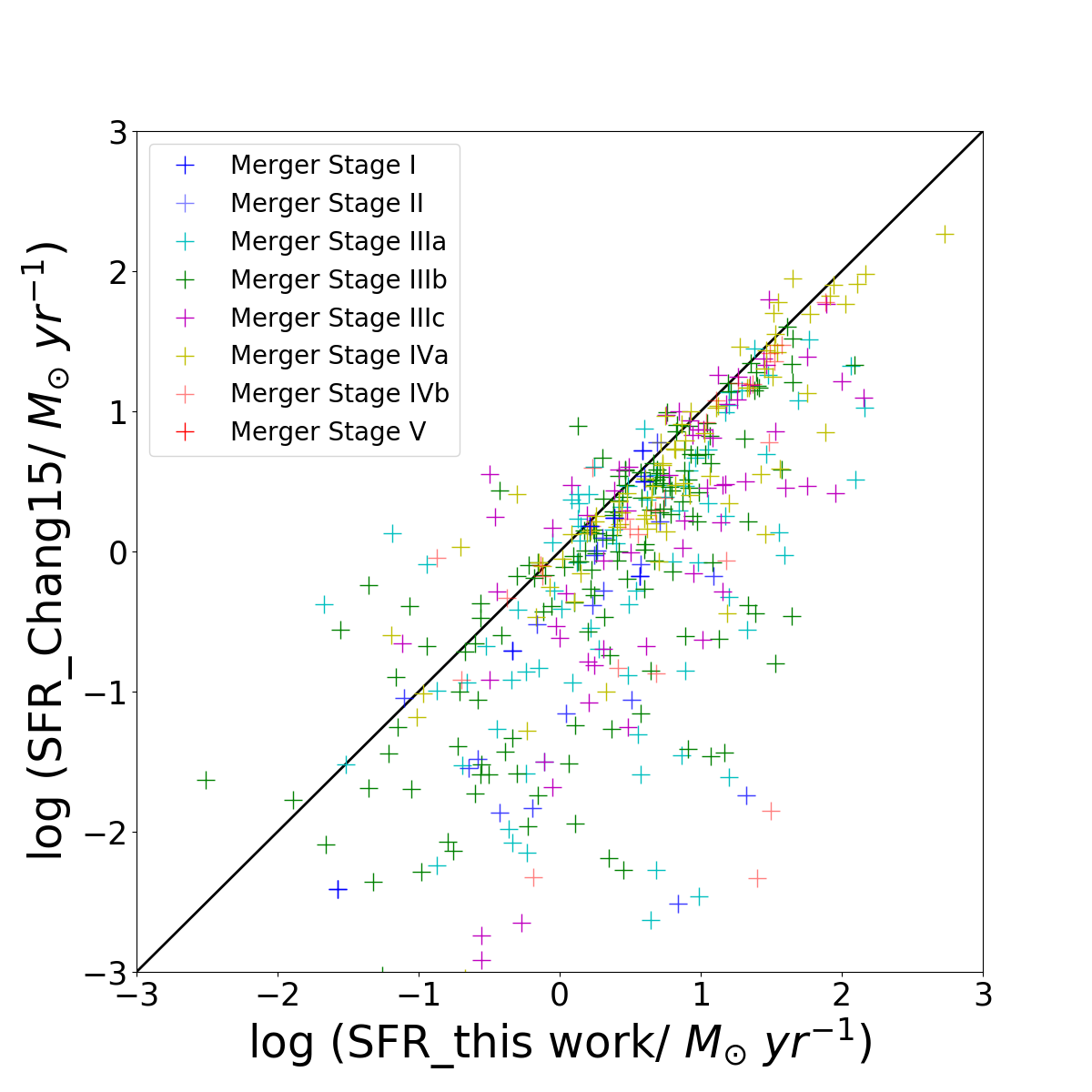}
  \includegraphics[width=0.35\textwidth,clip]{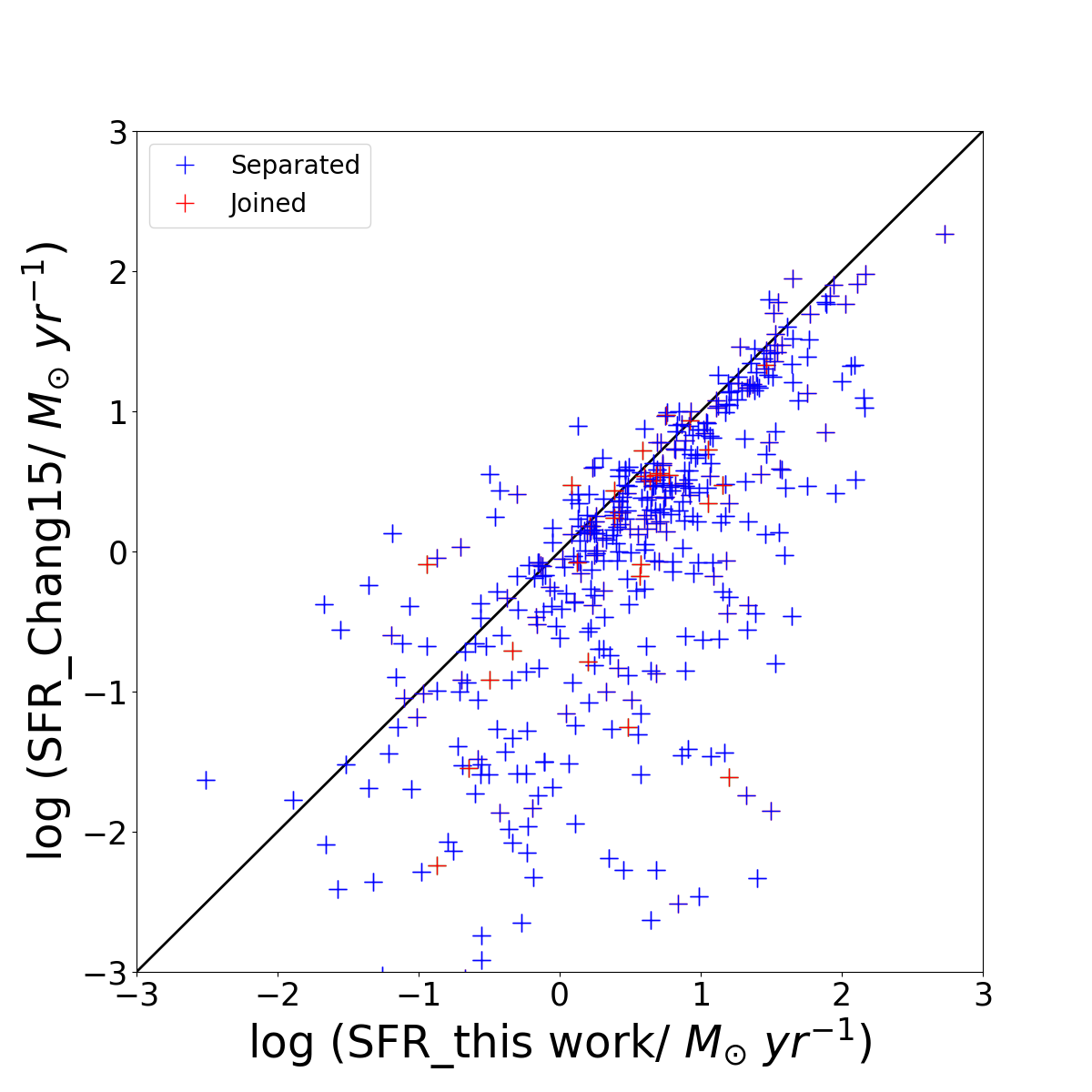}
\caption{Comparison between the M$_*$ (top panels) and SFR (bottom panels) estimated by MAGPHYS 
using the SDSS+WISE filters only and the  \textsc{Chang15} results, colour-coded according to  merger stage (left panels) 
and photometry flag (right panels).
}  
\label{fig:stellarmass_comps2}
\end{center}
\end{figure*}

We present some comparisons shown in Sect. \ref{sec:comparisons}; we colour-coded the different parameters 
to look for any dependences. Figure \ref{fig:stellarmass_comps2} shows the comparison between 
our M$_*$ (top) and SFR (bottom) results and those of  \textsc{Chang15},  colour-coded according to   
merger stage (left panel) and photometry flag (right panel). 
We can see that the correlations do not show any dependence in merger stage or 
photometry flag.

\end{document}